# Using nonlinear stochastic and deterministic (chaotic tools) to test the EMH of two Electricity Markets; the case of Italy and Greece

George P. Papaioannou [1,2], Christos Dikaiakos [1,3,†,*], Anargyros Dramountanis [3,†], Dionysios S. Georgiadis[4] and Panagiotis G. Papaioannou [5,†]

[1] Research, Technology & Development Department, Independent Power Transmission Operator (IPTO) S.A., 89 Dyrrachiou & Kifisou Str. Gr, 104 43 Athens, Greece; g.papaioannou@admie.gr

[2] Center for Research and Applications in Nonlinear Systems (CRANS), Department of Mathematics, University of Patras, Patras 26 500, Greece

[3] Department of Electrical and Computer Engineering, University of Patras, Patras 26 500, Greece; ece7777@upnet.gr

[4] Department of Management, Technology and Economics, ETH Zurich, Chair of Entrepreneurial Risks, Zurich 8092, Switzerland; E-Mail: dionysios.georgiadis@frs.ethz.ch

[5] Applied Mathematics and Physical Sciences, National Technical University of Athens, City zipcode, Greece; pgp2ntua@central.ntua.gr

† These authors contributed equally to this work.
\* Correspondence: c.dikeakos@admie.gr   Tel.: +30-210-946-6873; Fax: +30-210-519-2263

**Abstract:**

In this paper we utilize a set of nonlinear (stochastic and deterministic-chaotic) tools to characterize the state of development (developed or emerging) of the electricity markets in Italy and Greece. This is equivalent to testing the Efficient Market Hypothesis (EMH) on these markets. The tools include a variety of complexity measures like Maximal Lyapunov, $\lambda_{max}$ and Hurst, *H* exponents(estimated via R/S, DFA, GHE, AWC, PHG methods) and HHI index (Herfindahl - Hirshman Index) for market concentration and Entropy, E, a measure of uncertainty and complexity in a dynamical system, applied on the electricity wholesale marginal prices PUN and SMP of Italy and Greece. Our aim is to measure the complexity and dimensionality of the manifold on which the underlying stochastic (dynamical) system, govering the prices, evolve. We also use a $(\sigma, \lambda, H, HHI, E)$-analysis to connect the conditional volatility ($\sigma$) of prices which is a measure of the **market risk** with stability $\lambda_{max}$, a measure of the **potential market risk**, and Hurst exponent ($H$) used to investigate the properties of the fluctuations of the prices which are the "footprints" of the idiosygracies of each market. The conditional mean and volatility are modeled via ARIMA/EGARCH models fitted on SMP and PUN. Both markets are found to be inefficient ($H < 0.5$), antipersistent (mean reverting) with different speed of mean reversion. The state of development of the two markets  as well as their deviation from the benchmark $H = 0.5$ (for Brownian motion characterizing efficient markets) is monitored via a rolling weighted Generalized Hurst exponent (GHE). We calculated the correlation between Hurst Exponent and HHI and found that are negatively correlated, as expected. A causal "map" between $\sigma, \lambda, H$, $HHI$ and E is examined. We suggest possible explanations of the complex structures of the markets during their evolution stages by considering historic changes in fundamentals, macroeconomic and regulatory factors prevailing in these Countries during the entire period of analysis.

1. **Introduction and motivation**

We consider the convolution of electricity market characteristics including a) the instantaneous nature of this commodity, b) the structure of the supply function which is intrinsically steeply increasing, convex and discontinuous, as a result of the existence of diverse plant technologies in the market, c) the exercise of market power, a result of an almost demand inelastic price in the short term, participants or agents' asymmetries and incumbents behavior, d) complex market "architectures" e) significant participant learning due to the repetitive nature of the auctions as they adapt quickly their biding strategies to any regulatory changes as well as market restructurings. All the above characteristics create a complex competitive landscape, a result of their nonlinear coupled interaction. This nonlinear coupling of a large number of factors in the market creates a complex, difficult to forecast, dynamics of the spot electricity price. However, all the "information" generated by the above complexities in the electricity market is "contained" in the dynamics of the spot price or its stylized facts. The complexities "engrave» the dynamics of the spot price, or they are "encapsulated" in the dynamics of the price.

If we "see" the electricity market as a high dimensional nonlinear (stochastic) dynamical system (an evolving nonlinear manifold), all its degrees of freedom or dimensions (corresponding to the interaction at a large number of factors in the market) are "captured" in the dynamics at the spot price.

We use a number non-linear stochastic time series models of the ARIMA/GARCH family, instead of fundamental price specifications, for both the Italian and the Greek electricity markets, to capture the dynamic characteristics of conditional mean and conditional volatility of their wholesale prices. Based on the above mentioned assumptions, it would be a challenging work to try to "quantify" and compare the complexity of two electricity markets through modeling the complex dynamic evolution of their spot prices. This is the actual motivation behind this research paper, i.e. to construct the best possible models capturing the dynamics of wholesale prices of the Greek and Italian electricity markets in order to "quantify" the degree of complexity (i.e. dimensionality of the associated market "manifold"). We model the conditional volatility processes of their prices by trying a number of candidate ARMA/GARCH models, then we use forecasted conditional volatilities from the found models, corresponding to Italian and Greek markets as an input in correlation analysis and "cause-and-effect" mapping between the aforementioned variables..

We have decided not to use fundamental price models for modeling the electricity spot prices volatilities for the reason that they require a great amount of information related to economic, technical strategic and risk factors and their impacts on daily or intra-day prices as well as the dynamics of these effects over time (Karakatsani N. et al, 2008) [1]. This information although is relatively easy to find for the Greek market, is not the same for the Italian market (both time required to collect this information as well as corporate regulations, prohibiting the exchange of some kind of data, impose **unavoible** difficulties). Time series models on the other side do not need data on fundamental factors. As we have already mentioned, spot prices data "contain" all the necessary information related to the time evolution of structure of the market.

As an example of **how the structural complexity of a market is reflected by the dynamics of spot price volatility** we refer the reader to the cases of two markets, the Australian and Ontario electricity markets. These two markets are significant more volatile and spike-prone than most other markets, as this has been confirmed by various short-term price and spike forecasting studies. The reason is that these markets follow a single settlement real-term structure (a structural characteristic). For the Australian market see the works of Amjady & Keynia (2009) [2], Becker et al. (2007) [3], Christensen et al. (2012) [4]. For the Ontario Market the papers by Lei and Feng (2012) [5], Mandal et al. (2012) [6]. The link between power market structure and spot price dynamics, expressed as stylized facts is shown in the paper of Simonsen, I., et al. (2004) [7]. Nordic power (Nord Pool) spot market.

The theoretical fundamentals of chaotic nonlinear dynamic theory as well as of the nonlinear, chaotic, time series analysis, is out of scope of this paper. We refer the interested reader to Casdagli

M. and Eubank, S.(1992) **[8]**, Weigend, A. and Gershenfeld, N.A. (1993) **[9]**, Ott E. (1993) **[10]**, Abarbanel, H. (1996) **[11]**, Kantz, H and Schreiber T. (1997) **[12]**, Schuster, H. (1998) **[13]**, Strogatz, S. (2000) **[14]**. However, we provide below a very short description of the chaotic tools used in this paper to analyze SMP and PUN prices.

We investigate several interesting questions that relate to the link between **market structure** and the **behavior of electricity prices.** Because the aim of the deregulation of the electricity sectors in both Italy and Greece was to develop a competitive electricity market, which would benefit the consumers in these two countries, one might ask whether the degree of competition has increased over time during this process. Bask et al., (2009) **[15]**, answered this question finding that this is in fact the case. Another question is related to the behavior of wholesale prices i.e. whether an increase in competition affected systematically the volatility of electricity prices. Bask et al., (2011) **[16]** has found that the **volatility** of prices most often has **decreased** when the degree of competition has **increased** in the Nordic Power market.

That electricity markets generate complex nonlinear deterministic (or chaotic?) behavior and stochastic in their price volatility is further enhanced by the difficulty in out-of-sample accurate forecasting. As Weron (2006) **[17]** argues, *"…even if the forecasting accuracy is reported for the same market and the same out-of-sample test period, the errors of the individual methods are not truly comparable if different in-sample (calibration) periods are used"*. This is equivalent to say that the dynamic evolution of nonlinear high-dimensional (or multi-parametric) models are **very sensitive to initial conditions**, a well-known characteristic of chaotic systems (Serletis, 2007 **[18]**, Serletis et al., 2007 **[19]**, 2009 **[20]**, Papaioannou et al., 1995 **[21]**, 2016 **[22]**, Kantz et al., 2004 **[23]**).

It is now well demonstrated, by using the tools of nonlinear dynamics that seemingly random or irregular behavior in natural systems may arise from purely deterministic dynamics with unstable paths or trajectories. Even in appeared randomly distributed data an underlying order of pattern may exist. Dynamical systems that exhibit also sensitivity to initial conditions IC (even tiny change in IC can have tremendous effect on the future evolution of dynamics), are known as **chaotic system**. Therefore, if the electricity market as a system also processes a chaotic trait, manifested e.g. by a chaotic behavior of its wholesale price, generic stochastic models based on the theory of probability and statistics are not the appropriate and accurate ones to capture this chaotic feature of the market. As it is demonstrated by the references given above, chaos theory can reveal the intrinsic regularities of the electricity price or load, obscured by "noise" or randomness.

The aim of this work is to shed light on the mechanisms for this change of volatility of electricity prices. The underlying dynamics governing the evolution of electricity prices have change during the course of time, due to a number of internal and external factors (macroeconomic conditions, regulatory and government intervention etc.) which also have affected the structure of the market.

An innovative featured of this work, following the work of Bask, M. and Widerberg, A. (2009) **[15]**, Bask, M. (2010) **[24]**, is that we do not really need to know the dynamics of "all" fundamental variables that cover the price evolution, because by using a number of Nonlinear Time Series (NLTS) tools, we can reconstruct the dynamics using only electricity prices (see section 5.2). Actually, in this work we extent the work of Bask, M. and Widerberg, A. (2009) **[15]**, Bask, M. (2010) **[24]**, and propose an extensive ($\sigma,\lambda$,H,HHI,E)-analysis, which is a method to link invariant measures of the "market manifold" with the conditional mean and volatility of the price which evolves on this manifold. This link will provide us enough information to contrast for example the **stability** of the dynamics of each market, a measure of Potential Market Risk (Bask, M., 2010) **[24]**, (expressed via Lyapunov's largest exponent, $\lambda$) with the volatility of wholesale prices a measure of market risk (Bask, M., 2010) **[24]**. We show that the above mentioned analysis can be used as a tool to discriminate the degree of complexity of the Greek and Italian market. We examine whether an **increase in stability** is (always?) associated with a **decrease in volatility**, and if this is the case, on what degree this phenomenon is present in the two markets under investigation.

We consider that our work is useful for both the theoretical and practical reasons. First, because the "extraction" of the degree of complexity of an electricity market from the complexity of a model capturing the dynamics and volatility of its spot price introduces an innovative way of thinking in quantifying a market's organization or architectural complexity, and secondly because our work contributes towards a better understanding of the two electricity markets, Italian and Greek, a significant prerequisite especially nowadays that the two markets are about to be coupled (it is expected to take place in 2017). We think that having a good "picture" of the complexity of the two markets makes the work needed to prepare and implement the coupling more efficient and effective.

**2. A short literature review on Nonlinear Dynamical and Stochastic Systems**

There are ambiguous views among researchers about the nature of electricity load considered by some as random but also as a **chaotic process** due to the influence of many complicated factors such as weather variable, price of electricity etc. The reader is referred to the papers Choi, J-G et al. (1996) [25], Drezga, I. et al. (1999) [26], Hyndman, R., et al. (2008) [27], Kristoufek, L. (2009) [28], Liao, G-C (2006) [29], Wu, W. et al. (2004) [30]. The price, as well, depends on supply and demand of the market and the operating conditions of the transmission network that is affected by a large number of factors as weather, macroeconomic situation, planning for network expansion, power plant outages – failures and accidents – etc. (Wolak, F (1997) [31], Weron, R. et al. (2008) [32], Liao, G-C (2006)[29]). Therefore the complexity in load and price dynamics is the product of the joint effect (nonlinear interaction) of all mentioned factors. In order to analyze and forecast electricity prices and loads, the mainstream is to employ statistical methods.

In this paper we have tested on our data, a number of ARMAX/GARCH type models of the statistical cluster in order to model the conditional volatility of clearance prices of both markets that are required as inputs by other models used in this work to characterize the market. In order our work to be as self-contained as possible, regarding literature review, we provide below a short description of the above types of models, used here.

The necessary introductions and theoretical background of Autoregressive Moving Average, ARMA(p,q), Autoregressive Integrated Moving Average (ARIMA) or Box-Jenkins, and the Seasonal ARIMA (SARIMA) models are provided in Brockwell and Davis (1996) [33], Ljung (1999) [34], Shumway and Stoffer (2006) [35], Makridakis et al. (1998) [36] and an almost new, open access e-book by Hyndman and Athanasopoulos (2013) [37].

An excellent and updated literature review on electricity price forecasting is presented by Weron 2014 [38] and we strongly refer the reader to this work for a thorough review of the available methodologies. According to Weron 2014 [38], the electricity price models are categorized in five clusters, multi-agent, fundamental, reduced-form, statistical and computational intelligence, each one having its own sub-clusters.

We also proposed the application of an alternative entropy measure for assessing volatility in electricity markets. This measure of assessing non-Gaussian fluctuations in the return dynamics of financial market has been applied in the works of Bose, R and Hamacher, K. (2012) [39], and Pincus S. and Kalman, S. (2004) [40].

Although the concept of entropy is originated from physics, its relevant principles as maximum entropy and minimum entropy have been extensively applied in finance. For an update review of the applications of Entropy in finance, the reader is refered to the work of Zhou R.et al (2013) [41], well as in Zhou R. et al (2015) [42], Ormos M. and Zibriczky D. (2014) [43], Gradojevic and Gencay R. (2011) [44].

There are several measures of entropy: Shannon entropy, Renyi entropy, Tsallis entropy, Kullback Cross-entropy, Thallis relative entropy, fuzzy entropy etc. The main difference between Shannon, Renyi and Tsallis entropy is that both Renyi and Tsallis are adequate for anomalous systems while Shannon is best for equilibrium systems. Shannon (information) entropy (SE) is a possible form of entropy which is associated with tim-dependent propability distribution. However SE does not account for long-range interactions (i.e., it is extensive or additive), therefore it is better to utilize its generalized form, the non extensivity form of entropy proposed by Tsallis, the so-called Tsallis entropy TE (Tsallis C., 1989 [45], Tsallis C., 2009 [46]). In addition to its long memory characteristic, TE represents a powerful tool for the analysis of complex, nonlinear and nonstationary signals, therefore we adopt this measure of entropy in this study. The connection of ARCH and GARCH models, used extensively in Finance, with the nonextensive entropy of Tsallis justifies further our decision to adopt this kind of measure (Queiros S. and Tsallis C., 2005 [47]).

As far as the use of some kind of entropy applied to electricity markets is concerned, we refer the reader to three recent works, Perello J. et al. (2006) [48], Amjady N. et al. (2009) [49] and Ruiz M. et al (2012) [50].

The structure of this paper is as follows: Section 3 includes a detailed description of the Greek and the Italian electricity markets with emphasis on their structure regarding their integration towards the Target Model. This structure is inherent in the evolution of the system's national spot prices (SMP for the Greek Electricity Market and PUN for the Italian one) so modeling spot prices will enhance our insight into their structure. Section 4 includes the data description and performs all the necessary adjustment procedures on our timeseries for further processing. Section 5 is the core section of this work providing all necessary information on the various tools used. It encompasses the mathematical formulation of the proposed models and describes how they are applied in this study as well as the results. Section 6 (Conclusions) summarizes and proposes the next steps for further research.

**3. The Greek and Italian wholesale electricity markets in brief.**

In order to better assess the complexity of the two wholesale electricity markets under consideration, the Greek and the Italian one, we must first define their different operational principles. This will also be helpful in interpreting the dynamics of SMP and PUN prices and hence to enable us understand the structure of their respective electricity markets. Modelling aspects of Greek market and its intraction with the Italian one can be found in Papaioannou, G. et al (2015) [51].

*Greece's liberalized electricity market* was established according to the European Directive 96/92/EC and consists of two separate markets:
1. the Wholesale Energy and Ancillary Services Market and
2. the Capacity Assurance Market

The Greek wholesale electricity market (GEM) is currently in a transitional period, during which the market structure evolves towards its final design, namely the Target Model. The wholesale electricity market is a day ahead mandatory pool which is subject to inter-zonal transmission constraints, unit technical constraints, reserve requirements, the interconnection Net Transfer Capacities (NTCs) and in general all system constraints. More specifically, based on forecasted demand, generators' offers, suppliers' bids, power stations' availabilities, unpriced or must-run production (e.g., hydro power mandatory generation, cogeneration and RES outputs), schedules for interconnection as well as a number of transmission system's and power station's technical constraints, an optimization process is followed in order to dispatch the power plant with the lower cost, both for energy and ancillary services.

LAGIE (the independent market operator) ([www.lagie.gr](www.lagie.gr)) is responsible for the solution of the so-called Day Ahead (optimization) problem. This problem is formulated as a security constrained unit commitment problem, and its solution is considered to be the optimum state of the system at which the social welfare is maximized for all 24 h of the next day simultaneously. This is possible through matching the energy to be absorbed with the energy injected into the system, i.e., matching supply and demand (according to each unit's separate offers). The DA solution, therefore, determines the way of operation of each unit for each hour (dispatch period) of the dispatch day as well as the clearing price of the DA market's components (energy and reserves).

More specifically in this pool, market "agents" participating in the Energy component of the day-ahead (DA) market submit offers (bids) on a daily basis. Producers and importers submit energy offers with the limitation that the weighted average of the offer should be above the unit

Minimum Average Variable Cost. On the contrary exporters and load representatives submit load declarations. The bids are in the form of a 10-step stepwise monotonically increasing (decreasing) function of pairs of prices (€/MWh) and quantities (MWh) for each of the 24 h period of the next day. A single price and quantity pair for each category of reserve energy (primary, secondary and tertiary) is also submitted by generators. Deadline for offer submission is at 12.00 pm ("gate" closure time).

So, the DAS solution produces a 24 hour unit schedule and a unique price which is called the System's Marginal Price (SMP). The Dispatch Scheduling (DS) is used to define the time period between Day Ahead Schedule (DAS) and Real Time Dispatch (RTD) where the producers have the chance to change their declarations whenever has been a problem regarding the availability of their units. In the RTD the units are re-dispatched in real time in order to meet the actual demand. Finally in the IS stage an Ex Post Imbalance Pricing (EXPIP) is produced after the dispatch day which is based on the actual demand and unit availability. The capacity assurance market is a procedure where each load representative is assigned a capacity adequacy obligation and each producer issues capacity availability tickets for its net capacity. Actually this mechanism is facing any adequacies in capacity and is in place for the partial recovery of capital costs. The most expensive unit dispatched determines the uniform pricing in the day-ahead market. In case of congestion problems and as a motive for driving new capacity investment, zonal pricing is a solution, but at the moment this approach has not been activated. Physical delivery transactions are bounded within the pool although market agents may be entering into bilateral financial contracts that are not currently in existence. The offers of the generators are capped by an upper price level of 150€/MWh. Physical Transmission Rights (PTR) are explicitly allocated via auctions.

The monitoring of the GEM is performed by the Greek National Regulatory Authority (RAE). At the moment there is no intra-day market, balancing market and there are no any financial products. But as we have already mentioned the Greek Electricity Market (GEM) is in transition towards the Target Model. According to the Greek laws the GEM should be compliant with the Target Model starting from the 1st of January 2017. The former law states that the operation of the intra-day market as well as the publication of any financial products (forwards, futures) will be performed by LAGIE whereas the Balancing Market will be operated by the Independent Power Transmission Operator (IPTO) which is the Greek Transmission System Operator, ADMIE S.A. (www.admie.gr).

The dynamics of GEM including regulatory reforms are inherent in the evolution of SMP. The latter is highly correlated to Greece's energy mix where the dominant fuel used in the electricity sector is lignite. Apart from lignite generation units, the energy mix includes gas-fired units, large hydro plants, oil units, and Renewable Energy Sources such as Wind Parks, small Hydro, photovoltaic and biomass, as well as small cogeneration units. Even though capacity requirements spurred by environmental constraints were likely to reduce the dominance of lignite in the fuel mix, its share has actually increased since 2011 due to the high gas prices and low carbon prices. In figure 1 we display the annual shares of fuels in the generation mix and net imports for the year 2014, while in Table 1 is shown the evolution of Greece's energy mix for the time period 2005-2013.

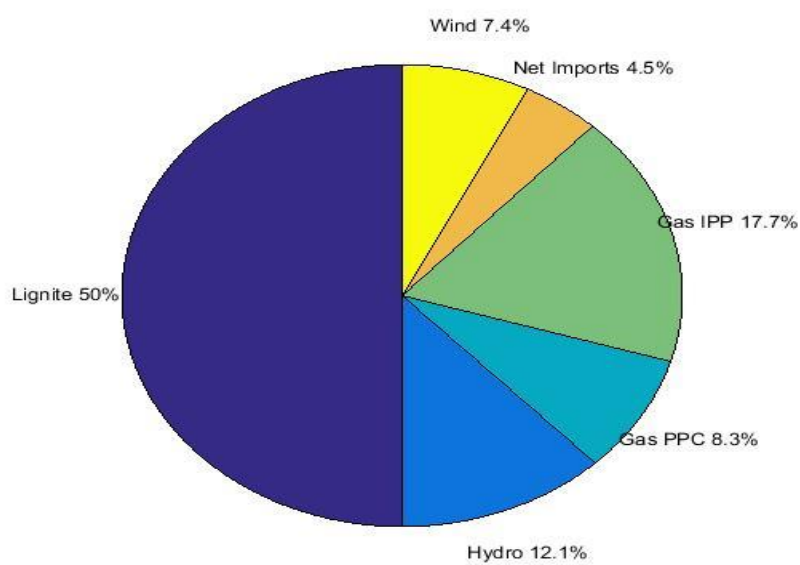

**Figure 1:** Greek Annual Share of Fuels and Net Imports for 2014 [52] (2014 National Report to the European Commission by Regulatory Authority for Energy in Greece). PPC stands for Public Power Corporation and IPP for Independent Power Producers.

**Table 1 :** Greece's Fuel Mix Generation (TWh) 2005-2013

|  | 2005 | 2006 | 2007 | 2008 | 2009 | 2010 | 2011 | 2012 | 2013 |
|---|---|---|---|---|---|---|---|---|---|
| **Lignite** | 32.050 | 29.160 | 31.090 | 29.870 | 30.540 | 27.440 | 27.570 | 29.280 | 23.470 |
| **Fuel Oil** | 3.300 | 3.300 | 3.260 | 3.510 | 1.690 | 0.110 | 0.009 | 0.064 | 0.000 |
| **Natural Gas** | 7.940 | 10.160 | 13.210 | 13.330 | 9.380 | 10.360 | 14.850 | 12.670 | 12.170 |
| **Large Hydro** | 5.420 | 6.230 | 3.140 | 2.970 | 4.950 | 6.700 | 3.680 | 3.540 | 5.440 |
| **RES** | 0.890 | 1.130 | 1.130 | 1.560 | 1.880 | 2.040 | 2.530 | 3.310 | 7.180 |
| **Total Net Generation** | 49.620 | 50.000 | 52.020 | 51.250 | 48.450 | 46.600 | 48.600 | 48.870 | 48.290 |
| **Net Imports** | 3.780 | 4.200 | 4.350 | 5.610 | 4.360 | 5.700 | 3.230 | 1.630 | 1.720 |

The *Italian wholesale electricity market*, commonly called the Italian Power Exchange (IPEX), is run by the Gestore del Mercato Elettrico (GME). GME takes care of the transactions in the day-ahead market and in the intra-day market whilst reserves and balancing markets are under the responsibility of the system transmission operator (TERNA). GME's day-ahead market is based on the economic merit-order criterion in order to construct the supply and demand curves. If there is a violation of the transmission constraints, the market is split into different zones. Producers then receive the zonal prices, whereas buyers pay the Prezzo Unico Nazionale (PUN) which results as an average of the zonal prices weighted by the zonal consumption levels.

The Italian Electricity Market consists of:
  ➢ The Spot Electricity Market (MPE) which includes:
- The Day-Ahead Market (MGP), where all market participants (producers, wholesalers, consumers) trade energy blocks for the next day. The market "players" submit offers/bids specifying both the quantity and the minimum/maximum price at which they are intending to sell/purchase. In general MGP is an auction market and the accepted bids are valued at the "Prezzo Unico Nationale" or PUN which is the national single price calculated as the average of the prices of geographical zones.

- The Intra-Day Market (MI) where the initial schedules can be modified by the market participants. The intraday market takes place in five stages: MI1, MI2, MI3, MI4 and MI5 which take place on different hours before the day of delivery and close also different hours of the day of delivery. Unlike in the MGP the accepted bids are valued at the zonal price.
- The Ancillary Services Market (MSD), which consists of the ex-ante MSD which is a scheduling substage of MSD and of the Balancing Market (MB). In the ex-ante MSD, Terna S.p.A which is the Italian Transmission System Operator (www.terna.it) procures all the necessary resources to the GME required for the creation of energy reserves, the relief of congestion incidents, real-time balancing, etc. Terna accepts energy demand bids and offers resulted from the ex-ante MSD.
  - The Forward Electricity Market (MTE), which is the venue where market participants can sell/purchase future electricity supplies via forward electricity contracts. MTE takes place on a continuous basis and the tradable contracts are of the following types: Base-load or Peak-load, delivered in an one month, four months or twelve (yearly) months period of time.
  - The platform for physical delivery of financial contracts concluded on IDEX (CDE). IDEX is the segment of the financial derivatives market operated by Borsa Italiana S.p.A where the electricity financial derivatives are traded.

In figure 2 we display the annual shares of fuels in the generation mix and net imports in Italy for the year 2014. It must be noted that the main source of electric power is the thermoelectric generation from fossil fuels (natural gas, coal, oil and other). As electricity demand grows, there is a respective increase in Italy's energy mix from renewable energy sources such as photovoltaic, wind energy and biomasses. The former conclusion is also enchanced by the evolution of the fuel mix in Italy during the period of interest 2005-2013 shown in Table 2. We must also note that there is a structural break (Di Cosmo, V., 2015 [53]) at the beginning of 2009, specifically it occurred on February 12, 2009. Before the structural break, Brent oil was the most significant fuel "driver" in determining PUN, which natural gas became the fuel that sets PUN oftr this date. The large investments in renewables have substantially a negative impact reduction on PUN spot price.

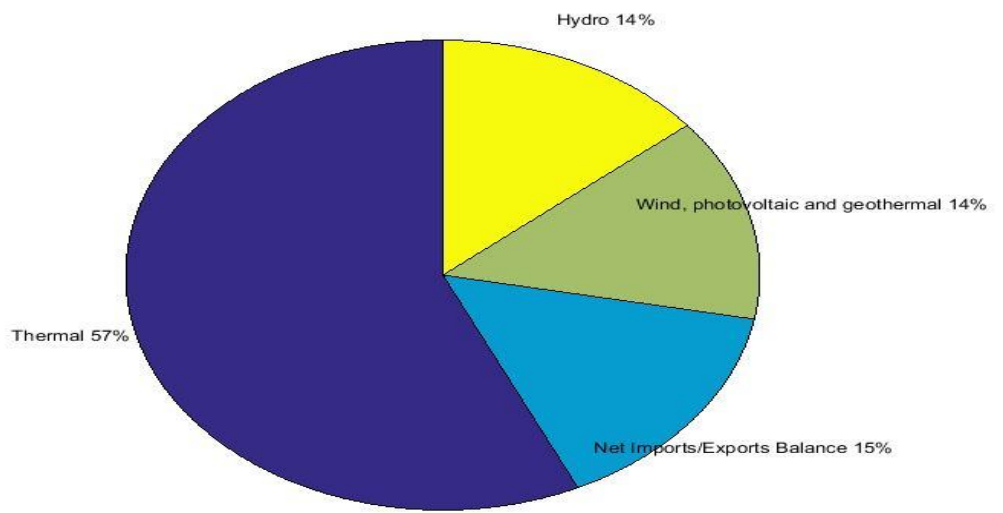

**Figure 2:** Italian Electricity net production per energy source and imports in 2014 [54-55]

**Table 2** : Italy's Fuel Mix Generation (TWh) 2005-2013

|  | 2005 | 2006 | 2007 | 2008 | 2009 | 2010 | 2011 | 2012 | 2013 |
|---|---|---|---|---|---|---|---|---|---|
| **Natural gas** | 149.3 | 158.1 | 172.6 | 172.7 | 147.3 | 153.8 | 139.87 | 123.75 | 109.400 |
| **Hydroelectric** | 42.9 | 43.4 | 38.4 | 47.2 | 53.4 | 53.7 | 47.20 | 43.25 | 54.070 |
| **Wind and photovoltaic** | 2.3 | 3.0 | 4.1 | 5.1 | 7.2 | 9.1 | 8.66 | 11.07 | 13.620 |
| **Other RES** | 12.5 | 13.8 | 15.1 | 16.5 | 20.2 | 23.8 | 26.31 | 37.86 | 42.144 |
| **Net Imports** | 49.2 | 45.0 | 46.3 | 40.0 | 45.0 | 43.9 | 45.73 | 43.10 | 42.137 |
| **Net Production** | 290.6 | 301.2 | 301.2 | 307.1 | 288.7 | 290.7 | 291.40 | 287.80 | 278.800 |

One of the most characteristic attributes of the Italian energy market is that its installed capacity is oversized compared to the national electricity demand. Electricity demand is still downward in 2013 and 2014 due to the economic ressesion in Italy and also some of these surplus power plants are old, low productive or even unused. However, this surplus in installed capacity does not guarantee Italy's energetic selfsufficiency since Italy's net imports every year are about 15% of its need in electricity.

More specifically the Italian transmission grid is interconnected to foreign power systems via 22 interconnected lines:
- Switzerland (twelve)
- France (four)
- Slovenia (two)
- Greece (one direct cable)
- Austria (one)
- Between Sardinia and Corsica

Further interconnection lines are being developed with France, Austria, Slovenia, Montenegro, Algeria and Tunisia. Italy imported about 13.2% of the electricity it used in 2012 which made Italy one of the largest importers of electric energy in the EU for this year.

Most electricity is generated in thermoelectric power plants using fossil fuels and a number of different technologies; steam plants, Combined Cycle Gas Turbine (CCGT) plants, and gas turbines. Thermoelectric generation (gas, oil and coal) decreased from overall 83% in 2007 to 62% in 2013. Of fossil fuels natural gas is the most commonly used (38% equal to 109.4 TWh), followed by coal (16% equal to 46.08 TWh) and other oil products (8% equal to 23.04 TWh).

Regarding the RES sector in Italy, after the introduction of the EU Energy Directives (2003/54/EC & 2009/72/EC), there was a huge rise in investments in this sector. This occurred mnainly due to a series of incentives promoted by the Italian state for the installation of domestic photovoltaics systems. During 2013 Italy produced 109.834 TWh of electricity from RES (including also the hydro production) which corresponds to 38.16% of the net production in this year.

**The Target Model and coupling of Greek and Italian markets.**

The European Target Model (TM) for day-ahead market visualizes a single European price coupling all over the continent. Cross-border capacities at the day-ahead stage are implicitly allocated via market coupling which means that the related to TM coupling algorithm should secure such commercial cross-border day-ahead exchange programs, and enables the right direction of Power flows, from the lower-to the higher-price zone (i.e. the "wrong-way flows" should diminished to zero).

The TM is designed to provide a single market rather a single price across Europe in all periods, although TM should deliver greater price convergence than the current structures. Indeed, the TM is structured around the concept that differences in zonal prices provide important locational signals for the operation of the investment in demand, generation and networks. It is expected that one single clearing price in the Day Ahead Market will apply in each Bidding zone. At the heart of the TM is the concept of price coupling, both at the Day Ahead and the intraday stages. Price coupling is a form of implicit auction, which means that available interconnection capacity and energy flows are effectively traded together. Price coupling is based on a single algorithm that uses bid/offer information from each zone and the available cross border capacities. The algorithm jointly establishes prices, generation volumes and interconnector flows for each coupled market taking into consideration all bids/offers from all markets.

Electricity Target Model aims to facilitate cross-border trade, a shared vision to improve the integration of the electricity market of Member States. In brief, ETM envisages a single European Price Coupling of the DA timeframe which, eventually, will replace explicitly auctions of cross-border capacity, a single continuous trading system in the intraday timeframe, a single European System for allocating and scheduling long-term transmission rights and finally a flow-based approach in extensively meshed networks.

Electricity Target Model therefore is the driver for the convergence of wholesale prices of different regions, defined according to Annex I of Regulation (EC) No 714/2009 [56] Greece and Italy belong to the Central South East, CSE, region, together with Slovenia and Switzerland.

Regarding the options for the Greek Wholesale electricity market to be compliant with the European Target Model (TM), a recent study has indicated electricity that the target should be that all borders of Greece will participate in one common market coupling. At this stage, however, it would be most likely the Greece-Italian Interconnection would participate in a price coupled mechanism while the others still will be based on explicit options. When the other interconnections are ready they will be added. Italy is the market that resembles the most to the Target Model, while the markets in the other Countries exhibit substantial gaps with TM. Hence, for Greece the obvious and easiest market coupling will be with Italy.

Italy is the most mature market and fully compliant with EU's TM. On February 2015 the Italian Border Market coupling was successfully launched. The Italian-Austrian, Italian-French and Italian-Slovenian Borders have been coupled with the Multi-Regional Coupling (MRC). This coupling provides evidence of the flexibility and reliability of the price coupling of Regions (PCR) solution[1].

Historically, because of the differential in wholesale electricity prices in the area, there are imports of electricity from adjacent countries north of Greece (mainly Bulgaria) and exports of electricity to Italy. Our view is that imports from Italy to Greece will continue to take place as long as photovoltaic in the south of Italy suppress prices close to zero, but gradually will be constrained to only critical situations. Gradually, a price convergence is expected to reduce the current profit margin of cross-border trading with Balkan Countries.

**4. Data sets Description and preparation**

We have collected data on both the Italian and Greek electricity markets. For the Italian market we have the hourly day-ahead system marginal price, so-called Prezzo Unico Nazionale, or PUN, quoted and publicly available on the Italian Power Exchange IPEX, website (www.mercatoelettrico.org). For the Greek market the hourly spot, day-ahead System wide Marginal Price, SMP. The data are available from the official site of the Greek Independent Power

---

[1] Price Coupling of Regions (PCR) is the initiative of seven European Power Exchanges (APX, Belpex, EPEX SPOT, GME, Nord Pool Spot, OMIE and OTE), to develop a single price coupling solution to be used to calculate electricity prices across Europe and allocate cross-border capacity on a day-ahead basis. This is crucial to achieve the overall EU target of a harmonized European electricity market. The integrated European electricity market is expected to increase liquidity, efficiency and social welfare. PCR is open to other European Power Exchanges wishing to join. Today, PCR is used to couple the Multi-Regional Coupling, covering 85% of European consumption, and the 4M Market Coupling between the markets of Czech Republic, Hungary, Romania and Slovakia.

Transmission Operator, IPTO, ADMIE S.A. (www.admie.gr). The study uses daily values of the SMP by taking the average of the 24 hourly prices of each day. The data span over the course of nine years, from January 1st 2005 to December 31 2013, consisting by 3287 measurements for each market.

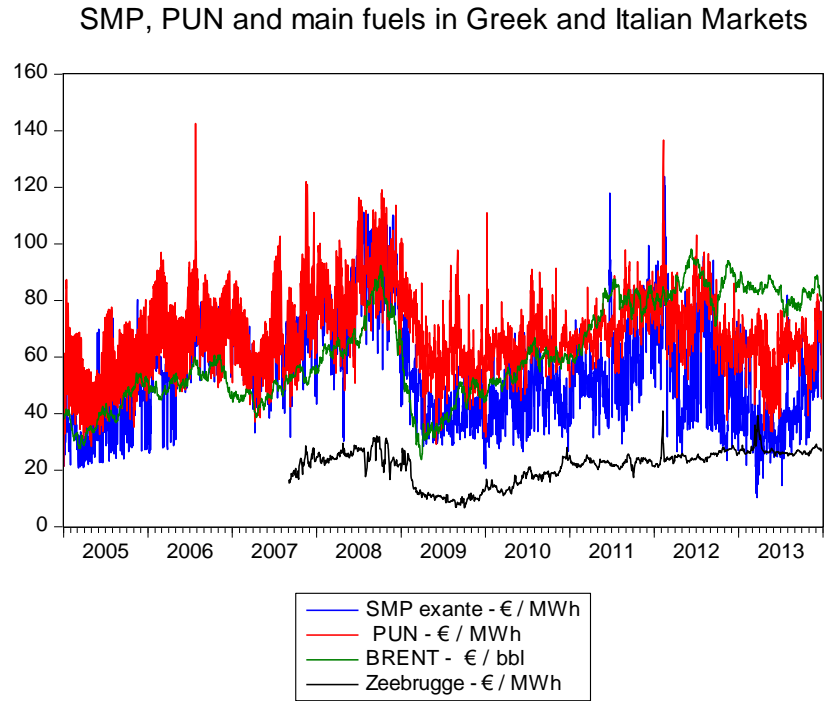

**Figure 3:** SMP, PUN daily electricity wholesale prices and the main fuels (Brent Oil and Natural Gas) used in generation, 2005-2013

**Figure 3** shows the co-movement of the two wholesale prices SMP and PUN, with the Brent Oil and Natural Gas (Zeebrugeer, ZEE hub). The units of measurements are given in the legend. For the period of 2005-2013 the mean value of Brent Oil, ZEE gas prices are 61.6 Euro/bbl, and 21.0 Euro/Mwh respectively. The estimated correlation coefficient between Brent Oil and Natural Gas ZEE, prices with SMP are 0.18 and 0.30, respectively while with PUN are 0.22, 0.34 respectively. So, both electricity prices are partially driven by Brent Oil and Nat Gas.

**Table 3:** Herfindahl-Hirschman Index (HHI) in Greek and Italian Power Generation 2005-2014 as appeared in the National reports of Regulators sent to EC.

| Year | HHI (Greek Market) | HHI (Italian Market) |
| --- | --- | --- |
| **2005** | 9409 | 1900 |
| **2006** | 8905 | 1643 |
| **2007** | 8390 | 1440 |
| **2008** | 8390 | 1380 |
| **2009** | 8427 | 1280 |
| **2010** | 6844 | 1097 |
| **2011** | 5746 | 953 |
| **2012** | 6983 | 884 |
| **2013** | 6553 | 830 |
| **2014** | 8091 | 908 |

**Table 3** provides information on the 'degree" of competition in the generation "market" in Greek and Italian Markets measured by the HHI index. HHI stands for Hefindahl-Hishman index, a measure of market concentration (United States Department of Justice, link: http://www.justice.gov/atr/public/guidelines/hhi.html). If HHI=10000 (upper limit) the market is considered a monopoly, while for HHI>5000 the market is over-concentrated, for H>1800 concentrated, for 1000<HHI<1800 efficiently or moderately competitive and finaly for HHI ≤1000 the market is low concentrated or competitive. The Index is calculated as follows:

$$HHI_n = 10000 \times \sum_{i=1}^{n} s_i^2 (\%) \qquad (1)$$

where $s_i$ is the market share of generator $i^2$.

According to **Table 3**, we can see that concentration levels for Italy and Greece have decreased, but the Italian market at the start of the period of analysis 2005, was a medium concentration market, while the Greek market was close to the monopoly top limit. However, the Greek market, even with a very slow rate, gradually becomes more and more less concentrated with the lowest HHI index value in 2011. However, it still remains a highly concentrated generation market. In the opposite direction, the Italian market decreased the level of concentration very fast and since 2007 went beyond the 1500 threshold. The reason for this fast moving can be attributed to various causes: 1) During 2001-2002, ENEL sold out 1500 MW, while at the same time ENEL's competitors increase their investment for installed power capacity, very intensively in the last years, causing a rise in green energy consumption. 2) Recent EU Energy policies, enhanced the energy purchasing from green energy generators. 3) The steady increase in the number of marginal generators (producers) which invested in wind and solar technologies, transforming the Italian generation "market" to the most competitive one.

**Figure 4** provides information on the average hourly profile, for SMP, PUN and GREC, for the period 2005-2014. GRECDA is the hourly day-ahead price for the Italian zone that is connected to Greece. It is interesting to mention here that while the hourly dynamic variation, within a day, is the same for all the time series (they exhibit peaks and troughs on the same hours), the Greek average hourly price, SMP, is lower than the other two over the entire period of examination, exhibiting as well a more smooth variation (smoother peaks and troughs), especially in the time span of 8:00 to 24:00 hr. From hr. 1:00 to 8:00 all prices behave the same, while in the period 9:00 to 24:00 the variation of both Italian prices is intense, with GREC always lower than PUN.

In **Figure 5** we show the *mean hourly volatility (standard deviation)*, for each separate hour of the day, over the period 2005-2013. This figure also enhances the smoother dynamic behaviour of SMP as compared to the strong deviation of the prices from the mean value of hourly prices over the entire period. Only for hours 6:00 to 8:00 the volatility is the same. The dynamics of volatility of prices of PUN and SMP follows a different pattern than their prices, i.e. GREC'S volatility is always larger than PUN's one.

---

[2] For example, for 2005 the first five generators had 88.54%, 11.48%, 7.38%, 7.46% and 3.75% shares, in the total annual generation, respectively. Thus the HHI is calculated as (the value differs slightly from the value given in table 2a)

$$HHI = 10000((0.8854)^2 + (0.1148)^2 + (0.783)^2 + (0.1746)^2 + (0.0375)^2) = 1749 = 1749$$

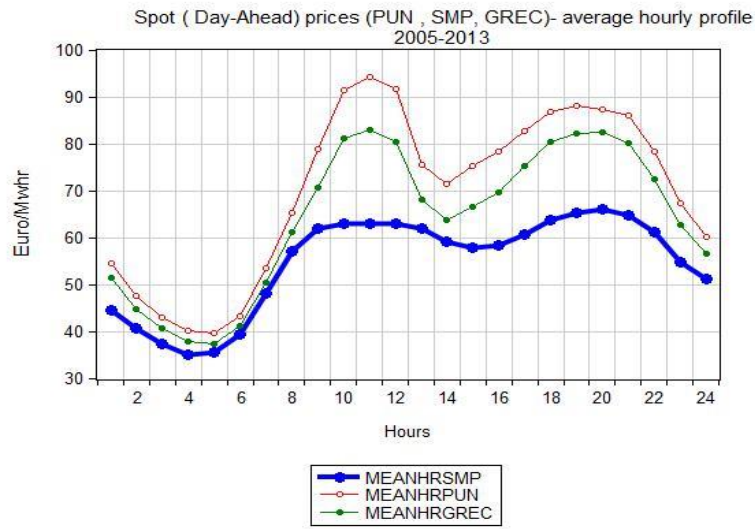

**Figure 4:** Average hourly profile of Spot DA prices PUN, GREC and SMP, 2005 - 2013.

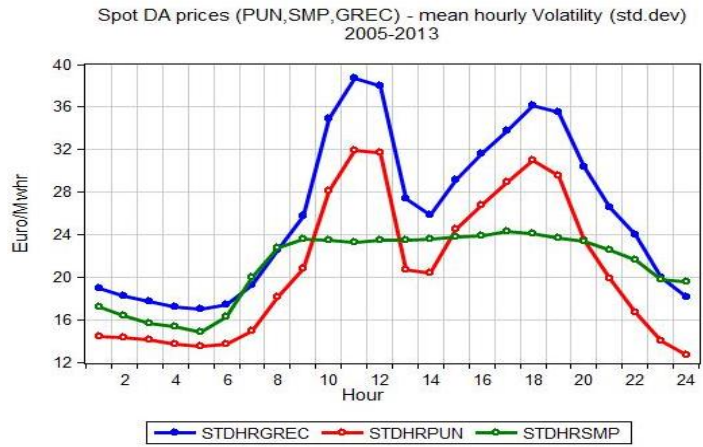

**Figure 5:** Mean hourly Volatility of Spot DA prices PUN, GREC and SMP, 2005 - 2013.

The price differentials between SMP and PUN are primarily due to the different generation mixes between the two countries. The short-run variable cost of the generation units essentially reflects the cost of fuel; so countries with an energy mix of low cost fuels (lignite, coal, hydro and nuclear) have an advantage compared with countries having an energy mix of high cost fuels (natural gas, fuel oils). By modeling the spot time series of the Greek and the Italian not only do we gain insight on the dynamics of their respective wholesale markets but also we have the ability to understand the constraints and barriers towards the market coupling between Greece and Italy.

**Figure 6** indicates that the volatility of daily returns of both series varies with time and that the series may be clustered into **regimes** with different volatility. Some periods are shown to have strong return fluctuations while other may be calm with only small fluctuations. Simonsen, Weron and Mo (Simonsen I., et al., 2004 [7]) argue that this **volatility clustering** phenomenon is reminisced of the intermittent patterns found in turbulent fluids velocities. We now take a closer look on the volatility clustering by calculated the **Rolling Volatility** with a moving window or filter of 90 days (one quarter) in Figure 7a and 365 days (one year) in figure 7b (Weron, 2006 [17]). We observe that the time-dependent unconditional volatilities are varying quite drastically in time, for both SMP and PUN. The volatility change may be due to many factors, for example the **non-stability in the**

**generating units** (supply shocks) cause enhanced volatility in prices shown as spikes and pronounced volatility clusters. Also, seasonal large fluctuations in **temperatures** could also lead to higher volatility. In **Figure 8** the rolling volatilities with a rolling or moving window=300 days are plotted as functions of SMP and PUN price levels. To better visualize any trend, we fit a linear trend to the data. It is seen that the volatility at the Greek and the Italian market tends to be at a higher level when the system price is low, almost with the same way.

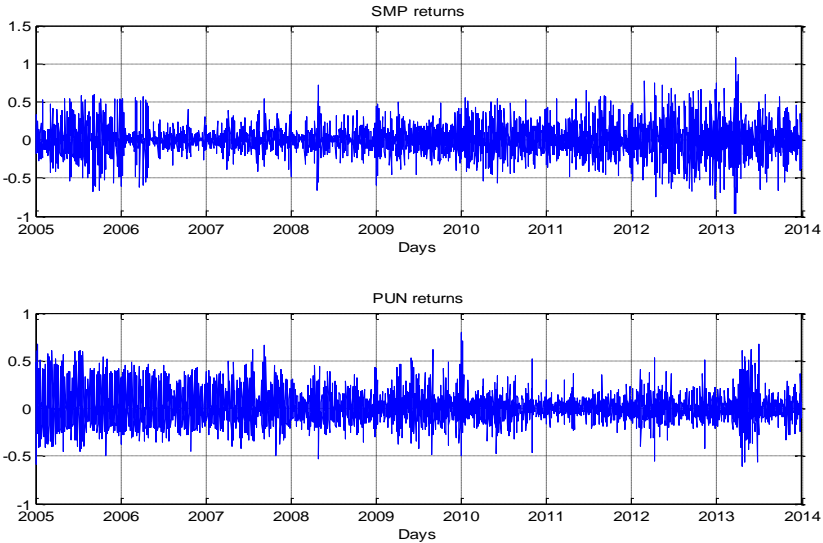

**Figure 6:** SMP and PUN returns for 2005-13

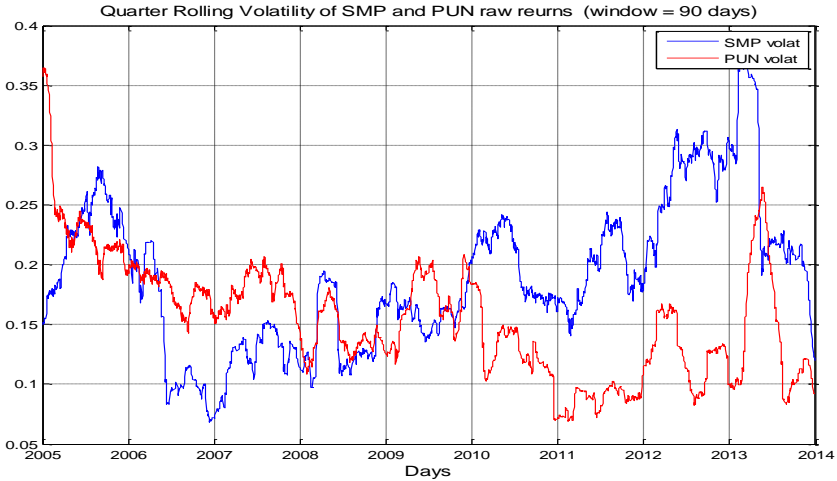

**Figure 7a:** Quarter Rolling unconditional volatility of SMP and PUN raw returns, 2005-13.

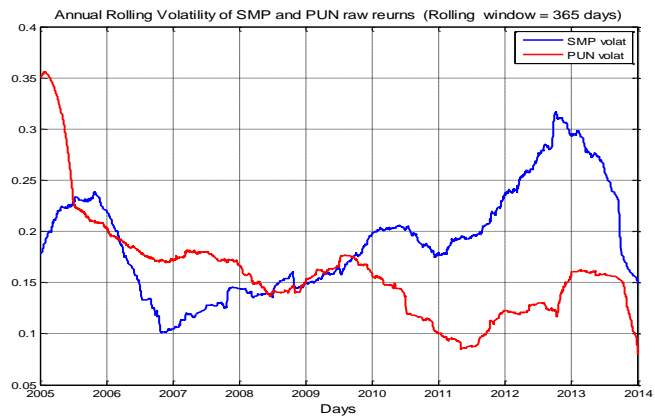

**Figure 7b:** Annual Rolling unconditional volatility of SMP and PUN raw returns, 2005-13.

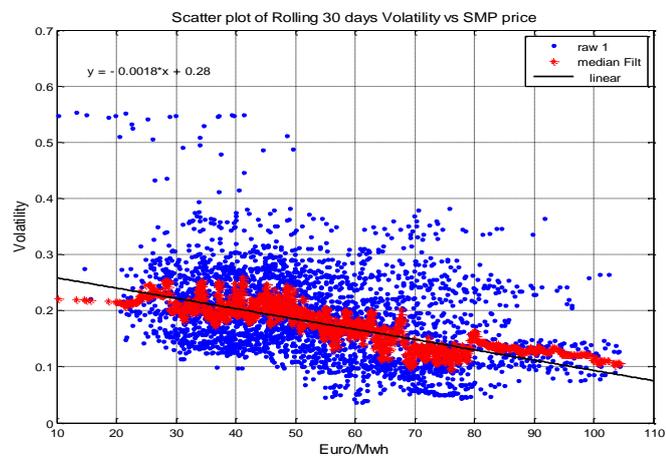

**Figure 8a:** Scatter plot of the rolling volatility vs. price levels for SMP. The red circles are the result of applying a median filter of length N=30 days to volatility.

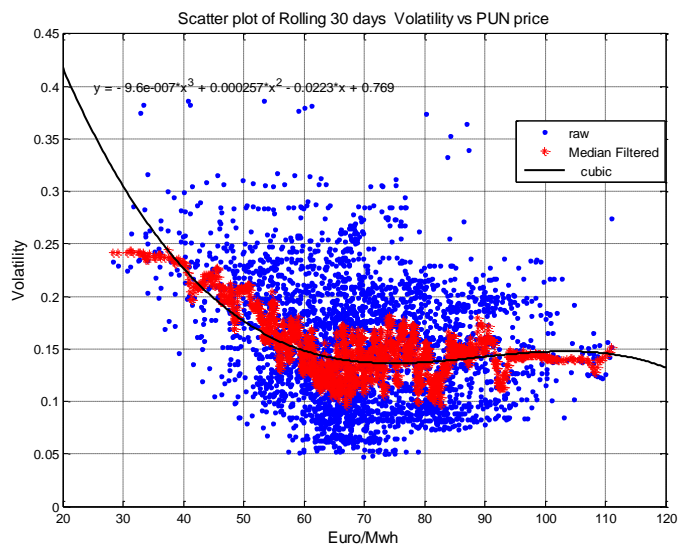

**Figure 8b:** Scatter plot of the rolling volatility vs. PUN price levels. The red circles are the result of applying a median filter of length N=30 days to volatility.

It is seen that the volatility at the Italian market tends to be at a higher level than at the Greek one, when the system price is low. It is interesting to understand the causes of this effect. Simonsen (Simonsen, I., 2005 [57]) argues that this effect might origin from **forced generation.** When the load is low which results in low wholesale prices, some generating units still need to produce a specified amount of energy. For example, in Greece, overfull water reservoirs will impose to a hydro station to either generate electricity (must-run production) or to drain the water without generating energy. Therefore, it is natural for a profit-optimizing company to option the production of more electricity even if the load (demand) and price is low. So, when refilling of the reservoirs is within safety limits, this generating unit could be to wait for periods of higher prices, forcing in this way the volatility up. We expect a similar mechanism that has this kind of effect (**higher volatility when prices are low**) in the Italian market. This can be found by considering the evolution of generating mix and other factors.

**Table 4:** Unconditional Volatilities of raw and deseasonalized SMP and PUN returns

| Year | Unconditional Volatility (St.dev) (raw returns) | | Unconditional Volatility (St.dev) (fully deseasonalized log returns) | |
|---|---|---|---|---|
| | SMP | PUN | SMP | PUN |
| 2005 | 0.226 | 0.230 | 0.172 | 0.251 |
| 2006 | 0.145 | 0.179 | 0.114 | 0.164 |
| 2007 | 0.123 | 0.179 | 0.092 | 0.173 |
| 2008 | 0.134 | 0.139 | 0.105 | 0.130 |
| 2009 | 0.161 | 0.164 | 0.137 | 0.213 |
| 2010 | 0.201 | 0.127 | 0.178 | 0.095 |
| 2011 | 0.194 | 0.087 | 0.160 | 0.132 |
| 2012 | 0.275 | 0.130 | 0.254 | 0.145 |
| 2013 | 0.227 | 0.145 | 0.237 | 0.087 |
| **2005-2013** | **0.199** | **0.160** | **0.167** | **0.164** |

In **table 4** we give the annual evolution of the unconditional volatility of raw returns as well as of filtered returns (as described below) of both wholesale prices. These values will serve as reference ones to compare with the conditional volatility that will be generated by the ARIMA/EGARCH model and they are also used in detecting their annual co-evolution with complexity and stability measures that will be introduced in the sections to follow. Descriptive statistics of both the Greek and Italian Electricity Markets such as liquidity and the total volumes traded for the period examined from 2004 – 2014, is given in Table 5.

**Table 5:** Summary statistics of key electricity market data of Italy and Greece 2004-2014.

| | ITALY | | | | | | GREECE | | | | |
|---|---|---|---|---|---|---|---|---|---|---|---|
| Year | PUN (€/MWh) | | | Total volumes (GWh) | Liquidity (%)[1] | Participants (31 Dec) | SMP (€/MWh) | | | Total volumes (GWh)[2] | Participants (31 Dec)[3] |
| | average | min | max | | | | average | min | max | | |
| 2004 | 51.6 | 1.1 | 189.19 | 231.572 | 29.1 | 73 | 28.21 | 14.32 | 48.77 | 51721 | 8 |
| 2005 | 58.59 | 10,42 | 170,61 | 323.185 | 62,8 | 91 | 38.82 | 18.25 | 120 | 53400 | 13 |
| 2006 | 74.75 | 15,06 | 378,47 | 329.790 | 59,6 | 103 | 62.32 | 23.81 | 86.58 | 54207 | 10 |
| 2007 | 70.99 | 21,44 | 242,42 | 329.949 | 67,1 | 127 | 62.62 | 28.89 | 90.78 | 55253 | 17 |
| 2008 | 86.99 | 21,54 | 211,99 | 336.961 | 69 | 151 | 82.27 | 29.6 | 138 | 55675 | 19 |
| 2009 | 63.72 | 9,07 | 172,25 | 313.425 | 68 | 167 | 43.36 | 0 | 96.5 | 52436 | 30 |
| 2010 | 64.12 | 10 | 174,62 | 318.562 | 62,6 | 198 | 45.66 | 0 | 98.3 | 52365 | 38 |
| 2011 | 72.23 | 10 | 164,8 | 311.494 | 57,9 | 181 | 59.36 | 0 | 150 | 51872 | 37 |
| 2012 | 75.48 | 12,14 | 324,2 | 298.669 | 59,8 | 192 | 56.45 | 0 | 150 | 50558 | 33 |
| 2013 | 62.99 | 0 | 151,88 | 289.154 | 71,6 | 214 | 41.48 | 0 | 97 | 50717 | 28 |
| 2014 | 52.08 | 46.51 | 62.00 | 281.980 | 66.0 | 254 | 62.469 | 13.740 | 84.020 | | 24 |

Source: GME, 2013 and ADMIE S.A., 2013

Data for 2004 are from April to December

1. Liquidity is the ratio of Volume (in GWh) traded to the total Volume. For example, in 2006 the Volume traded was 196555 GWh, so liquidity is 196555/329790=59.6%.
2. Due to the mandatory physical trading in the Greek wholesale market, the traded volume of electricity is equal to the annual demand (including the 256 interconnection balance)
3. Total number of Traders

### 5. Methodology-tools and Empirical results

#### 5.1 BDS Test in the analysis of SMP and PUN

The BDS test by Brock et al (1996) [58] is a statistical test for nonlinearity dependence, although not designed as a direct test for this purpose, it can distinguish between deterministic nonlinear systems, random process, and stochastic (Opong et al, 1999 [59]). It is a nonparametric technique that allows us to distinguish whiteness from an unspecified alternative (Yousefpoor et al., 2008 [60], Barnett and Serletis, 2000 [61]). However we must stress here that the BDS represents *a necessary but not sufficient condition for chaos* (Barnett et al, 1997 [62]; Barnett and Hinich, 1992[63]; Barnett and Serletis, 2000 [61]). It is a widely used test in most studies analyzing chaotic behavior in financial time series (Papaioannou, G, and Karytinos, A., 1995[21]). The null hypothesis $H_0$ is that the series is an independent and identically distributed (i.i.d) process and the BDS statistic is given (se previous papers) by

$$W(N, m, \varepsilon) = \sqrt{N} \frac{C(N, m, \varepsilon) - C(N, m, \varepsilon)^m}{\hat{\sigma}(N, m, \varepsilon)} \tag{2}$$

where $C(N, m, \varepsilon)$ is the correlation integral of the embedding dimension $m$, expressed by the following relation:

$$C(N, m, \varepsilon) = \frac{1}{M(M-1)} \sum_{i=1}^{M} \sum_{j=i+1}^{M} H(\varepsilon - \|X_i - X_j\|) \tag{3}$$

where $X = [X_1 X_2 \ldots X_M]^T$ is the constructed trajectory and $X_i$ the state of the system at discrete time $i$. Let our series have $N$ data $\{x_1, x_2, \ldots x_N\}$, $X_i$ is given by $[X_i X_{i+J} \ldots X_{i+(m-1)J}]$, where L is the lag or the reconstruction delay. Thus, we symbolize by $X$ the $M \times m$ matrix, where $M$ is given by $M = N - (m-1)L^3$. Let also denoted by $\hat{\sigma}(N, m, \varepsilon)$ the estimate of the asymptotic standard deviation of the expression $C(N, m, \varepsilon)^m - C(N, 1, \varepsilon)$. Under H₀ of i.i.d or whiteness, the quantity $\hat{\sigma}^2(N, m, \varepsilon)$ is calculated as follows:

$$\hat{\sigma}^2(N, m, \varepsilon) = 4 \left\{ K^m(N, \varepsilon) + 2 \sum_{j=2}^{m-1} K^{m-j}(N, \varepsilon) C^{2j}(N, 1, \varepsilon) + (m-1)^2 C^{2m}(N, 1, \varepsilon) \right.$$
$$\left. - m^2 K(N, \varepsilon) C^{2m-2}(N, 1, \varepsilon) \right\} \tag{4}$$

where

$$C_m(N, \varepsilon) = \frac{2}{N(N-1)(N-2)} \sum_{i=1}^{N} \sum_{j=i+1}^{N} \sum_{K=j+1}^{N} H(\varepsilon - \|X_i - X_j\|) H(r - \|X_j - X_K\|) \tag{5}$$

and the Heaviside function is

$$H(\varepsilon - \|X_i - X_j\|) = \begin{cases} 1 & \text{if } \varepsilon - \|X_i - X_j\| > 0 \\ 0 & \text{otherwise} \end{cases} \tag{6}$$

Brock (1986) [64] have demonstrated that the $W(N, m, \varepsilon)$ statistics follows, asymptotically, the standard normal distribution. A rejection of the null i.i.d hypothesis in this test is an indication of *nonlinear dependence* in the data. Furthermore, a ***positive and significant BDS*** statistics hints that certain patterns within the reconstructed trajectory have the probability to *emerge more frequently*, a behavior not shown in a random process. On the opposite, a **negative and significant BDS** indicate the *infrequent appearance of certain patterns* compared with a random process.

Brock (1986) [64] proposed a residual test in order to validate the results of the BDS test as well as to make the statistic a valuable tool in distinguishing between deterministic chaos and nonlinear stochastic processes. This test is applied to the standardized residual series from an ARCH-type model (Brock et al, 1993 [65]; Adrangi et al, 2001 [66]), as we will see in section 5.6. To test the null hypothesis of i.i.d. and whiteness, BDS is a powerful tool. However, the rejection of the null hypothesis does not lead directly to the conclusion that the series may be chaotic, since this rejection can be due to many alternatives like nonstationarity (e.g. due to changing macroeconomic, regulatory etc. factors), low complexity chaotic dynamics etc. (Youseproor et al., 2008 [60]; Hsieh, 1991 [67]). Therefore, we point out here that rejecting the $H_0$ of i.i.d. does not lead us necessarily to accept the chaotic dynamic behaviour of our data. We restate the null and alternative assumptions as follows :

- Null Hypothesis $H_0$: The observations of the sample are i.i.d.
- Alternative $H_1$: The observations are not i.i.d.

The BDS test statistic results for the stationary log returns of SMP and PUN are given in **Tables 6 and 7** respectively. Here, *m* is the embedding dimension and *e* is the distance between two observations, defined in terms of *σ* (St.dev). We took for *e/σ* the value 0.5 and *m=1* to *6*. Because the BDS statistics are positive and the p values are very significant (p=0.000) the null hypothesis is rejected (also the BDS coefficients are less than 2.0 and 3.0 and, according to Opong et al. (1999) [59], the test rejects the $H_0$ of i.i.d. process ). So both the SMP and PUN returns, with 95% and 99% confidence , cannot be considered i.i.d, and clearly exhibit a **nonlinear dependence**, which is one of the four "footprints" of chaos. This finding indicate that SMP and PUN returns **do not reflect the historical price information**, i.e. the efficiency of the two markets is against the weak-form EMH. The BDS test is also applied on the standardized residuals from the ARIMA/EGARCH models of conditional mean and volatilities, in order to check for any remaining structure in the residuals so assessing the quality of the ARIMA/EGARCH models (see section 6). The results (not reported here) provide little support for the presence of nonlinear dependence in the residuals enhancing the good quality of the model. The combination of results of BDS and R/S-Hurst analysis, prove that both returns deviated strongly from random walk. We stress here, however, that the BDS is a **necessary but not sufficient condition** of the chaotic behaviour and also that it is consistent with other direct test for chaos, e.g. maximum Lyapunov exponent (see section 5.4). we have deduced the same result as Dakhlaoui I., et al. (2013) [68] that, contrary to existing literature (Opong et al., 1999 [59]) the R/S-Hurst and BDS test work complementarily in the verification of the necessary conditions of chaos.

**Table 6:** The BDS statistics applied to the daily returns of SMP time series

| BDS Test for SMPRET | | | | | |
|---|---|---|---|---|---|
| Date: 17/10/16   Time: 08:48 | | | | | |
| Sample: 1/01/2005 31/12/2013 | | | | | |
| Included observations: 3287 | | | | | |
| Dimension | BDS Statistic | Std. Error | z-Statistic | Prob. | |
| 2 | 0.023611 | 0.001126 | 20.96881 | 0.0000 | |
| 3 | 0.025776 | 0.000891 | 28.91836 | 0.0000 | |
| 4 | 0.018807 | 0.000530 | 35.46008 | 0.0000 | |
| 5 | 0.011913 | 0.000277 | 43.04621 | 0.0000 | |
| 6 | 0.007327 | 0.000134 | 54.75693 | 0.0000 | |
| Raw epsilon | | 0.099448 | | | |
| Pairs within epsilon | | 3722168. | V-Statistic | 0.344716 | |
| Triples within epsilon | | 5.36E+09 | V-Statistic | 0.151097 | |
| Dimension | C(m,n) | c(m,n) | C(1,n-(m-1)) | c(1,n-(m-1)) | c(1,n-(m-1))^k |
| 2 | 767177.0 | 0.142229 | 1857733. | 0.344409 | 0.118618 |
| 3 | 359438.0 | 0.066678 | 1857343. | 0.344547 | 0.040902 |
| 4 | 177181.0 | 0.032888 | 1855821. | 0.344474 | 0.014081 |
| 5 | 90321.00 | 0.016775 | 1855631. | 0.344649 | 0.004863 |
| 6 | 48468.00 | 0.009008 | 1855418. | 0.344819 | 0.001681 |

**Table 7:** The BDS statistics applied to the daily return of PUN returns

| BDS Test for PUNRET |
|---|
| Date: 17/10/16   Time: 09:02 |
| Sample: 1/01/2005 31/12/2013 |

| Included observations: 3287 | | | | | |
|---|---|---|---|---|---|
| Dimension | BDS Statistic | Std. Error | z-Statistic | Prob. | |
| 2 | 0.024502 | 0.001258 | 19.47594 | 0.0000 | |
| 3 | 0.021843 | 0.001045 | 20.90815 | 0.0000 | |
| 4 | 0.011911 | 0.000652 | 18.26255 | 0.0000 | |
| 5 | 0.005480 | 0.000357 | 15.34633 | 0.0000 | |
| 6 | 0.003659 | 0.000181 | 20.20040 | 0.0000 | |
| Raw epsilon | | 0.080376 | | | |
| Pairs within epsilon | | 3903186. | V-Statistic | 0.361480 | |
| Triples within epsilon | | 5.92E+09 | V-Statistic | 0.166721 | |
| Dimension | C(m,n) | c(m,n) | C(1,n-(m-1)) | c(1,n-(m-1)) | c(1,n-(m-1))^k |
| 2 | 835937.0 | 0.154976 | 1948368. | 0.361212 | 0.130474 |
| 3 | 371701.0 | 0.068952 | 1946911. | 0.361162 | 0.047109 |
| 4 | 155864.0 | 0.028931 | 1945899. | 0.361194 | 0.017020 |
| 5 | 62551.00 | 0.011618 | 1944117. | 0.361083 | 0.006138 |
| 6 | 31653.00 | 0.005883 | 1943969. | 0.361276 | 0.002223 |

**5.2 Nonlinear Time series methods in the analysis of SMP and PUN**

Our analysis makes use of non-linear methods based on chaos theory (the theory of **deterministic systems** with apparently random evolution due to **sensitive dependence on small changes in the initial conditions**) (Kugiumtzis D. et al, 1994 [69]). In terms of **chaos theory**, geometrical objects formed by the system trajectories (so-called **attractors**) are characterized by **fractal dimension**. For attractors from regular deterministic systems, e.g. limit cycles or tori, the fractal dimension is equal to their topological dimension, but for attractors from chaotic systems (**strange attractors**) it is typically a non-integer that indicates their fractal structure. Among various measures of the **fractal dimension** the most common is the **correlation dimension** due to its computational simplicity (Kugiumtzis D. et al, 1994 [69], Theiler J., 1990 [70]).

*5.2.1 Reconstruction of the state space*

Given a time series, an attractor can be embedded in a multidimensional state space. It is known that under certain conditions the reconstructed attractor preserves the topology of the original attractor of the system that generated the data (Takens F., 1981 [71]., Mane R., 1981 [72], Sauer T. et al, 1991 [73]). Working with SMP and PUN time series data from the two Electricity Markets (high dimensional systems), the expression "the original attractor" refers to the interactivity of the **global variables** of the system. Carefully chosen reconstruction schemes are needed in order to maintain the equivalence of the two attractors.

In our work with SMP and PUN data, we have evaluated the two most common reconstruction methods, the **Method of Delays (MOD)** and the **Singular Spectrum Approach (SSA)**. For both methods, the chosen parameter setting is crucial for the estimation of ν. In the following we shortly outline the two methods and their parameters (details can be found in e.g. Broomhead D.S. et al, 1986 [74], Albano A.M. et al, 1988 [75], Casdagli M. et al, 1991 [76], Kugiumtzis D, 1996 [77]).

The sampled SMP signal is denoted $\{s_i\}_{I=1}^{N} = \{s(i\tau_s)\}_{i=1}^{N}$, where $\tau_s$ is the sampling time and N is the length of the time series.

*5.2.2 Method of Delays*

The reconstructed state vector with MOD is formed directly from the **scalar measurements**

$$s_i = [s_i, s_{i+\tau}, \ldots s_{i+(m-1)\tau}]^T \quad (7)$$

where τ is the delay time, given as a multiple integer of the sampling time τs and m is the embedding dimension of the reconstructed space. The parameters τ and m give the time window τw of length (m-1)τ. The N-(m-1) ×m matrix

$$s_i = [s_1, s_2, \ldots s_{N-(m-1)\tau}]^T \quad (8)$$

is then the constructed trajectory matrix.

The delay time is usually estimated as the value giving the least correlation among the components of si. Linear decorrelation is obtained by choosing the first zero of the **autocorrelation function** $R(\tau)$ ($R(\tau 0)=0$) or the correlation time $\tau_c$ corresponding to $R(\tau)=1/e$, while general decorrelation is determined as the **first local minimum** of the **mutual information function** $I(\tau)$ (min $I(\tau)= I(\tau m)$) (Fraser A.M. et al, 1986 [78]). However, these two functions often provide very large estimates of τ for the SMP time series. This may result in a poor description of the details of the geometric shape of the attractor.

For the embedding dimension m, a lower limit for the reconstruction to be valid is given by Takens' theorem m≥2d+1 (Takens F., 1981 [71]). Takens' originally proposed d to be the topological dimension, but recent theoretical results relax this criterion assigning d to the fractal dimension of the underlying attractor(Sauer T., et al., 1991 [73]). In practice, however, lower values for m are often sufficient to reconstruct the attractor successfully, and one typically searches for the minimum embedding dimension m*. A method often used to estimate m* is the **False Nearest Neighbors (FNN)** (Kennel M.B., 1992 [79]). This method applies a geometrical criterion – based on the behavior of spatially near neighbors – to the attractor embedded in successively higher dimensions until a limit m* is reached where the criterion is fulfilled. The implementation of this method to SMP data did not give unique m* but showed a dependence of m* on τ as shown in fig. 9.

In the correlation dimension estimation with MOD reconstruction, only τ is the critical parameter since m is increased in order to investigate if there is a saturation of the scaling of logC(r) vs. logr.

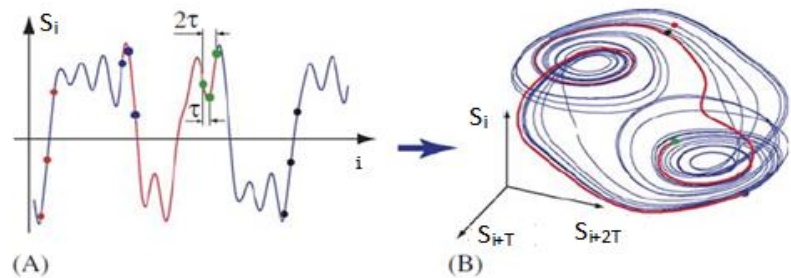

**Figure 9:** A pictorial representation of the phase space reconstruction

### 5.2.3 *Singular Spectrum Approach*

Using SSA for the reconstruction, the state vector

$$s_i = [s_i, s_{i+\tau}, \ldots s_{i+(p-1)}]^T$$

is formed for p large and τ=1, where the mean value of the time series is first subtracted from si. The trajectory in Rp is then the (N-p+1)*p matrix

$$S = [s_1, s_2, \ldots s_{N-p+1}]^T$$

A new basis for Rp is formed by the p singular vectors of S computed with the Singular Value Decomposition (SVD) (Broomhead D.S., et al., 1986 [74]). The trajectory is then projected onto the m first singular vectors ranked according to the variance they explain, which correspond to the m largest singular values σi of S. The final (N-p+1)×m trajectory matrix is

$$Y = [s_1^T C \ s_2^T C \ .... \ s_{N-p+1}^T C]^T \tag{9}$$

where the p×m matrix C has the m first singular vectors as columns.

The parameters of this method are the initial embedding dimension p and the final embedding dimension m. the initial embedding dimension p determines the time window length τw=(p-1). For the estimation of m, the cut-off of the spectrum of singular values (σ1 >…>σm*>> σm*+1>… σp) has been proposed, but for signals from nonlinear systems such a cut-off is not guaranteed.

### 5.2.4 Finding the time delay and embedding dimension

Determining the **time delay** and the **embedding dimension** is considered as one of the most important steps in nonlinear time series modeling and prediction. A number of methods have been developed in determining the time delay and the minimum embedding dimension since the early beginning of nonlinear time series study. Here we will describe and apply several of them to the foreign exchange time series data sets.

### Time delay

The first step in phase space reconstruction is to choose an optimum delay parameter $\tau$. different prescriptions have appeared in the literature to choose $\tau$ but they are all empirical in nature and do not necessarily provide appropriate estimates.

- *First passes through zero of the autocorrelation function:* in earlier works it was suggested to use the value of $\tau$ for which the autocorrelation function

$$C(\tau) = \sum_n (s_i - \bar{s})(s_{i+\tau} - \bar{s}) \tag{10}$$

  first passes through zero which is equivalent to requiring linear independence.
  The application of the zero crossing of the autocorrelation function gives **quite high** values for both series.

- *First minimum of the Average mutual information:* This is our preferred method Fraser and Swinney (1986) [80] suggested to use the average mutual information (AMI) function $I(\tau)$, as a kind **of nonlinear correlation** function to determine when the values of $s_i$ and $s_{i+\tau}$ are independent enough of each other to be useful as coordinates in a time delay vector but not so independent as to have no connection which each other at all. For a discrete time series, $I(\tau)$ can be calculated as,

$$I(\tau) = \sum_{n,n+T} P(s_i, s_{i+\tau}) \log_2 \left[ \frac{P(s_i, s_{i+\tau})}{P(s_i) P(s_{i+\tau})} \right] \tag{11}$$

  where $P(s_i)$ refers to individual probability and $P(s_i, s_{i+\tau})$ is the joint probability density. Following the method developed by Abarbanel (1996) [11], to determine $P(s_i)$ we simply project the values taken from $s_i$ versus $i$ back onto the $s_i$ axis and form a histogram of the values. once normalized, this gives us $P(s_i)$. For the join distribution of

$s_i$ and $s_{i+\tau}$ we form the two-dimensional histogram in the same way.

In general, the time lag provided by $I(\tau)$ is normally lower than the one calculated with the $C(\tau)$, $\tau_{AMI} \geq \tau_{correl}$ and provides the appropriate **characteristic time scales** for the motion. Even through $C(\tau)$ is the optimum linear choice from the poin of view of predictability in a least square sense of $s_{i+\tau}$ from knowledge of $s_i$, it is not clear why it should work for nonlinear systems and it has been shown that in some cases it does not work at all.

Figures 10 and 11 show the results of the AMI for SMP and PUN returns. The first minimum for both series is about 4 days (see below). This value is used as input to calculate the minimum embedding dimension in the algorithm of FNN.

*Embedding dimension*

The dimension, where a time delay reconstruction of the system phase space provides a necessary number of coordinates to unfold the dynamics from overlaps on itself caused by projection, is called the **embedding dimension, $d_E$**. This is a **global dimension**, which can be different from the **real dimension**. Furthermore, this dimension depends on the time series measurement, and hence, if we measure **two different variables of the system, there is no guarantee that the $d_E$ from time delay reconstruction will be the same from each of them.**

The usual method for choosing the minimum embedding dimension used for the computations, one notes when the value of the invariant stops changing. Since these invariants are geometric properties of the dynamics, they become independent of $d$ for $d \geq d_E$, i.e. after the geometry is unfolded.

In this work, we have used the method of False Nearest Neighbours (FNN).

- *False Nearest Neighbours (FNN):* the method of False Nearest Neighbours (FNN) was developed by Kennel et al. (1992). In this case, the **condition of no self-intersection** states that if the dynamics is to be reconstructed successfully in $R^d$, then all the neighbor point in $R^d$ should be also neighbours in $R^{d-1}$. The method checks the neighbours in successively higher embedding dimensions until it finds only a negligible number of false neighbours when increasing dimension from $d$ to $d+1$. This $d$ is chosen as the embedding dimension.

    It was found by Kennel et al. (1992) [79] that **if the data set is clean from noise, the percentage of false nearest neighbours will drop from nearly 100% in dimension one to strictly zero when $d_E$ is reached.** Further, it will remain zero from then on since the dynamics is unfolded. If the signal is contaminated with noise (infinite dimension signal) we may not see the percentage of false nearest neighbours drop to near zero in any dimension. In this case, depending on the signal to noise ratio the determination of $d_E$ will degrabe.

    For both time series, the FNN method **suggest an embedding dimension of 6** setting the "active" dimensionality of the manifold on which the system evolves, see fig.12, 13. The increase of the number of FNN after a certain $d_E$ is an indication of the presence of noise in the signal.

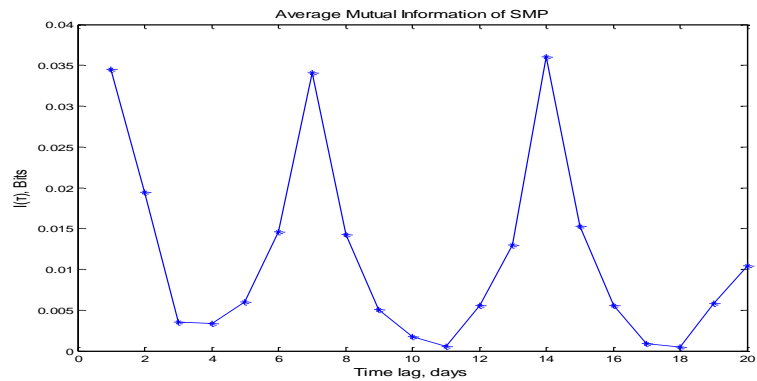

**Figure 10:** Average Mutual Information (AMI) of SMP returns. The first minimum is 4 days, defining the time delay $\tau$ in the reconstruction vector by MOD.

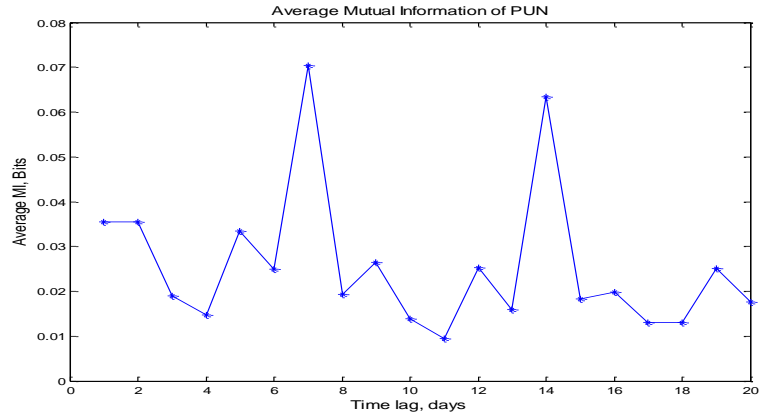

**Figure 11:** Average MI (AMI) of PUN returns. The first minimum is 4 days, defining the time delay $\tau$ in the reconstruction vector by MOD.

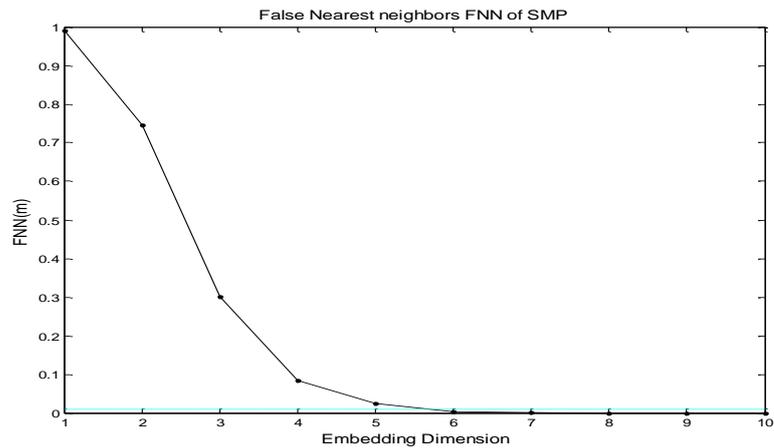

**Figure 12:** False Nearest Neighbors for estimating the minimum embedding dimension of the reconstructed phase space for SMP returns.

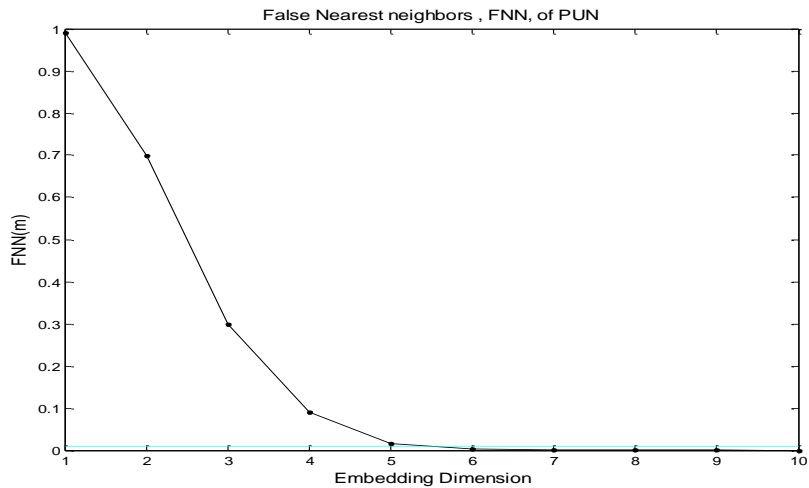

**Figure 13**: False Nearest Neighbors for estimating the minimum embedding dimension of the reconstructed phase space for PUN returns.

*5.5 EMH (Efficient Market Hypothesis) and its relation to Hurst*

In an **anti-persistent** time series (also known as a mean-reverting series) an increase in values will most likely be followed by a decrease or vice-versa (i.e., values will tend to revert to a mean). This means that future values have a tendency to return to a long-term mean.

In a **persistent** time series an increase in values will most likely be followed by an increase in the short term and a decrease in values will most likely be followed by an increase in the short term.

If the Hurst exponent $H > 0.5$, then it reveals that the price process is correlated or that **a price increase in the past is more likely to be followed by an additional increase than a price decrease.** Hense, daily price movements are **persistent** and subject to trends. We can also interpret $H$ as a measure of the bias in the fractional Brownian motion (fBm). Thus, **the deviation from random walk provides interesting information on the volatility and risk inherent in electricity market.** A relatively high $H$ underpins the relatively smooth trend and **less or controlled volatility**. The persistent effect when $H > 0.5$ is stronger than when $H < 0.5$, i.e. the anti-persistent or anti-correlation or mean-reversion process of price movements. All the above information is related to the market efficiency. **Higher $H$ values correspond to emerging or developing markets** in which the **EMH is not satisfied.** We know that EMH roughly stated, "argues" that in competitive markets all information is instantaneously reflected in prices and it is possible to systematically beat the market. A consequence of this postulate is that prices should follow a random walk ($H = 0.5$) and market cannot be forecasted. **The longer $H$ deviates from 0.5, the less noise in system or equivalently the more mean-reverting (or anti-correlated) is the price time series, therefore the ability of a model to forecast it is larger.** Expressed differently, the closer the H value is to 0.00, the stronger is the tendency for the time series to revert to its long-term mean value. We can use Hurst exponents to classify power markets: a) those where electricity price processes exhibit a strong **mean-reverting mechanism** ($H < 0.5$) and b) those where electricity prices behave almost like **Brownian motion** ($H \approx 0.5$).

Due to mean reversion behavior, the electricity market is **incomplete** since there may exist infitely many risk-neutral probability measures (Oum, Orem and Deng, 2006 [81]). Market incompleteness arises due to real-world market frictions. Spot electricity markets, due to the fact that physical electricity is not a tradable asset, are incomplete, therefore not efficient in the sense of EMH (Vehvilainen, 2004 [82], Karatzas and Shreve, 1998 [83]).

*5.3 Hurst and Detrending Fluctuation Analysis (DFA) of SMP and PUN*

In this section we shortly describe five methods that are usualy applied in time series for detecting long-term correlations. The results are presented in Table 7 below.

Hurst developed a method called the **rescaled range (R/S)** for **distinguishing completely random time series from correlated time series** (Hurst, E., 1965 [84]). The procedure in Hurst analysis consists of the steps, starting with dividing a **time series (of returns)** of length $L$ into $d$ subseries of length $n$. Next for each subseries $k = 1, \ldots d$; a) find the mean ($E_k$) and standard deviation ($S_k$); b) normalize the data ($Z_{i,k}$) by subtracting the sample mean $X_{ik} = Z_{i,k} - E_k$ for $i = 1, \ldots, n$; c) create a cumulative time series $Y_{i,k} = \sum_{j=1}^{i} X_{j,k}$ for $i = 1, \ldots, n$; d) find the range $R_k = max\{Y_{1,k}, \ldots Y_{n,k}\} - min\{Y_{1,k}, \ldots Y_{n,k}\}$; e) rescale the range $R_k/S_k$. Finally, the mean value of the rescaled range for subseries of length $n$ is $(R/S)_n = (1/d)\sum_{k=1}^{d} R_k/S_k$.

Equivalently, we can plot the $(R/S)_n$ statistics against $n$ on a double-logarithmic paper. **If the returns process is white noise the plot is roughly a straight line with slope 0.5.** if the process is **persistent** then **the slope is > 0.5; if it is anti-persistent then the slope is <0.5**. the significance level is usually chosen to be $\sqrt{1/N}$ – the standard deviation of a Gaussian white noise. However, it should be noted that for small $n$ there is a significant deviation from the 0.5 slope. For this reason the theoretical (i.e. for white noise) values of the $(R/S)$ statistics are approximated by Weron, R. et al. (2000) [85].

$$E(R/S)_n = \begin{cases} \dfrac{n-\frac{1}{2}}{n} \dfrac{\Gamma((n-1/2))}{\sqrt{\pi}\Gamma(n/2)} \sum_{i=1}^{n-1} \sqrt{\dfrac{n-i}{i}} & for\ n \leq 340, \\ \dfrac{n-\frac{1}{2}}{n} \dfrac{1}{\sqrt{n}(\pi/2)} \sum_{i=1}^{n-1} \sqrt{\dfrac{n-i}{i}} & for\ n > 340\,. \end{cases} \quad (12)$$

$R/S$ is shown to follow asymptotically the relation $(R/S)_n \sim cn^H$ thus by taking logs of both sides are have $log(R/S)_n = log c + H log n$. Therefore the value of $H$ can be estimated by simple regression.

**The Hurst component $0 \leq H \leq 1$ is equal to 0.5 for random walk time series < 0.5 for anticorrelated series, and >0.5 for positively correlated series.** the Hurst exponent is directly related to the **"fractal dimension"**, which gives a measure of the roughness of a surface. The relationship between the **fractal dimension**, $D$ and the Hurst exponent, $H$, is:

$$D = 2 - H \quad (12\,a)$$

Hurst exponents **quantify the correlation of a fractional Brownian motion.** A fractional Brownian motion (fBm) is a **random walk** with a **Hurst exponent different from 0.5** and thus **with a memory.** The decaying of **spectral density** of an **fBm** has a relationship with the Hurst exponent as follows:

$$P(f) = spectral\ density \propto \dfrac{1}{f^\beta} \quad (13)$$

where 
$$\beta = 2H + 1 \quad (13a)$$

is the power spectrum exponent. In Figure 13a we give the estimation of $\beta$ from the slope of the linear fitted on the data of $f$ vs spectral density. The found indices $\beta$ (0.97 and 0.84) indicate that the prices of both markets do not follow an fBm or fGN (see section 5.3.3) processes.

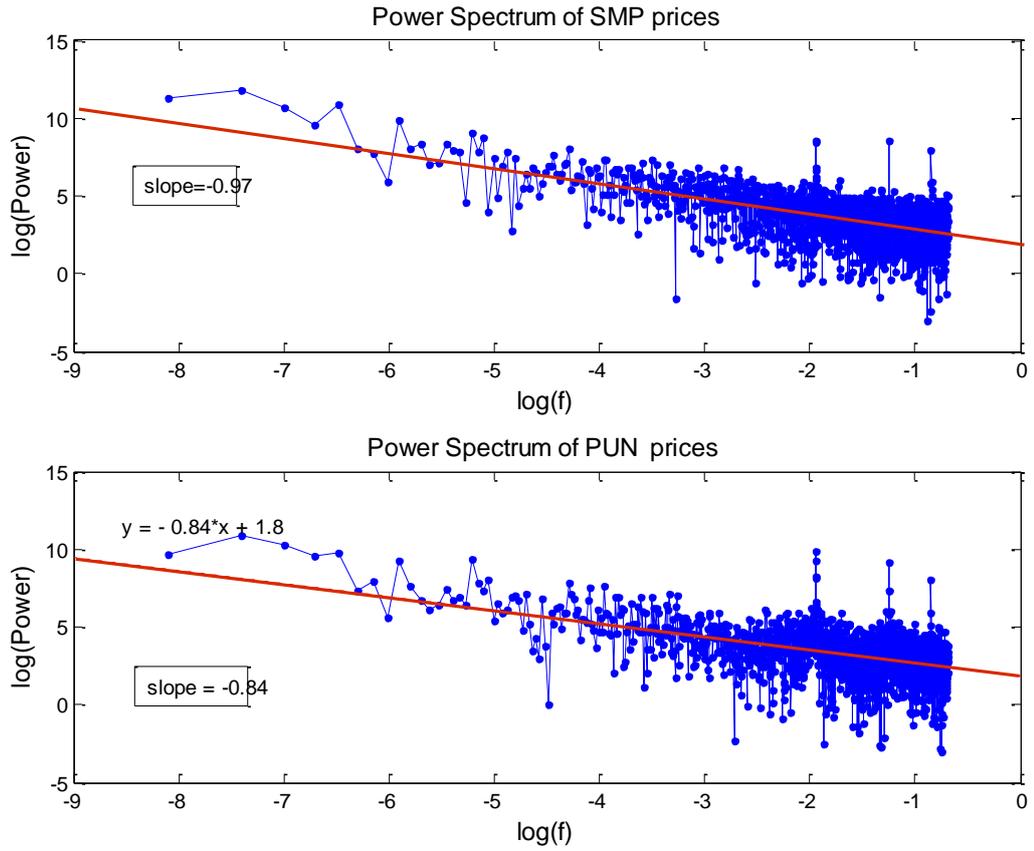

**Figure 14:** Power spectrum of SMP and PUN returns. The slopes 0.97 and 0.84 are the index $\beta$ in (13).

Financial time series have been found to exhibit some universal characteristics that resemble the scaling laws typical of natural systems in which **large number of units** interacts. For instance, the Hurst exponent has been extensively applied by Peters (1996) [86] to various **capital markets and in most of the cases he has found persistent memory. Table 7** provide information on the time evolution of Hurst exponent, estimated by the classical and modified R/S approach, for SMP and PUN returns. The anti-persistent, mean reverting behavior, a deviation from the 'reference' value H=0.5 of the Brownian motion is clear, for both series for every period.

*5.3.1 Generalized Hurst Exponent GHE and weighted GHE.*

In order to study we employ a powerful tool to study directly the **scaling properties** of our data via the **qth-order moments** of the distribution of the increments. This is called the **generalized Hurst exponent** (GHE) and it is related to the long-term statistical dependence of a certain time series $X(t)$, with $t = (1,2,\ldots,k,\ldots,\Delta t)$, defined over a **time-window** $\Delta t$ with time-steps of one-unit. GHE measures the correlation persistence, thus it has to be related to fundamental statistical quantities, actually to the **qth-order moments** of the distribution of the increments of the time series, defined as

$$K_q(\tau) = \frac{\langle |X(t+\tau) - X(t)|^q \rangle}{\langle |X(t)|^q \rangle} \tag{14}$$

(Di Matteo T., 2007 [87], Barabasi A.L et al., 1991 [88]) where $\tau$ can vary between 1 and $\tau_{max}$ and $\langle \cdot \rangle$ symbolizes the **sample average over the time-window.** We point out here that for $q = 2$, $K_q(\tau)$ is analogous to the **autocorrelation function:** $C(t,\tau) = \langle X(t+\tau)X(t) \rangle$. From the scaling behavior of $K_q(\tau)$, the **generalized Hurst exponent** is then defined given that the following relation holds:

$$K_q(\tau) \propto \tau^{qH(q)} \tag{15}$$

Based on the aforementioned behavior, Processes exhibiting this scaling can be categorized into two groups: (a) **Processes with** $H(q) = H$, i.e. independent of $q$, i.e they are **uniscaling (or unifractal)** and their behavior is uniquely determined by the **constant $H$ (Hurst exponent or self-affine index,** (Di Matteo T., 2007 [87])**)**; (b) Processes with varying $H(q)$, called **multiscaling (or multifractal)** and each moment scales with a **different exponent.** Figure 16 shows the variation of qH(q) versus q for SMP and PUN returns. The nonlinear behavior in the case of Italian market is more pronounced, indicating that in this market a higher degree of multifractality is expected to exist.

Di Matteo, T., (2007) [87], Di Matteo, T. et al. (2005) [89] have pointed out how financial data exhibit **multi-fractal** scaling behaviors. The GHE is calculated from an average over a set of values corresponding to different values of $\tau_{max}$ in relation (15) (Barabasi, A.L et al.,1991 [88], Di Matteo, T., 2007 [87]). Due to the fact that all the information about the scaling properties of a time series is contained in the scaling exponent $H(q)$, the analysis based on GHE is quite simple.

If we consider that the recent past is more important than the remote past we can use the "**exponential smoothing**" technique to take into account that the informational relevance of observations decays exponentially the further we move to the past. This is obtained by defining smoothing as

$$w_s = w_0 \exp\left(-\frac{s}{\theta}\right), \qquad \forall s \in \{0,1,2,\ldots,\Delta t - 1\} \tag{16}$$

where $\theta$ is the weights characteristic time and $a = \frac{1}{\theta}$ is the exponential decay factor. According to Pozzi F., et al., (2012)[90], the parameter $w_0$ is

$$w_0(a) = \frac{1 - e^{-a}}{1 - e^{-a\Delta t}} \tag{17}$$

In general, for any quantity $f(u_t)$, t**he weighted average over the time-window** $[t - \Delta t + 1, t]$ is given by

$$\langle f \rangle_w(t) = \sum_{s=0}^{\Delta t - 1} w_s f(u_{t-s}) \tag{18}$$

and the **weighted GHE (wGHE)** is therefore obtained by substituting the normal averages in relation (15) with weighted averages:

$$K_q^w(t,\tau) = \frac{\langle |X(t+\tau) - X(t)|^q \rangle_w(t)}{\langle |X(t)|^q \rangle_w(t)} \tag{19}$$

Using the scaling law in relation (16) we reach to the linear equation

$$\ln\left(K_q^w(t,\tau)\right) = qH^w(q)\ln(\tau) + const. \tag{20}$$

which provides the **wGHE.** Figure 14 depicts the variation of $qH(q)$ versus $q$ for both markets. The nonlinear behavior in the case of Italian market is more pronounced.

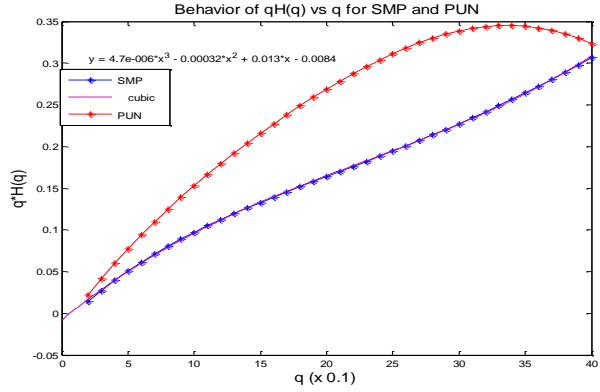

**Figure 15 :** Variation of qH(q) versus q for SMP and PUN returns.

*5.3.2 Detrended fluctuation Analysis (DFA)*

DFA is an alternative tool to estimate long-range dependence, proposed by Peng et al. (1994) [91], and its advantage over $R/S$ analysis is that it avoids the artifacts of **non-stationarity** expressed as apparent long-range correlation. To estimate DFA we devide again a time series (of returns) of Length $L$ into $d$ subseries of length $n$. For each subseries then, $k = 1, ..., d$; a) we create a commutative time series $Y_{i,k} = \sum_{j=1}^{i} X_{j,k}$ for $i = 1, ..., k$; b) we fit a least squares line $Y_k(x) = \widetilde{a_k x} + b_k$ to $\{Y_{1,k}, ... Y_{n,k}\}$ and finally c) compute the RMS fluctuation (i.e. standard deviation) of the integrated and detrended time series

$$F(k) = \sqrt{\frac{1}{n}\sum_{i=1}^{n}(Y_{i,k} - a_k i - b_x)^2} \tag{21}$$

The mean value of RMS fluctuation for all subseries of length $n$ is

$$\bar{F}(k) = \frac{1}{d}\sum_{k=1}^{d} F(k) \tag{22}$$

As is the case of Hurst analysis, we plot on a double-logarithmic "paper" $\bar{F}(k)$ against $n$ and we get the power-law scaling $F(k) \sim k^\alpha$ where $\boldsymbol{\alpha}$ **is the fractal DFA** exponent. If the returns process is **white noise** then the **slope is roughly 0.5.** If the slope is > 0.5 the process is **persistent** while for an **antipersistent** process the slope is < 0.5. For various processes, index $a$ takes the following values: $a < 1/2$ (anti-correlated), $a \approx 1/2$, (white noise, uncorrelated), $a > 1/2$ (correlated), $a \approx 1$ ($1/f-$ noise, pink noise), $a > 1$ (non-stationary, unbounded) and $a \approx 3/2$ (Brownian noise). The classical Hurst exponent corresponds to the second moment for stationary cases $H = a$ and to the second moment -1 for nonstationary cases $H = a - 1$ (Stanley H. et al.,1999 [92]). Both measures H and DFA have a major drawback, the fact that no asymptotic distribution theory has been derived. However, Weron (2002b) [93] has obtain empirical confidence intervals for both measures via a Monte Carlo study.

*5.3.3 Autocorrelation function ACF and Hurst*

A **long memory process** is a process with a random component, where a past event has a decaying effect on future events. The process has some memory of past events, which is "forgotten" as time moves forward. The mathematical definition of long memory processes is given in terms of autocorrelation. When a data set exhibits autocorrelation, a value $x_i$ at time $t_i$ is correlated with a value $x_{i-d}$ at time $t_{i-d}$ where $d$ is some time increment in the future. In a **long memory process autocorrelation decays** over time and the **decay** follows a power law, i.e.

$$p(k) = Ck^{-\delta} \qquad (23)$$

where $C$ is a constant and $p(k)$ is the autocorrelation function with $lag k$. The **Hurst exponent** is related to the exponent $\delta$ by

$$H = 1 - \frac{\delta}{2} \qquad (24)$$

DFA algorithm is used to test the efficient market hypothesis (EMH) (section 5.5) and is appropriate for detecting weak autocorrelations (AC) in nonstationary data. It has the ability to distinguish intrinsic AC associated with memory effects in the underlying dynamics from effects that are attributed to exogenous factors.

*5.3.4 Periodogram Regression (GPH)*

This technique denoted as GPH was originally proposed by Geweke and Porter-Hudak[94] in order to estimate the Hurst exponent. This method is based on observations of the slope of the spectral density around the angular frequency (w=0) of a fractionally integrated model. The spectral density of a general fractionally integrated model with differencing parameter d is identical to that of a fractional Gaussian noise with Hurst exponent H = d+0.5.

This estimation procedure of the Hurst exponent begins with calculating the periodogram. Then a linear regression model is developed such as:

$$\log\{I_L(\omega_\kappa)\} = \alpha - d\, x_t + \varepsilon_\kappa = a - d \log\left\{4 \sin\left(\frac{\omega_\kappa}{2}\right)^2\right\} + \varepsilon_\kappa \qquad (25)$$

at low frequencies $\omega_k$, k=1,2,…,K ≤ [L/2]. The least squares estimate of the slope yields the differencing parameter d, hence the Hurst exponent is given by the following formula:

$$H = 0.5 + \hat{d} \qquad (26)$$

The choice of k differs in the sense that some authors recommend K = [$L^{0.5}$], however other values have been also suggested; K = [$L^{0.45}$] or [$L^{0.2}$] ≤ K ≤ [$L^{0.5}$].

Parameter $\hat{d}$ is calculated according to the formula:

$$\hat{d} \sim N\left(d, \frac{\pi^2}{6 \sum_{K=1}^{K}(x_t - \overline{x})^2}\right) \qquad (27)$$

Tables 8 and 9 provide the estimation of Hurst exponent by using different methods.

**Table 8 :** Annual evolution of Hurst exponent computed by different methods (2005-2013)

| Year | R/S - Hurst (traditional) | | R/S - Hurst with Anis-Lloyd modification for sample size* | | DFA Hurst | | GPH Hurst | | GHE Hurst (q=1) | |
|---|---|---|---|---|---|---|---|---|---|---|
| | SMP | PUN | SMP | PUN | SMP | PUN | SMP | PUN | SMP | PUN |
| 2005 | 0.222 | 0.205 | 0.134 | 0.253 | 0.116 | 0.043 | 0.121 | 0.335 | 0.150 | 0.092 |
| 2006 | 0.207 | 0.252 | -0.069 | 0.223 | 0.193 | 0.048 | 0.054 | 0.386 | 0.223 | 0.097 |
| 2007 | 0.279 | 0.258 | 0.282 | 0.248 | 0.153 | 0.100 | 0.055 | 0.579 | 0.198 | 0.148 |
| 2008 | 0.315 | 0.258 | 0.277 | 0.179 | 0.186 | 0.081 | 0.006 | 0.090 | 0.192 | 0.125 |
| 2009 | 0.271 | 0.323 | 0.225 | 0.151 | 0.109 | 0.128 | 0.088 | 0.267 | 0.133 | 0.134 |
| 2010 | 0.227 | 0.247 | 0.125 | 0.213 | 0.123 | 0.095 | -0.015 | 0.380 | 0.140 | 0.119 |
| 2011 | 0.290 | 0.248 | 0.257 | 0.152 | 0.147 | 0.100 | 0.199 | 0.086 | 0.148 | 0.107 |
| 2012 | 0.204 | 0.347 | 0.142 | 0.326 | 0.126 | 0.106 | -0.244 | 0.000 | 0.113 | 0.071 |
| 2013 | 0.302 | 0.230 | 0.232 | 0.220 | 0.122 | 0.095 | 0.046 | 0.337 | 0.154 | 0.094 |
| **2005-2013** | **0.267** | **0.254** | **0.211** | **0.231** | **0.143** | **0.097** | **0.329** | **0.254** | **0.166** | **0.107** |

**\* Note:** Anis-Lloyd correction for small sample bias (see Weron, R. (2002a) [95], Anis, A.A. et al. (1976) [96], Peters, E. (1994) [86])

**Table 9:** Hurst exponents computed by various methods for SMP, PUN raw and deseasonalized returns (2005-2013)

| SMP | | |
|---|---|---|
| Method | Original returns | Deseasonalized returns* |
| Hurst GPH | 0.330 | 0.263 |
| Hurst R/S | 0.211 | 0.212 |
| Hurst DFA ($\alpha$) | 0.143 | 0.158 |
| Hurst GHE (prices) (q=1) | 0.166 | 0.189 |
| **PUN** | | |
| Method | Original returns | Deseasonalized returns* |
| Hurst GPH | 0.254 | 0.190 |
| Hurst R/S | 0.231 | 0.246 |
| Hurst DFA ($\alpha$) | 0.097 | 0.124 |
| Hurst GHE (prices) (q=1) | 0.107 | 0.138 |

**\*Note:** for the Hurst values computed by GHE method we use prices and deseasonalized prices, not returns

It is apparent from the tables above that both SMP and PUN are generated by anti-persistent, mean-reverting processes far away from the Geometric Brownian Motion for which H=0.5. Consequently the price process is non-Markovian (Shreve, S., (2004) [97], Eydenland, A., Wolyniec, K. (2003) [98], Benth, F. et al. (2008) [99]) which means that if we move away from in principle time-varying fundamental price level, we tend to return back to this level again (mean reverting process). Thus, spot electricity price process is a **(non-Markovian) anti-correlated, mean-reverting process.** This result is of great importance in pricing options or futures in electricity markets. If this behavior of spot price is not taken into account it increases the risk exposure for someone which

"writes" the Option contract. It is also well known, at least in the financial markets, that the fundamental assumption in the **Black-Scholes-Merton (BSM)** model for option pricing (Black, F. (1973) [100], Merton, R. (1973) [101], Eydenland, A., Wolyniec, K. (2003) [98]) is that the underline asset follows a **Geometric Brownian Motion, GBM (H=0.5)**. For the electricity spot price process, this assumption is far from being satisfied. It is very important, therefore, to understand why this anti-persistent, mean-reverting dynamics is shown to be so robust with time. In a stock market we also observe such a process, but it is shortly lived because the agents take advantage of this, they "learn" about it quickly, so ultimately this arbitrage opportunity is vanished. In order to shed light on this peculiar feature, we must note some specific characteristics of a commodity (electricity is a peculiar commodity). To hold a **long position** in a commodity, in general, we have to pay certain price that normally is much higher than, say, a stock, since we need to rent storage for it. Also, as time passes by, the commodity may be deteriorated, etc. In case of electricity, its storage is not economically efficient or is not possible at all (no efficient, at a reasonable price, technology exist today for storing large amounts of electricity). Thus, this commodity has to be consumed immediately, after generation. As a consequence, no economical motives exist for exploiting the non-vanishing anti-correlations, i.e. the mean-reverting process of power prices does not represent an arbitrage opportunity as in the financial markets.

We now show the relations between indeces $\beta$ (power spectral index), DFA index $\alpha$ (section 5.3.2), index in decaying ACF $\delta$ (23), and the Hurst exponent $H$. Using Wiener-Khinchin theorem we have

$$\begin{aligned}\delta &= 2 - 2\alpha \\ \beta &= 2\alpha - 1 \\ \delta &= 1 - \beta\end{aligned} \quad (28)$$

Therefore, $\alpha$ is connected to the slope of the power spectrum $\beta$, useful in describing the **color noise** by the following relation

$$\alpha = (\beta + 1)/2 \quad (28a)$$

For **fractional Gaussian noise (FGN)**, $\beta \in [-1, 1]$ therefore $\alpha = [0, 1]$ and $\beta = 2H - 1$.

For **fractional Brownian motion (FBM)**, $\beta \in [1, 3]$ so $\alpha = [1, 2]$ and $\beta = 2H + 1$. The index $\alpha$ is equal to $H + 1$ for FBM. Therefore, FBM is the cumulative sum or the integral of FGN, which equivalently means that they have power spectra $\beta \neq 2$. As we have seen in section 5.3, the two markets do not belong to either fGN or fBm categories.

### 5.4 The maximal Lyapunov Exponent

Chaos arises from the exponential growth of infinitesimal perturbations associated simultaneously with a global folding process in order to secure boundedness of the trajectory evolution. Lyapunov exponents characterize this exponential instability (Eckman et al. ,1985 [102]). Suppose that we assume a local decomposition of the phase space into directions with different stretching or contraction rates. It is natural then to consider that the spectrum of all exponents is the proper average of these local rates over the whole invariant set (there are as many exponents as there are space directions).

Since we do not really know the exact number of degrees of freedom of the dynamics we investigate (more often the physical phase space is unknown), it creates a difficult problem in time series analysis, so instead the spectrum of Lyapunov exponents is computed in some embedding space (see section 5.2.2), meaning that the number of exponents depends on the dimensions of the reconstructed space, and might be larger than in the actual or physical phase space. Therefore, the additional exponents (corresponding to extra dimensions) are spurious. Stoop et al. (1991) [103] provide practical suggestions to either avoid or to identify them. We point out here that Lyapunov exponents are invariant under smooth transformations and are thus independent of the measurement function or the embedding procedure. They carry a dimension of an inverse time and have to be normalized to the sampling interval (Kantz H., 1994 [104]).

We can determine the maximal Lyapunov exponent without the explicit construction of a model for the time series. Checking explicitly the independence of embedding parameters and the exponential law for the growth of distances secures a reliable characterization. The presence of noise in a data series is the main problem in detecting Chaos. Therefore in any test used in searching for chaos the noise effect must be taken into consideration.

We consider the time series data as a trajectory in the *reconstructed embedding space*, of dimension m and delay time τ (see relation (8), section 5,2,2). Let us assume also that we observe in a neighbor $u_i$, of diameter $\epsilon$, a very close return $s_i'$, to a previously visited point $s_i$. We now can consider the distance $\Delta_0 = s_i - s_i'$ as a small perturbation, which should grow exponentially in time. Its future dynamic evolution some time t=*l* ahead can be monitored by plotting the time series: $\Delta_l = s_{i+l} - s_{i+l}'$. If we find that $|\Delta_l| \approx \Delta_0 e^{\lambda l}$ then $\lambda$ is (with probability one) the maximal Lyapunov exponent. Because of many effects, in practice there will be fluctuations. Therefore, due to fluctuations we can derive a robust, consistent and unbiased estimator for the maximal Lyapunov exponent given as follows

$$S(\epsilon, m, t) = \left\langle \ln\left(\frac{1}{|u_i|} \sum_{s_{i'} \in u_i} |s_{i+t} - s_{i'+t}|\right) \right\rangle \quad (29)$$

If $S(\epsilon, m, t)$ exhibits a linear increase with identical slope for all m larger than some $m_0$ and for a reasonable range of $\epsilon$, then the slope can be taken as an estimate of the maximal exponent $\lambda_{\max}$. We point out here that the r.h.s of equation (30) is the *average divergence or separations of Nearest Trajectories NT*, and is the basis of the method developed by Rosenstein et al. (1993) [105] in estimating $\lambda_{\max}$.

The Lyapunov exponent measures the average exponential divergence (in case of (+) exponent) or convergence (in case of (-) exponent) rate between two adjacent trajectories or Nearest Trajectories (NT) within a time distance that differ in initial conditions just by an extremely tiny amount. If $\lambda < 0$ then the orbits attracts to a stable **fixed point or stable periodic orbit**. Dissipative or non-conservative dynamical systems have negative Lyapunov exponents, exhibiting asymptotic stability. **The more negative the exponent, the greater the stability.** At the extreme side, $\lambda = -\infty$ corresponds to super-stable fixed or periodic points. If $\lambda = 0$ the orbit is actually a **neutral fixed** point so the system is in a **steady state** mode and it is called a **conservative** system, exhibiting Lyapunov stability. Such a system with $\lambda \approx 0$ is near the "transition to chaos". If $\lambda > 0$, the trajectory is **unstable and chaotic**. Neighboring points, no matter how "close" to each other are, will diverge, in time, with an arbitrary distance. Therefore, the entire phase space will be "filled" with visiting points that are unstable. The maximum Lyapunov exponent test is considered to be the natural way for chaos detection and constitutes the most direct approach for its estimation (Yousefpoor et al, 2008 [60]). According to Barnett and Serletis (2000) [61], Banks et al, (2003) [106], Iseri et al (2008) [107] and Yousefpoor et al (2008) [60], maximum Lyapunov exponent, $\lambda_{\max}$, is the most appropriate statistical invariant in detecting the sensitive dependence on initial conditions characteristic of a chaotic signal. As we seen abobe, max Lyapunov exponent, $\lambda_{\max}$, measures the average rate of exponential separation (positive value) or contraction (negative value) between two nearby trajectories in the initial conditions (Barnett and Serletis, 2000 [61]; Kyrtsou and Terraza, 2002 [108]; Yousefpoor et al, 2008 [60]; Orzeszko, 2008 [109]). In general, a (+) $\lambda_{\max}$, indicates a chaotic system while a (-) $\lambda_{\max}$ indicates a stochastic system.

For the calculation of $\lambda_{\max}$ three methods are available : a) The *direct method* developed by Wolf et al (1985) [110] is the older one, also used by Rosenestein et al (1993) [111] (in which, as said above, the average of NT is used as given in relation (3) above), b) *the Jacobian or regression method*, developed by Nychka et al (1992) [112], and McCaffrey et al (1992) [113], according to which $\lambda_{\max}$ is calculated by first estimating the Jacobians and then using neural network models (Barnett and Serletis, 2000 [61]; Yousefpoor et al, 2008 [60]) and finally c) the third method based on a new algorithm using neural network methodology suggested by Gençay and Dechert (1992) [114], Kyrtsou and Terraza, 2002 [108].

The direct method requires long time series containing a large number of observations (Brock and Sayers, 1988 [115], Barnett and Serletis, 2000 [61]) and it is very sensitive in the presence of noise, so the detection of chaotic dynamics requires a delicate and tedious work and the result cannot be easily accepted as a reliable one, especially for noisy financial series (Kyrtsou and Terraza, 2002 [108]), since in most cases gives overestimated exponents (Kyrtsou and Terraza, 2002 [108], Barnett and Serletis, 2000 [61]). The problem of noise sensitivity of the first method is solved by the third method. Overall, according to existing literature in empirical studies of max Lyapunov exponent, the Jacobian approach provides the most reliable estimation despite the presence of noise in the time series.

In this work we adopt the Jacobian method which is briefly described below following the work of Barnet and Serletis (2000, p. 712) [61]; based on the method as suggested by Nychka, D., et al, (1992) [112], Nychka, D., et al. (1997) [116], Bailey, B., et al. (1998) [117].

According to Jacobian method, in order to estimating $\lambda_{max}$ we must first estimate the individual Jacobian matrixes:

$$J_t = \frac{\partial F(S_t)}{\partial S^T} \tag{30}$$

where $S_t$ is the state-space representation of $s_t$, given as

$$S_t = F(S_{t-\tau}) + E_t, \quad F: \mathbb{R}^m \to \mathbb{R}^m \tag{31}$$

$S_t$ can be expressed as a function of $s_t$ as $S_t = (s_t, s_{t-\tau}, \ldots, s)$, where $s_t$ is the sequence of variables $\{s_t\}$ produced by a nonlinear AR model, given below.

For $1 \leq t \leq N$, let parameter $\tau$ is the time-delay and $m$ is the length of the autoregression, so we have

$$s_t = f(s_{t-\tau}, s_{t-2\tau}, \ldots, s_{t-m\tau}) + e_t \tag{32}$$

Therefore,

$$F(S_{t-\tau}) = (f(s_{t-\tau}, \ldots, s), s, \ldots, s_{t-m\tau+\tau})^T \tag{33}$$

and

$$E_t = (e_t, 0, \ldots, 0)^T \tag{34}$$

where $f$ is a smooth unknown function and $\{e_t\}$ is a random process, independent random variables with zero mean and unknown constant variance.

The optimal combination of the parameters in the calculation of max Lyapunov exponent, of the triplet $(\tau, m, q)$, is found by minimizing the Bayesian information criterion (see Nychka et al, 1992 [112]). So, the estimate of the maximum Lyapunov exponent $\hat{\lambda}_{max}$ is given by (35)

$$\hat{\lambda}_{max} = \frac{1}{2N} \log|\hat{v}_1(N)| \tag{35}$$

where $\hat{v}_1(N)$ is the largest eigenvalue of the matrix $\hat{T}_N^T \hat{T}_N$ and

$$\hat{T}_N = \hat{J}_N \hat{J}_{N-1} \ldots \hat{J}_1 \tag{36}$$

$\hat{J}_t$, the nonparametric estimator of $J_t$, can be written as

$$\hat{J}_t = \frac{\partial \hat{F}(X_t)}{\partial X^T} \tag{37}$$

A relatively fast algorithm in estimating $\lambda_{max}$ based on the above method and Neural Networks is provided by BenSaida A. (2015) [118]. In his algorithm, the test hypothesis H are : null hypothesis $H_0$: $\lambda \geq 0$, which indicate presence of Chaos ; and alternative hypothesis $H_1$: $\lambda < 0$, which indicates absence of chaos. The crucial parameters in applying this method on noisy real data are the embedding dimension, m, the delay time, τ, (as described and estimated in section 5, 5.2) and the optimum (according to some criteria described in BenSaida's paper) number of hidden layers, q, in the Neural Network. The best triplet (τ,m,q) is found by the algorithm, based on preliminary max orders of these parameters, estimated by trial and error until a saturation of the effects of changes of their combination on λ is attained. The algorithm then finds the optimum orders of (τ,m,q) that maximizes λ, computed from all τ x m x q estimates (see BenSaida A. et al., 2013 [119], BenSaida A., 2015 [118]). In appendix C we provide the trial and errors results. The preliminary max orders of the triplet found are (2,7,3) for both SMP and PUN, while the optimum ones are as shown in the table 10 below :

**Table 10 :** Max Lyapunov Exponent, $λ_{max}$, for SMP and PUN daily raw returns, 2005-13, using BenSaid algorithm, which implements Nychka's et al. (1992) [112] method.

| Series | Optimum Orders (τ,m,q) | $λ_{max}$ | p-Value | Confidence Interval, CI | Hypothesis |
|---|---|---|---|---|---|
| SMP | (2,7,3) | **-0.1663** | 0.000 | [-0.1774, inf[ | $H_1$ |
| PUN | (1,7,1) | **-0.0694** | 0.000 | [-0.0793,inf[ | $H_1$ |

**Table 11:** Maximum Lyapunov Exponent, computed for each separate year, for SMP and PUN returns hourly data (based on 8760 data and estimated by BenSaid A. algorithm (BenSaida A.,2015 [118]), using the optimum orders of parameters (τ,m,q).

| Year | Time delay, τ (by Mutual Information) | | Embedding, m (by False Nearest Neighbors, FNN) | | Maximum Lyapunov Exponent, $λ_{max}$ | | Dynamical System stability or chaos | |
|---|---|---|---|---|---|---|---|---|
| | SMP | PUN | SMP | PUN | SMP | PUN | SMP | PUN |
| **2005** | 3 | 2 | 10 | 6 | **0.1102** | -0,257 | chaos | stability |
| **2006** | 3 | 2 | 10 | 6 | **0.6550** | -0.2090 | chaos | stability |
| **2007** | 3 | 2 | 10 | 6 | -0.0180 | -0.1600 | stability | stability |
| **2008** | 3 | 3 | 10 | 6 | -0.0076 | 0.1158 | stability | stability |
| **2009** | 3 | 3 | 10 | 6 | 0.3957 | -0.1280 | stability | stability |
| **2010** | 3 | 3 | 10 | 6 | -0.0577 | -0.1560 | stability | stability |
| **2011** | 3 | 3 | 10 | 6 | **0.1140** | -0.1266 | chaos | stability |
| **2012** | 3 | 3 | 10 | 6 | -0.1034 | 0.0740 | stability | stability |
| **2013** | 3 | 3 | 10 | 6 | **0.0369** | -0.0835 | chaos | stability |

For the calculations in table 11, we used hourly data (8760 values per year) providing an adequate date set size, so securing good statistics in $λ_{max}$ estimation. We observe that $λ_{max}$ is negative for the entire period 2005-13, indicatind a stable dunamical system, while for the yearly computations PUN is a stable system while SMP returns for years 2005, 2006, 2011 and 2013 show positive $λ_{max}$, indicating that the market was, in that period, in a "chaotic" state, however not permanent. Overall, we can state that *both markets are not chaotic ($λ_{max} < 0$)*, since there is no evidence of *persistent* chaotic dynamics. This result is consistent with the *mean-reverting* dynamic behavior indicated by Hurst exponent and the SARIMA/EGARCH model for the conditional mean and variance of both wholesale prices.

In figure 17 we plot the maximum Lyapunov exponent as a function of embedding dimension, computed by the Jacobian method, using BenSaida's algorithm. The figure reveals that $\lambda_{max}$ for both returns converge to the interval [-0.05, -0.15], implying the the market systems generating the returns converge to an *asymptotic stability*, more pronounced in the case of Italy.

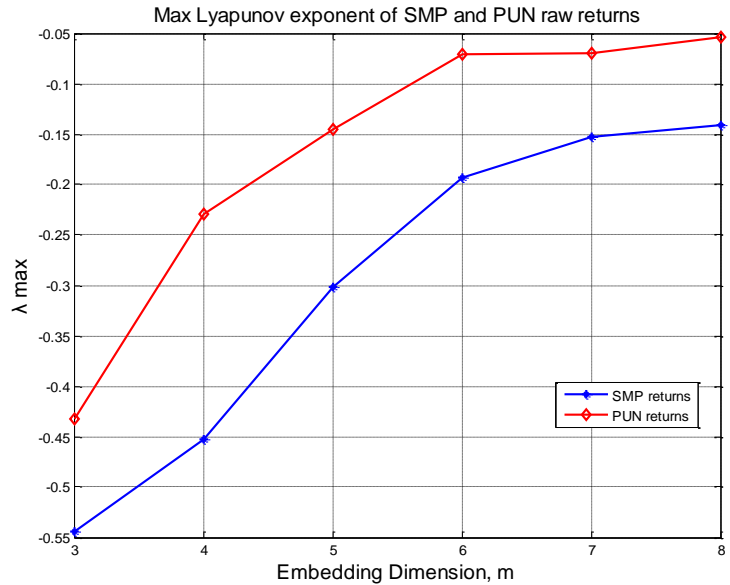

**Figure 16:** max Lyapunov exponent versus the embedding dimension, m, for SMP and PUN raw returns, computed by BenSaida's algorithm.

From the above tables and figure 16 it is clear that there is no unstable (chaotic) dynamics in both SMP and PUN returns since the largest Lyapunor exponent is negative. However it is also clear that the SMP returns have *a more negative $\lambda_{max}$* than the PUN returns, *suggesting a more stable dynamic process ( a more stable 'market' )* that generates the returns.

We now compute the changes in max Lyapunov exponent from year to year.

**Table 12 :** Directions of changes of $\lambda_{max}$, $H$ and $\sigma$ for SMP and PUN returns

| Year | $\lambda_{max}$ | $\Delta\lambda_{max}$ | Stability | Conditional Variance (Conditional Volatility $\sigma_t$) | $\Delta\sigma_t$ | Direction of change of Volatility |
|---|---|---|---|---|---|---|
| SMP | | | | | | |
| 2005 | 0.1102 | - | - | 0.0890    (0.298) | - | - |
| 2006 | 0.655 | LN | decrease | 0.00746   (0.086) | -0.212 | decrease |

| Year | λmax | ΔλMax | Stability | Conditional Variance (Conditional Volatility σt) | Δσt | Direction of change of Volatility |
|------|------|-------|-----------|--------------------------------------------------|------|-----------------------------------|
| 2007 | -0.0180 | MN | increase | 0.01153 (0.107) | 0.021 | increase |
| 2008 | -0.0076 | LN | decrease | 0.00904 (0.095) | -0.012 | decrease |
| **2009** | 0.3957 | LN | decrease | 0.02177 (0.147) | 0.052 | increase |
| **2010** | -0.0577 | MN | increase | 0.0108 (0.103) | -0.044 | decrease |
| 2011 | 0.1140 | LN | decrease | 0.004618 (0.067) | -0.036 | decrease |
| 2012 | -0.1034 | MN | increase | 0.009197 (0.0959) | 0.029 | increase |
| **2013** | 0.0369 | LN | decrease | 0.01047 (0.1023) | 0.006 | increase |
| PUN | | | | | | |
| Year | λmax | ΔλMax | Stability | Conditional Variance (Conditional Volatility σt) | Δσt | Direction of change of Volatility |
| 2005 | -0.257 | - | - | 0.0317 (0.1780) | - | - |
| 2006 | -0.209 | LN | decrease | 0.0163 (0.127) | -0.051 | decrease |
| 2007 | -0.160 | LN | decrease | 0.0075 (0.086) | -0.041 | decrease |
| **2008** | -0.116 | LN | decrease | 0.0077 (0.087) | 0.001 | increase |
| 2009 | -0.128 | MN | increase | 0.014 (0.118) | 0.031 | increase |
| 2010 | -0.156 | MN | increase | 0.0225 (0.15) | 0.032 | increase |
| 2011 | -0.126 | LN | decrease | 0.01851 (0.136) | -0.014 | decrease |
| **2012** | -0.074 | LN | decrease | 0.0439 (0.209) | 0.073 | increase |
| **2013** | -0.0835 | MN | increase | 0.03109 (0.176) | -0.033 | decrease |

From **Table 12** we can track the direction of changes of $\lambda_{max}$, and conditional volatility according to the following rational:

- If $\lambda_{max}$ becomes more negative (MN) then the stability is increased (I) so the volatility $\sigma$ must decrease (D).
- The further Hurst exponent, $H$ moves from 0.5, either towards to 0 or to 1, the *less efficient* the market becomes (the more mean-reverting or more persistent, respectively, is the time series), which is equivalent to say, **more predictable but also more volatile, due to inefficiency.** So, stability and volatility must move in opposite directions (i.e they are negatively correlated). From the Table above the change directions of $\lambda_{max}$ and $\sigma$ are "reasonable" for only a number of periods indicated by bold.

It is proved that as the dominant and stable eigenvalue decreases (hence its distance from zero increases), the system becomes more stable. Closely related to stability is the notion of volatility. We can draw conclusions on the stability of the market by observing the behavior of its respective volatility. It is clear that in almost all cases of the Greek Electricity Market, a *larger negative eigenvalue leads to reduced volatility (and vice versa)*, meaning that as the system becomes **more stable**, it also becomes **less volatile**. In the case of PUN however this induction cannot be applied. We observe an almost opposite behavior, showing increased volatility as the dominant eigenvalue sinks further into the negative plane. This could be explained as increased betting from the side of the electricity traders during periods of established certainty and conversely, as more conservative activity during periods of uncertainty. This can be combined with the observation drawn earlier, that the Italian market experiences long memory effect, so traders know the previous state of the system and try to profit from its current status. Additionally, we know that the Italian market is well interconnected with external markets and during the period of financial crisis, it is possible that the phenomenon of volatility spillover occurred and resulted to unwanted increase of the volatility of electricity prices. There are also the first two years, when the PUN showed a positive eigenvalue, which represents an unstable system. The regulatory authorities quickly took countermeasures to dispel the uncertainty linked with instability and drag the market to stability and after two years the market showed negative eigenvalues.

*5.4.2 Application of tools on Conditional Volatility*

The concept of volatility has been used extensively in financial markets in order to define the degree of uncertainty of a time series over time. Different statistical approaches are currently used to measure volatility. A GARCH(P,Q) model of concitional volatility variance is:

$$\varepsilon_t = \sigma_t z_t, \qquad z_t \text{ is i.i.d process}$$

$$\sigma_t^2 = k + \sum_{i=1}^{P} \gamma_i \sigma_{t-i}^2 + \sum_{j=1}^{Q} \alpha_j \varepsilon_{t-j}^2$$

The most simplest and used GARCH process is the GARCH(1,1) which is described by the following equations:

$$\varepsilon_t = \sigma_t z_t$$

$$\sigma^2(t) = a_1 + b\, z_{t-1}^2 + c\, \sigma_{t-1}^2$$

Where $z_t$ is an i.i.d random variable with zero mean and unity variance, i.e. $<z_t> = 0$ and $<z_t^2> = 1$. The quantity $\sigma_t$ (volatility) is time varying and as shown from the previous equations is dependent on the past values of the return $z_t$. The GARCH(1,1) process is completely defined when parameters k, z and a as well as the noise nature $P_n(w_t)$ are specified.

The best model found based on AIC and BIC criteria (Appendix A) for the conditional variance for the entire period 2005-2013 is an EGARCH(1,1), for both series, which now takes the form:

$$\log(\sigma_t^2) = k + \gamma_1 \log(\sigma_{t-1}^2) + a_1 \left[\frac{|\varepsilon_{t-1}|}{\sigma_{t-1}} - E\left\{\frac{|\varepsilon_{t-1}|}{\sigma_{t-1}}\right\}\right] + \xi_1 \left(\frac{\varepsilon_{t-1}}{\sigma_{t-1}}\right) \tag{40}$$

In relation (40), the term $E\left\{\frac{|\varepsilon_{t-1}|}{\sigma_{t-1}}\right\} = E\{z_{t-1}\} = \sqrt{\frac{2}{\pi}}$, for Gaussian distribution, so it is a mean zero shock. The last term contains $\varepsilon_{t-1}$, which is also a mean zero shock. The two shocks behave differently (the $\varepsilon_{t-1}$ terms). The second produces a symmetric rise in the log variance while the last term creates an asymmetric effect. $\xi_1$ is typically estimated to be **less than zero** and **volatility rises more subsequent to negative shocks than to positive ones**. In the usual case where $\xi_1 < 0$, the magnitude of the shock can be decomposed by conditioning on the sign of $\varepsilon_{t-1}$.

$$shock\ coefficient = \begin{cases} \alpha_1 + \xi_1 & \text{when } \varepsilon_{t-1} < 0 \\ \alpha_1 - \xi_1 & \text{when } \varepsilon_{t-1} > 0 \end{cases} \tag{41}$$

Therefore, both shocks are mean zero and the current log variance is linearly related to past log variance via $\gamma_1$, the EGARCH(1,1) model is an AR model. We point out here that EGARCH models provide superior fits when compared to standard GARCH models. The essence of the asymmetric term is largely for the superior fit since many asset return series have been found to exhibit a "leverage" effect and the use of standardized shocks ($\varepsilon_{t-1}$) in the evolution of the log variance tend to dampen the effect of large shocks. Tables 13 & 14 summarize conditional mean and conditional volatility of the optimal model for both SMP and PUN returns respectively.

**Table 13:** Conditional Mean & Conditional Volatility of the optimal model for the SMP returns

| Conditional Mean | | | |
|---|---|---|---|
| **Variable** | **Symbol** | **Coefficient** | **Probability** |
| AR(1) | $\varphi_1$ | 0.2230 | 0.0000 |
| SAR(7) | $\Phi_1$ | -1.0420 | 0.0000 |
| SAR(14) | $\Phi_2$ | -0.7990 | 0.0000 |
| SAR(21) | $\Phi_3$ | 0.0343 | 0.0000 |
| MA(1) | $\theta_1$ | -0.7200 | 0.0000 |
| SMA(7) | $\Theta_1$ | 0.1610 | 0.0000 |
| SMA(14) | $\Theta_2$ | -0.1610 | 0.0000 |
| SMA(21) | $\Theta_3$ | -0.8041 | 0.0000 |
| **Conditional Volatility** | | | |
| **Variable** | **Symbol** | **Coefficient** | **Probability** |
| K | $k$ | -0.4320 | 0.0000 |
| GARCH(1) | $\gamma_1$ | 0.9085 | 0.0000 |
| ARCH(1) | $\alpha_1$ | 0.3477 | 0.0000 |
| Leverage(1) | $\xi_1$ | -0.0662 | 0.0000 |

**Table 14:** Conditional Mean & Conditional Volatility of the optimal model for the PUN returns

| Conditional Mean | | | |
|---|---|---|---|
| **Variable** | **Symbol** | **Coefficient** | **Probability** |
| AR(1) | $\varphi_1$ | 1.2090 | 0.0000 |
| AR(2) | $\varphi_2$ | -0.2810 | 0.0000 |
| SAR(7) | $\Phi_1$ | 0.6780 | 0.0000 |
| SAR(14) | $\Phi_2$ | 0.2030 | 0.0000 |
| MA(1) | $\theta_1$ | -1.6240 | 0.0000 |
| MA(2) | $\theta_2$ | 0.6260 | 0.0000 |
| MA(3) | $\theta_3$ | 0.0070 | 0.0000 |
| SMA(7) | $\Theta_1$ | -0.6350 | 0.0000 |
| SMA14) | $\Theta_2$ | -0.1840 | 0.0000 |

| Conditional Volatility | | | |
|---|---|---|---|
| Variable | Symbol | Coefficient | Probability |
| K | $k$ | -0.1110 | 0.0000 |
| GARCH(1) | $\gamma_1$ | 0.9710 | 0.0000 |
| ARCH(1) | $\alpha_1$ | 0.2380 | 0.0000 |
| Leverage(1) | $\xi_1$ | -0.0853 | 0.0000 |

Substituting the values of coefficients $K, \gamma_1, \alpha_1, \xi_1,$ the final models for the Conditional Volatility of PUN and SMP are written respectively:

$$\log(\sigma_t^2) = -0.1432 + 0.9085 \log \sigma_{t-1}^2 + 0.347 \left[\frac{|\varepsilon_{t-1}|}{\sigma_{t-1}} - E\left\{\frac{|\varepsilon_{t-1}|}{\sigma_{t-1}}\right\}\right] - 0.066 \left(\frac{\varepsilon_{t-1}}{\sigma_{t-1}}\right) \tag{42}$$

$$\log(\sigma_t) = -0.111 + 0.971 \log \sigma_{t-1}^2 + 0.238 \left[\frac{|\varepsilon_{t-1}|}{\sigma_{t-1}} - E\left\{\frac{|\varepsilon_{t-1}|}{\sigma_{t-1}}\right\}\right] - 0.0853 \left(\frac{\varepsilon_{t-1}}{\sigma_{t-1}}\right) \tag{43}$$

where the required condition for the **persistence** $\gamma < 1$ holds for both models. Since $\xi_i < 0$ for both series the *leverage effect* and the *asymmetric impact conditions* are confirmed. This result for PUN is in the opposite direction with that found in the work of Petrella and Sapio(2012) [120], possibly due to different sampling periods or due to the fact that the PUN series, in our case, has undergone a strong filtering before fitted to the model. Figures 21 and 22 below, show the conditional volatilities and residuals from the models provided in tables 18 and 19 respectively.

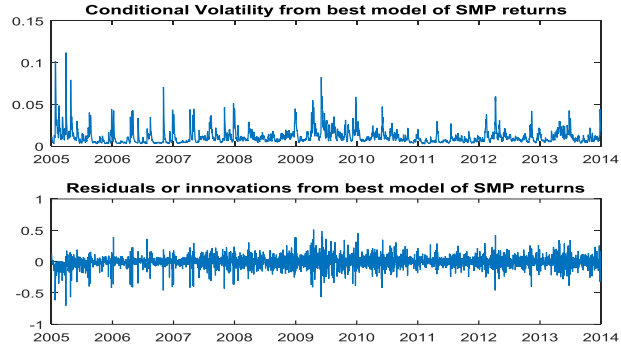

**Figure 17:** The graphs of conditional volatility and residuals of SMP returns from model given in table 18.

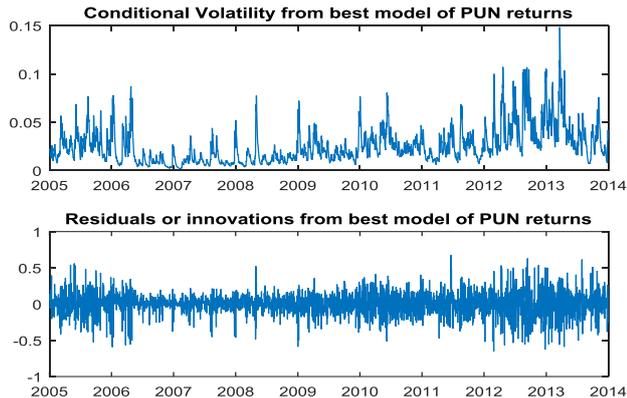

**Figure 18:** The graphs of conditional volatility and residuals of PUN returns from model given in table 19.

Table 15, shows the inter-annual GARCH coefficients as well as the theoretical Unconditional Variance $\sigma_\varepsilon$.

**Table 15:** Inter-Annual GARCH Coefficients $\gamma_1$ and theoretical Unconditional Variance $\sigma_\varepsilon$

| Period | SMP returns | | | | PUN | | | |
|---|---|---|---|---|---|---|---|---|
| | $\gamma_1$ | $\Delta\gamma_1$ | $\sigma_\varepsilon$ | $\Delta\sigma_\varepsilon$ | $\gamma_1$ | $\Delta\gamma_1$ | $\sigma_\varepsilon$ | $\Delta\sigma_\varepsilon$ |
| **2005** | 0.928 | - | 0.055 | - | 0 | - | 0.038 | - |
| **2006** | 0.000 | -0.928 | 0.007 | -0.0543 | 0.8077 | 0.807 | 0.009 | -0.026 |
| **2007** | 0.1414 | 0.1414 | 0.013 | 0.006 | 0.2136 | -0.594 | 0.071 | 0.062 |
| **2008** | 0.852 | 0.710 | 0.010 | -0.003 | 0.9628 | 0.749 | 0.006 | -0.065 |
| **2009** | 0.778 | -0.074 | 0.027 | 0.017 | 0.6436 | -0.319 | 0.014 | 0.008 |
| **2010** | 0.0204 | -0.757 | 0.011 | -0.016 | 0.753 | 0.109 | 0.023 | 0.009 |
| **2011** | 0.651 | 0.630 | 0.004 | -0.007 | 0.950 | 0.197 | 0.017 | -0.006 |
| **2012** | 0.865 | 0.214 | 0.008 | 0.004 | 0.183 | -0.767 | 0.042 | 0.025 |
| **2013** | 0 | -0.865 | 0.012 | 0.004 | 0.897 | 0.714 | 0.032 | -0.01 |

Theoretical Unconditional Variance $\sigma_\varepsilon$ can be easily calculated from GARCH coefficients with good accuracy. In general, this assumption is valid as long as the preferred model has inherent symmetry; this is not applicable in our case where the best found model is the EGARCH(1,1). Table 16 summarizes the unconditional volatility on raw returns and log returns for both SMP and PUN series. Also HHI, Hurst and Lyapunon exponents have been calculated for both markets.

**Table 16:** Annual mean values of volatility HHI, Hurst and maximum Lyapunov exponents for the Greek and the Italian markets

| Year | Unconditional Volatility (St.dev.) (raw returns) | | Unconditional Volatility (St.dev.) (fully deseasonalized log returns) | | HHI (Greek Market) | HHI (Italian Market) | Hurst, R/S | | Lyapunov, $\lambda_{max}$ | |
|---|---|---|---|---|---|---|---|---|---|---|
| | SMP | PUN | SMP | PUN | | | SMP | PUN | SMP | PUN |
| **2005** | 0.226 | 0.230 | 0.172 | 0.251 | 9409 | 1900 | 0.2220 | 0.2050 | 0.1102 | -0.2570 |
| **2006** | 0.145 | 0.179 | 0.114 | 0.164 | 8905 | 1643 | 0.2070 | 0.2520 | 0.6550 | -0.2090 |
| **2007** | 0.123 | 0.179 | 0.092 | 0.173 | 8390 | 1440 | 0.2790 | 0.2580 | -0.0180 | -0.1600 |
| **2008** | 0.134 | 0.139 | 0.105 | 0.130 | 8390 | 1380 | 0.3250 | 0.2580 | -0.0076 | -0.1158 |
| **2009** | 0.161 | 0.164 | 0.137 | 0.213 | 8427 | 1280 | 0.2710 | 0.3230 | 0.3960 | -0.1280 |
| **2010** | 0.201 | 0.127 | 0.178 | 0.095 | 6844 | 1097 | 0.2270 | 0.2470 | -0.0577 | -0.1560 |
| **2011** | 0.194 | 0.087 | 0.160 | 0.132 | 5746 | 953 | 0.2900 | 0.2480 | 0.1140 | -0.1266 |

| | | | | | | | | | |
|---|---|---|---|---|---|---|---|---|---|
| **2012** | 0.275 | 0.130 | 0.254 | 0.145 | 6983 | 864 | 0.2040 | 0.3470 | -0.1034 | -0.0740 |
| **2013** | 0.277 | 0.145 | 0.237 | 0.087 | 6553 | 830 | 0.3020 | 0.2300 | 0.2130 | -0.0835 |
| **2005-2013** | **0.199** | **0.160** | **0.167** | **0.164** | - | - | - | - | - | - |

Table 17 compares volatility (conditional and unconditional) against system's stability expressed through Lyapunov exponents.

**Table 17:** Correlation matrix between volatility $\sigma$ and $\lambda_{max}$

| Volability $\sigma$ | $\lambda_{max}$ | |
|---|---|---|
| | SMP_lyap | PUN_lyap |
| **Unconditional Volatility (raw)** | -0.260 | -0.734 |
| **Unconditional Volatility (deseasonalized)** | -0.287 | -0.643 |
| **Conditional Volatility (deseasonalized)** | -0.0049 | 0.122 (not expected) |

From table 17 we conclude that less price volatility in a market means that the dynamical system generating the price series is **more stable** i.e the max Lyapunov exponent $\lambda_{max}$ is **more negative**, therefore the correlation coefficient between $\sigma$ and $\lambda_{max}$ is expected to be **negative**

$$\text{corr}(\sigma_t, \lambda_{max}) < 0$$

If the market competition increases which is equivalent to say that market concentration reduces or HHI decreases, then the distance Hurst exponent from the value of 0.5 (EMH) is expected also to decrease (since the market becomes more and more efficient, and Hurst moves toward 0.5). Therefore the correlation between Hurst distance from 0.5 and HHI is expected to be positive i.e

$$\text{Corr (Hurst distance, HHI)} > 0$$

The more efficient becomes the market (i.e Hurst distance from 0.5 is getting less and less) the less volatile is the market, so Hurst_distance and $\sigma$ are positively correlated

$$\text{Corr (Hurst distance}, \sigma) > 0$$

The former conclusions are summarized on table 18; from this table we extract useful conclusions regarding the correlation of conditional and unconditional volatility, HHI, Hurst and, Lyapunov exponents as well as the correlation between Hurst distance from 0.5, for both markers; these are designated as: uv_smp for the unconditional volatility of the SMP, uv_des_smp for the deseasonalized unconditional volatility of the SMP, C_VOLAT_SMP for the conditional volatility of the SMP, HHI_G is the HHI for the Greek Market, Hurst_RS_smp is the Hurst exponent of the SMP, H_smp_dist is the Hurst distance from the 0.5 of the SMP, and smp_lyap is the Lyapunov exponent of the SMP. The respective designations for the Italian Market use PUN instead of SMP and IT instead of G. For instance uv_pun is the unconditional volatility of the PUN whereas, HHI_IT is the HHI for the Italian Electricity Market.

**Table 18 :**

| | uv_smp | uv_pun | uv_des_smp | uv_des_pun | C_VOLAT_SMP | C_VOLAT_PUN | HHI_G | HHI_IT | Hurst_RS_smp | H_smp_dist | Hurst_RS_pun | H_pun_dist | smp_lyap | pun_lyap |
|---|---|---|---|---|---|---|---|---|---|---|---|---|---|---|
| uv_smp | 1,00 | | | | | | | | | | | | | |
| uv_pun | -0,15 | 1,00 | | | | | | | | | | | | |
| uv_des_smp | 0,98 | -0,26 | 1,00 | | | | | | | | | | | |
| uv_des_pun | -0,24 | 0,75 | -0,32 | 1,00 | | | | | | | | | | |
| C_VOLAT_SMP | 0,17 | 0,79 | 0,04 | 0,76 | 1,00 | | | | | | | | | |
| C_VOLAT_PUN | 0,95 | -0,01 | 0,93 | -0,06 | 0,27 | 1,00 | | | | | | | | |
| HHI_G | -0,49 | 0,89 | -0,55 | 0,75 | 0,60 | -0,32 | 1,00 | | | | | | | |
| HHI_IT | -0,52 | 0,84 | -0,62 | 0,75 | 0,65 | -0,33 | 0,91 | 1,00 | | | | | | |
| Hurst_RS_smp | -0,27 | -0,33 | -0,29 | -0,31 | -0,28 | -0,56 | -0,23 | -0,25 | 1,00 | | | | | |
| H_smp_dist | 0,27 | 0,33 | 0,29 | 0,31 | 0,28 | 0,56 | 0,23 | 0,25 | -1,00 | 1,00 | | | | |
| Hurst_RS_pun | 0,06 | -0,31 | 0,20 | 0,03 | -0,35 | 0,11 | -0,13 | -0,41 | -0,19 | 0,19 | 1,00 | | | |
| H_pun_dist | -0,06 | 0,31 | -0,20 | -0,03 | 0,35 | -0,11 | 0,13 | 0,41 | 0,19 | -0,19 | -1,00 | 1,00 | | |
| smp_lyap | -0,26 | 0,30 | -0,29 | 0,26 | 0,00 | -0,16 | 0,35 | 0,34 | -0,15 | 0,15 | -0,07 | 0,07 | 1,00 | |
| pun_lyap | 0,34 | -0,73 | 0,45 | -0,65 | -0,66 | 0,12 | -0,68 | -0,90 | 0,43 | -0,43 | 0,58 | -0,58 | -0,33 | 1,00 |

Based on table 18 we highlight the most important correlations of the abovementioned parameters:

- As Hurst increases its distance from H=0.5, moving upwards or downwards from this value the volatility σ **decreases** since we have either persistence or anti-persistence (mean-reverting).

$$Hurst\_RS\_smp \leftrightarrow uv\_smp: correlation = -0.27$$
$$Hurst\_RS\_pun \leftrightarrow uv\_pun: correlation = -0.31$$

- As HHI is reduced, i.e. the market becomes less concentrated or more competitive or more efficient, the Hurst is moving toward H=0.5 (efficient market). If it starts from a value <0.5, it increases while, when it starts from H>0.5, it is heading toward H=0.5, i.e. the market becomes more complex, less predictable i.e. σ **increases.**

$$If\ HHI \uparrow\ then\ Hurst \downarrow\ and\ \sigma \downarrow$$

Indeed:
$$Hurst\_RS\_smp \leftrightarrow HHI\_G: correlation = -0.23$$
$$Hurst\_RS\_smp \leftrightarrow uv\_smp: correlation = -0.27$$
$$Hurst\_RS\_pun \leftrightarrow HHI\_IT: correlation = -0.41$$

- If $\lambda_{max}$ becomes **more negative** then the system becomes **more stable** so unconditional volatility σ **decreases.**

$$smp\_Lyap \leftrightarrow uv\_smp: correlation = -0.260$$
$$smp\_Lyap \leftrightarrow uv\_des\_smp: correlation = -0.29$$
$$smp\_Lyap \leftrightarrow C\_VOLAT\_SMP: correlation = 0.0$$
$$pun\_Lyap \leftrightarrow uv\_pun: correlation = -0.73$$
$$pun\_Lyap \leftrightarrow uv\_des\_pun: correlation = -0.65$$
$$pun\_Lyap \leftrightarrow C\_VOLAT\_PUN: correlation = 0.12$$

- We must note that the negative correlation between Lyapunov exponent and volatility is not valid for the conditional volatility.

*5.4.3 Tsallis Entropy as an alternate measure of volatility*

In section 5.3 we examined a number of analyzing techniques of Hurst exponents (classical R/S, GHE, DFA and AWC). All these methods rely on the scaling of some kind of fluctuation measure, as the standard deviation or variance, within a rolling window and measure the correlation scaling exponent, based on the assumption that the series analyzed follows a Gaussian distribution, as for example in the case of fractional Brownian motion, FBN. However, in the case of **wholesale electricity price series**, the Gaussanity assumption is not valid. As described in section 4, SMP and PUN have distributions with fatter tails and exhibit both correlated and non-Gaussian increamenets. Therefore the correct correlation scalling exponent (Hurst Exponent) can be reliably estimated by a method that consider not only Gaussian but also non-Gaussian (e.g Levy walks) as processes that both contributing or driving the dynamics of electricity prices. Therefore a method that determines and separates the contribution to the scaling from both correlations and non-Gaussian process is needed. Entropy analysis is such a method and is based on the thermodynamics view of a time series.

Tsallis Entropy is a useful tool for the analysis of a broad range of signals including complex, nonlinear as well as nonstationary signals. Although these concepts are most often named after Tsallis due to his work in the area, they had been studied by others long before him.

In general, entropy in physics is defined as the measure of "disorder" or uncertainty, but in finance entropy is associated with a time dependent probability distribution function.

Entropy is defined as a quantitative representation of "disorder" (or measure of uncertainty). It is actually a relationship between macroscopic and microscopic quantities that quantify the dispersion of energy. The equivalence of the dispersion in financial and energy markets is the way the number of states (or regimes) translates themselves into a probability distribution of the aggregate sentiment in these markets.

The most well-known entropy is the **Shannon information measure** $S_n(P)$ of a probability

measure $p$ on a finite set $X$, given by (in discrete setting)

$$S_n(P) = -\sum_{i=1}^{n} p_i \ln p_i \tag{1}$$

where $\sum_{i=1}^{n} p_i = 1$, $p_i \geq 0$ and $0 \ln 0 = 0$ and $i = 1, \ldots, n$ the number of states, $p_i$ the propability of outcome $i$.

The entropy is the sum over the product of probability of outcome ($p_i$) times the logarithm of the inverse $p_i$, which is also called $i$'s surprisal and the entropy of x is the expected value of its outcome's surprisal. We point out here that if two states A and B are independent from each other, then p(AUB)=p(A)p(B), then $S_n$(A)+$S_n$(B)=$S_n$(AUB).

Let $x$ is a stochastic variable as the wholesale price of an electricity market (e.g SMP or PUV etc) within the general context of anomalously diffusing systems, in which the above mentioned price "live", a time-dependent probability distribution $f(x,t)$ of $x$. In the non-extensive TE the parameter $a$, as we have noted, characterizes the non-extensivity of the entropy, with which the following constraints are associated

$$\int f(x,t)dx = 1 \tag{1}$$

$$\langle x - \bar{x}(t) \rangle_\alpha \equiv \int [x - \bar{x}(t)] f(x,t)^\alpha \, dx = 0 \tag{2}$$

$$\langle (x - \bar{x}(t))^2 \rangle \equiv \int [x - \bar{x}(t)]^2 f(x,t)^\alpha dx = \sigma_\alpha(t)^2 \tag{3}$$

Relation (1) is just the normalization of probability. In (2) and (3) $f(x,t)$ is raised to the power $a$. Unless $a =1$ these are not the usual constraints that load to the mean and variance, so $\sigma_\alpha^2$ (the "$a$-variance") is not the usual variance. Notice that the Tsallis parameter a does not depend on time. Using the constraints above, the maximization of TE for fixed a yields (Tsallis C.,1989 [45])

$$f(x,t) = \frac{1}{z(t)} \{1 + \beta(t)(\alpha - 1)[x - \bar{x}(t)]^2\}^{-\frac{1}{\alpha-1}}$$

where $z(t)$ a normalization constant and $\beta$ the Langrage multipliers, related to (1) and (3) constraints. The **ordinary variance** of $f(x,t)$ given by (4) is

$$\sigma^2(t) = \langle (x - \bar{x}(t))^2 \rangle_{\alpha=1} = \begin{cases} \frac{1}{(5-3\alpha)\beta(t)}, & a < \frac{5}{3} \\ \infty & a \geq \frac{5}{3} \end{cases}$$

Therefore for application to stochastic data, as SMP, PUN etc, with finite variance, the Tsallis parameter $\alpha$ must be within the range

$$1 < \alpha < \frac{5}{3}$$

In continuous setting (1) is written as

$$H = -\int_{-\infty}^{\infty} f(x) \ln f(x) dx, \quad \int_{-\infty}^{\infty} f(x) dx = 1, \quad f(x) \geq 0,$$

where $f(x)$ is the density function of a continuous probability distribution, evaluated for all x values.

Such a case is the Shannon entropy applied to a probability density function f(x) or in a discrete set where is defined as the product of the probability of the outcome $p_i$ times the logarithm of the inverse of $p_i$. But, Shannon entropy cannot be extended to account for **long-range interactions**.

**Nonextensive entropy** also known as **Tsallis Entropy (TE)** is used instead, because it generalizes the classical and quantum statistics. More specifically if we consider any positive real number $a$, the Tsallis entropy of order $a$ of a probability measure $p$ on a finite set $X$ is defined as (Tsallis,C., 1988 [121]):

$$H_a(p) = \begin{cases} \frac{1}{a-1}(1 - \sum_{i \in X} p_i^a), & if\ a \neq 1 \\ -\sum_{i \in X} p_i\ lnp_i, & if\ a = 1 \end{cases} \quad (3)$$

As we have mentioned before, the characterization of the Tsallis entropy is the same as that of the Shannon entropy except that for the Tsallis entropy, the degree of nonadditivity is $a$ instead of $1$. The exponent $a$ is a measure of nonadditivity such that

$$H_\alpha(A \cup B) = H_\alpha(A) + H_\alpha(B) - (1-\alpha)H_\alpha(A)H_\alpha(B)$$

Larger values of $\alpha$ emphasize longe-range interactions between states and can be interpreted as long memory-parameter $H_\alpha(P)$ recovers the shannon entropy if $\alpha \rightarrow 1$.

From the above, TE is a bounded measure, where the lower corresponds to a total lack of beliefe dispersion while the upper to the maximum dispersion at a given point in time.

TE is used in this work to study the dynamics of market expectations of SMP and PUN. The idea behind using TE in financial markets is that the sentiments in them can be captured through the aggregation of the subjective expectations of market agents. Extreme price movements (large volatility) are less likely to occur in case the expectations of market agents are highly dispersed and independent (i.e the entropy is low). The opposite happens when market agents have highly dependent and less dispersed expectations, in which case the aggregate market sentiment can drive the prices to very high levels. By using this particular measure we place emphasis on long-range, time-dependent, interactive instability in the market.

Considering now the following constraints:

$$\int f(x,t)dx = 1$$
$$\frac{\int x^2 f(x)^a dx}{\int f(y)^2 dy} = \sigma^2$$

We can yield the a-Gaussian probability density function from the maximum entropy principle for TE:

$$f(x) = \frac{exp_a(-\beta_\alpha x^2)}{\int exp_a(-\beta_\alpha x^2)dx} \propto \frac{1}{z}[1 + (1-a)(-\beta_\alpha x^2)]^{\frac{1}{1-a}}$$

where $\beta_\alpha$ and $Z$ are a function of $a$ and the $\alpha$-exponential function is defined as:

$$exp_\alpha(x) = \begin{cases} [1 + (1-a)x]^{\frac{1}{1-a}} & if\ 1 + (1-a)x > 0 \\ 0, & otherwise \end{cases}$$

For $a \longrightarrow 1$ a-Gaussian distribution is simply the usual Gaussian distribution. The parameter $a$ itself measures the non-extensivity of the system and it is not a measurement of system's complexity.

The relationship between TE and ARCH-GARCH processes has been reported in literarure by linking the dynamic parameters $\gamma_1$ and $a_1$ as well as the noise mature with the entropic index $a$ (Queiros, S., Tsallis, C., 2005 [47]). There are cases (Balasis, G., et al,2008 [122]) where high entropy values indicate a low degree of organization which is reflected in the low volatility of the system, whilst lower entropy values indicate a high degree of organization or equivalently high volatility of the system. In this sense, TE is regarded as an alternative risk indicator capable of measuring and forecasting market volatility.

Thus without loss of generality (Balasis, G., et al,2008 [122]) there is a stong link between GARCH processes and Tsallis Entropy which is also apparent in our research and better illustrated in figures 21 and 22.

Figures 19 and 20 show the estimated TE on SMP and PUN time series, via a moving window approach in order to track the evolution of $H_\alpha$ over time. This rolling TE emphasizes the relationship between the $\alpha$-Gaussian and ARCH-GARCH processes as mentioned above, used in financial and Energy markets.

The rolling TE depends on the following: a) the number of states: if small it may not be possible to characterize reliably the underline market sentiment while if large it is difficult to track fine changes b) method of partitioning: fixed and adaptive (it changes over time) which can track transitive stages c) estimation of $\alpha$: the degree of long memory in time series when we plot time-dependent TE versus time, i.e the TE is more sensitive to possible changes in the probability distribution function d) rolling window size (K) and time step (dt). K is the number of data used in estimating TE while dt is the number of data by which the moving window is shifted forward across time. For tracking local changes dt=1 while for dt≥k a general trend of a time series is monitored. Based on the previous definitions, TE's parameters have been assigned the following values: a= 1.575, n=1 and K=365 days. The number of observations is 2922. The $\alpha$ value is between 1 and 2 which reveals high volatile activities in the signal or equivalently the entropy is more sensitive to possible disturbances in the probability density function. It has been found (Gradojevic, N., Garic, M.,2015 [123]) that the optimal $\alpha$ is equal to 1.62 for the daily data which is close enough to our calculations which considers also daily dara. A value of 1.63 and 1.64 are used for the weekly and monhtly data respectively.

The best model already defined on section 5.4.1 found (based on AIC and BIC criteria) for the conditional variance for the entire period 2005-2013 is an EGARCH(1,1), for both series. This is repeated for conveniency by the following equation:

$$\log(\sigma^2(t)) = k + \gamma_1 \log(\sigma_{t-1}^2) + a_1 \left[ a_1 \left[ \frac{|\varepsilon_{t-1}|}{\sigma_{t-1}} - E\left\{ \frac{|\varepsilon_{t-1}|}{\sigma_{t-1}} \right\} \right] + \xi_1 \right] + \xi_1 (\frac{\varepsilon_t}{\sigma_t})$$

with, k=-0.432, $\gamma_1$=0.9085, $\alpha_1$=0.3477 and $\xi_1$=-0.0662.

Next we monitor the annual rolling conditional volatility for both time series, and the resulting entropy in either case. Entropy values are depicted on the top side of the figures 19 and 20 and the conditional volatility is on the bottom side of the figure for both PUN and SMP time series.

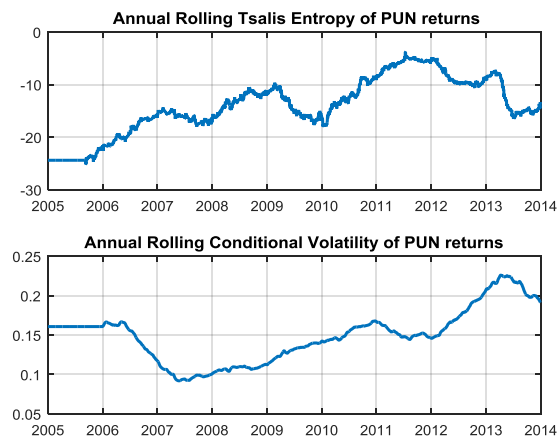

**Figure 19:** (Top) Annual rolling (K=365 days) Tsallis Entropy (TE) and (bottom) annual rolling of conditional volatility of PUN returns.

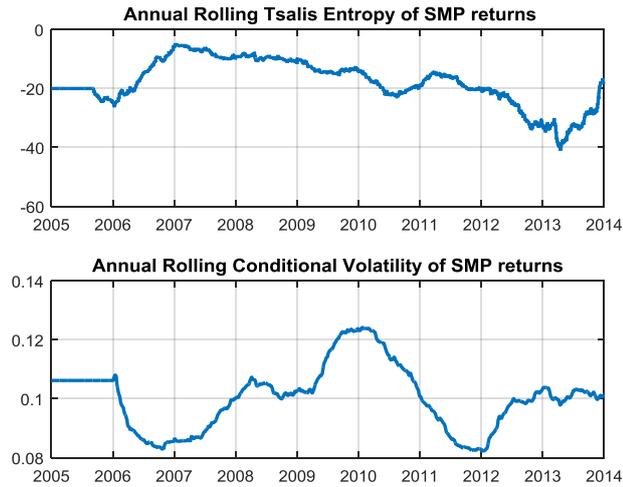

**Figure 20:** (Top) Annual rolling (K=365 days) Tsallis Entropy (TE) and (bottom) annual rolling of conditional volatility of SMP returns.

From both figures we observe that the conditional volatility fluctuates smoothly and does not exhibit large bursts; More specifically in figure 19 which corresponds to the Italian wholesale Electricity Market, we observe that the fluctuation of volatility is larger than the Greek one which is justified by the fact that the Italian Wholesale Electricity Market is more mature and also because each coutry has a different energy mix and dependence on imports/exports. What is common for both Elecricity Markets is that a smooth fluctuation in conditional volatility leads also to smooth fluctuations in entropy around a tight band but in the opposite direction; it is clear that a drop in entropy is caused from an increase in the conditional volatility and vice versa. It has been also reported in literature (Gradojevic, N., Garic, M.,2015 [123]) cases where entropy falls dramatically in high volatility periods. But a similar result is not observed in our case since the our conditional volatility model has been estimated on clean data, after extracting spikes and seasonal effects from the initial time series. Our findings also confirm the predictability (Gradojevic, N., Garic, M.,2015 [123]) inherent in the structure of the TE; In the case of SMP we observe that the entropy, exhibits a decreasing trend with a clear warning signal on 2010 reflected as a constant decrease in volatility thus expressing the consequences of the economic ressesion in Greece started this period of time. On the other hand PUN time series exhibits an increasing trend (until 2013) but the warning signal was evident earlier and more specifically on 2006; the conditional volatility of PUN time series evolved constantly decreasing until 2007.

Rolling Tsallis Entropy (TE) is used in this work as an alternative approach to rolling Lyapunov algorithm, due to inherent weakness of the later (non-robustness and heavy computational burden). Moreover TE successfully captures the higher moments in the entire spectrum of our analysis and thus can be also compared against the unconditional volatility of both time series;

This is illustrated in the following figures 21 and 22 where we monitor the annual rolling unconditional volatility for both time series (PUN and SMP), and the resulting entropy in either case. As before TE's parameters have been assigned the following values: n=1, K=90 days (one quarter) and the number of observations is 2922. Since we apply the rolling TE algorithm, we get different values of the optimum index a for each time window during the time period 2005-2014.

Entropy values are depicted on the top side of the figure and the unconditional volatility is on the bottom side of the figure for both PUN and SMP time series.

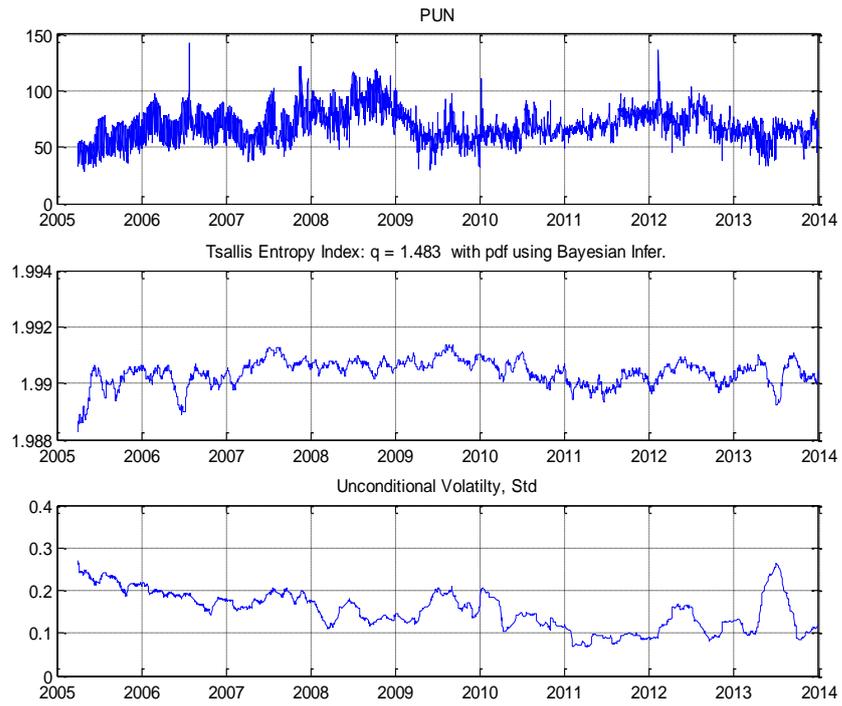

**Figure 21:** (Top) Annual rolling (K=90 days) Tsallis Entropy (TE) and (bottom) annual rolling of unconditional volatility of PUN returns.

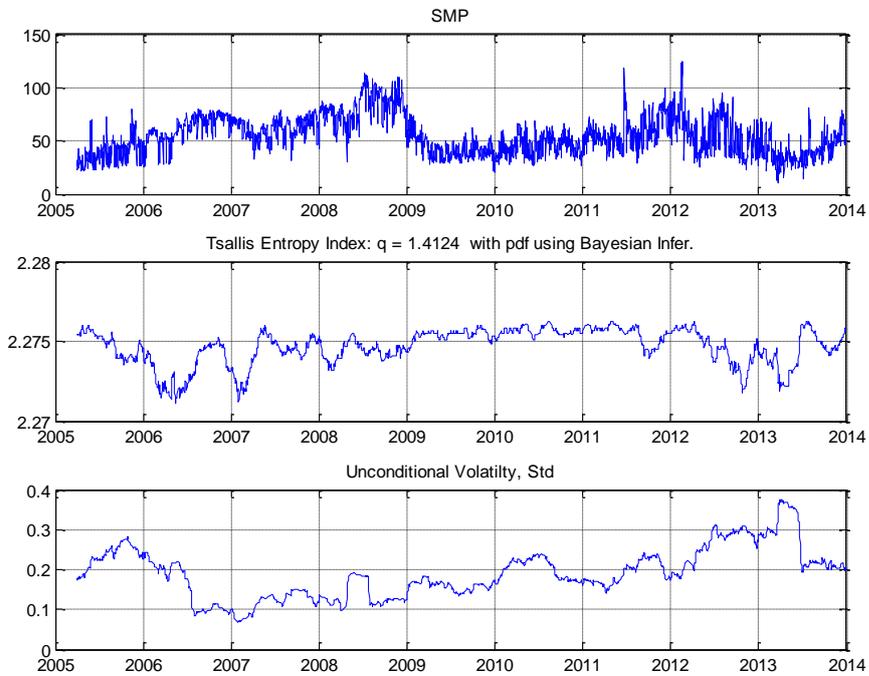

**Figure 22:** (Top) Annual rolling (K=90 days) Tsallis Entropy (TE) and (bottom) annual rolling of unconditional volatility of SMP returns.

Contrary to the results obtained from figures 19 and 20 for the conditional volatility, we observe (figures 21 and 22) that the unconditional volatility does not fluctuate smoothly but exhibits large bursts; this is more evident in the case of the Greek Wholesale Electricity Market, for the reasons already mentioned above. According to (Sheraz, M.,et al, 2014 [124]) entropy is used to capture both the linear and the nonlinear trends of volatility; the nonlinear character of the underlying datasets can be extracted from the fact that values of entropy are positive for both PUN and SMP. What is common in both cases is that conditional volatility exhibits large swings denoted as highly volatile or as highly unpredictable. This high volatility is strongly linked to low values of TE thus TE can be used as a leading indicator in the evolution of the unconditional volatility of both of the above mentioned time series. Our findings confirm this behavior; more specifically for both PUN and SMP time series TE exhibits warning signals shown as large excursions. For instance during 2013 a sudden drop in the TE entropy was followed by a large spike in the unconditional volatility for both time series. This reveals the inherent predictability of TE but also depicts the strong correlation between the two markets since one signal could be used as a leading indicator for the other one.

The mathematical definition of entropy reveals also the strong and clear link between the entropy and the predictability of a system. Positive entropy means that the future state of the system is unpredictable, thus independently of the time period of the previous observations we cannot obtain knowledge of the future state(s) of the system.

On the other hand, Lyapunov's exponents can be used to determine the rate of predictability in the sense that positive Lyapunov's exponents indicate chaos; the bigger the largest Lyapunov exponent is, the less predictable is the system under consideration. Predictability thus, can be expressed as the reciprocal of Lyapunov exponent according to the following formula:

$$Predictability = \frac{1}{\lambda}$$

where $\lambda$ is the Lyapunov exponent.

There is a strong relation between entropy and Lyapunov's exponents according to the following formula:

$$H(T) \leq \sum_{i, \lambda_i > 0} \lambda_i$$

We see that the entropy is upper bounded by the sum of the positive Lyapunov exponents and is therefore finite and positive. Since positive Lyapunov's exponents reveal instability the same intepretation is also valid for the positive entropy. On the other hand, zero entropy means that if we observe long enough a system's state(s) we are certain about its future one(s).

Entropy as a measure of uncertainty has been extensively used in literature in order to quantify the behavioral aspects of systemic risks (Gradojevic, N., Garic, M.,2015 [123]) as well as to gain insight into the evolution of the aggregate market expectations (Gencay, R.,Gradojevic, N.,2006 [125]).

In conclusion, both the Italian and Greek electricity markets show they are scale dependent structure. The fluctuations of PUN and SMP are, as well, significantly different (they diverge) from a Brownian Walk (the required price process for a market satisfying EMH) and exhibit large-scale correlations, a "footprint" of complex nonlinear systems characterized by cooperative interactions of many participants or agents in the market. As mentioned before, Greek market is "hardly" less inefficient than the Italian market. This difference could reveal a higher complexity of the Greek market. However, this higher degree of complexity of the Greek market cannot be attributed as it naturally should be, to a considerably larger number of power generators (networks) and suppliers acting in the Greek market, but rather to a large number of frequently made State or/and Regulator (driven usually by various and conflicting interest of the agents interventions) that increase the complexity of the market not in a way that enhances competition but rather enhancing market friction and inefficiency. Another reason of this less efficiency of the Italian market could be its generation mixture in comparison with that of Greek market. The nature of non-storability of electricity as a traded commodity reveals the significant role a hydroelectric Power Plant can play which in combination with the incumbent's (PPC in Greece) strategic management of its Lignite Units and its market power (it is also the larger Supplier in the country) is further enforced. If SMP is not left completely "free" to adjust to "true" marginal costs of generating units but it is "artificially" set (due to above mentioned mallfunctions) equilibrium point of supply and demand, results in a higher but "not physical" complexity driving both $\alpha_{SMP}$ and $\beta_{SMP}$ closer to the efficient market values, misleading in that way one's interpretation of the above results provided by nonlinear tools.

### 5.6 Rolling window Hurst Analysis

Many factors of short-term or seasonal in duration act concurrently leading to complex and time-varying electricity price processes. Following the work of Uritskaya and Serletis (2008) [126], we apply a rolling window Hurst analysis to zoom in on the time evolution of H, since H is a sensitive indicator to changing conditions or time. So, instead of just providing one value for H to capture the global properties of electricity time series, the scaling exponent or correlation dynamics is put in a rolling time window giving us the ability to monitor the time-variability of H.

We have examined the evolution in time of the Hurst exponent for both markets. This information provides insight over the state of the system for past values and shows clearly the evolution of their mean-reverting property over the course of time. Figures 23 and 24 show the rolling Hurst Exponent with a window of 365 days, estimated by R/S (on raw returns) and GHE (on raw price levels) methods respectively. We observe that for both markets the Hurst exponent remains always between 0<H<0.5. Definitely they are not persistent and not Brownian, instead they are clearly *anti-persistent (mean reverting)* with their Hurst exponent being within a zone ranging from 0.05 to 0.3 (see figure 23), having regions where the PUN's Hurst exponent is larger (closer to 0.5) than SMP's exponent and regions where it happens the opposite, revealing the different time evolution of the structural complexities of the two markets. A better picture, however, of how the structural complexities are evolved through time is provided by using the rolling GHE figure 24. As it is mentioned before, GHE due to its inherent "smoothing" technique (section 5.3.1) that gives "emphasis" on recent values provides a more effective tool in "distinguishing" or separating the two signals. It is now clear that PUN exhibits a stronger-faster mean-reverting process than the SMP, in the period 2006 to middle of 2009, after which the two signals show Hurst exponents approaching each other but remaining indistinguishable up to the 3rd quarter of 2011. Since then strong downward trend of exponents with different slopes is shown up to the end of 2013.

It is also interesting to note the effect of the financial crisis started on September 2008 on the Hurst exponent of two markets which is significantly reduced after this date, but this downward trend for Italy ends in middle of 2009 while for Greece ends 2 years later, a fact revealing that Greek market was hit by the crisis more severely than the Italian one. Italian energy exchange activity is more vivid and agitated which could translate that Italian trading industry is more active than its Greek counterpart. Indeed this is the case in between those two countries and this behavior is captured by the numerous tests applied for the purpose of the analysis of the two markets. The major similarities the markets share, is related to their interconnection, as they both represent neighboring countries that import and export energy to one another. Regional factors also need to be taken into account. The seasonality property of load demand for both countries is the same since they are both located in the Mediterranean region.

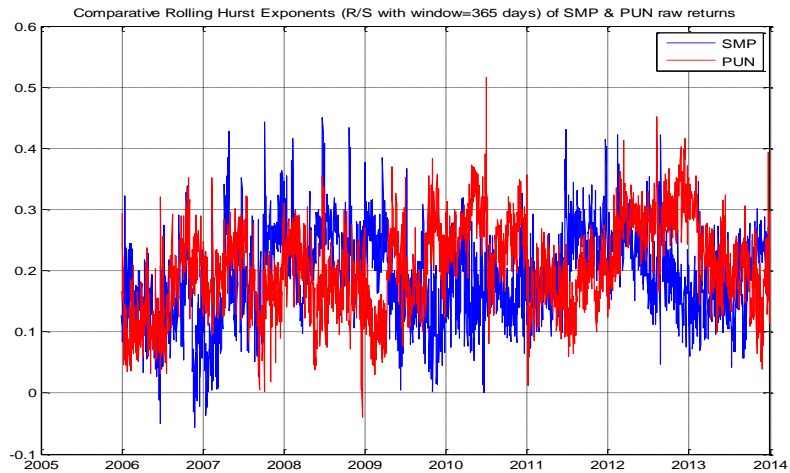

**Figure 23:** Historic evolution of Hurst exponent of SMP and PUN returns for the time period 2005 – 2013, estimated by R/S method with **rolling window** 365 days.

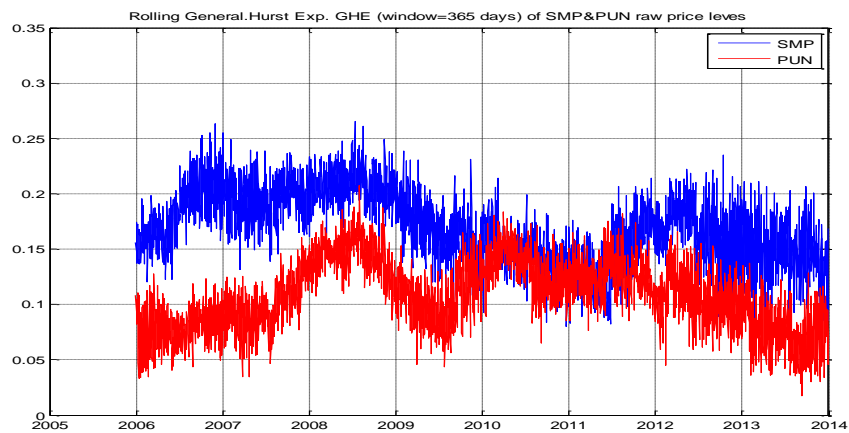

**Figure 24** : Generalized Hurst exponent GHE of SMP and PUN price levels for the time period 2005 – 2013, with **rolling window** 365 days.

**Figure 25** indicate the aforementioned trends in HHI index, along with the time evolution of the Hurst exponent (calculated as described in **section 5.3**, **table 9**), for the same period. The figure reveals that as **HHI is reduced**, i.e. the **"generation market"** is becoming less concentrated, the **Hurst exponent** moves in the opposite direction, it is **increased** heading toward to benchmarking H=0.5 (efficient market), i.e. **the market becomes more complex, less predictable so more efficient.** This relation is clearer for the Italian case. To enhance this interesting result, we have computed the correlation coefficient between the annual HHI index of SMP and PUN (Table 8) and the annual values Hurst exponent for these two series (table 9). The result is a negative correlation -0.264 for the Greek market and a stronger negative value -0.402 for the Italian market. A very interesting point is also the "peak" in the Hurst exponent in PUN, an abrupt increase in complexity that corresponds to the structural break of the series, on February 12, 2009, the date after which the natural gas is the crucial fuel setting PUN level.

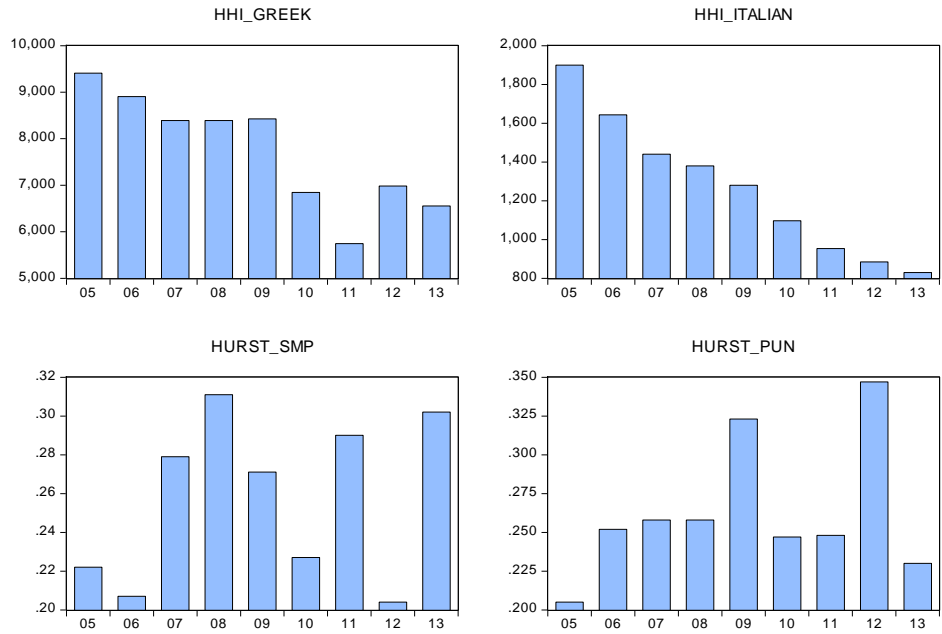

**Figure 25:** The HHI index and Hurst exponent 2005-13 for the two markets

Results of the long-range dependence analysis for returns of PUN and SMP are reported in the table 19. Hurst exponent H estimates are significant at the (two-sided) 90%, 95% significance levels (Weron, 2000 [127]). Mature liquid markets have values of H(2) smaller than 0.5 whereas less developed markets shows a tendency to have H(2)>0.5 (Di Matteo et al., 2003 [128]). Both SMP and PUN belong clearly to mature markets.

The results show also that R/S on average overestimate Hurst exponent when compared with DFA while the overestimation decreases with growing time series (Kristoufek L. 2009 [28]).

**Table 19:** Comparative H exponents between electricity and financial markets (in parenthesis, levels of significance)

| Market | Country | Hurst exponent | |
|---|---|---|---|
| GEM | Greece | 0.21 | (95%) |
| GME | Italy | 0.24 | (95%) |
| Nord Pool | Scandinavia | 0.38 | (99%) |
| OMEL | Spain | 0.24 | (99%) |
| TGE | Poland | 0.21 | (99%) |
| EEX | Germany | 0.13 | (99%) |
| CalPX | USA (California) | 0.32 | (99%) |
| ASE (Stock Exchange) | Greece | 0.60 | (95%) |
| Dow Jones (Stock Exchange) | USA | 0.52 | (<90%) |

### 6. Conclusions

The main objective of this empirical research work is to shed light on the interaction or possible "causal" relations among a variety of quantitative measures usually applied to a market, like the degree of competition (HHI), stability of the "market system" through maximum Lyapunov exponent ($\lambda_{max}$), volatility ($\sigma$, conditional and unconditional), the complexity measures of Hurst exponent and the modern concept of entropy, focused on the electricity markets in Italy and Greece. By quantifying the way these interactions have evolved in time, we have also the opportunity to reveal how the degree of maturity of these two markets has evolved towards the ideal of an efficient market.

For this purpose we applied a number of nonlinear tools from stochastic and chaos theory of time series to investigate empirically the impact of the internal mechanisms of both markets on the time evolution of the volatility of SMP and PUN prices. Tools like Hurst exponent (estimated by different methods), BDS, maximum Lyapunov exponent and Tsallis entropy were used in order to obtain an understanding on the dynamics of the structural complexity of the Greek and Italian electricity markets. The results on the above tests were associated with the dynamic behavior of the unconditional and conditional volatilities, computed via two SARIMA/EGARCH models of different specifications reflecting the different complexities in the structure of the two markets.

Our findings show that all estimations for Hurst exponent (by different methods) are less than the ideal value of H=0.5 that corresponds to the Geometric Brownian motion required by the Efficient Market Hypothesis. Since H<0.5 for both markets they are anti-persistent, mean-reverting and non-Markovian. The BDS test indicated also that both SMP and PUN returns are not i.i.d., clearly exhibit a nonlinear dependence and are against the **weak-form efficient market hypothesis (EMH)**. Indeed, the SMP and PUN prices found not to reflect all the past information.

To some degree therefore, the endogenous complexities shown by the two markets are due to the interdependency of some unexpected causes which enhanced by the randomness, depresses or even obscures the memory of the markets. This can create arbitrage opportunities to the participants which are more informed to beat the market. As mentioned before both markets were found to be mean reverting however with different "speeds" of mean reversion, and according to the mathematical theory of finance, they are **incomplete** markets, indicating the existence of real-world market frictions, causing almost all electricity markets to **deviate** significantly from the requirements of EMH.

Using the **Lyapunov maximum Exponent** $\lambda_{max}$ to measure the stability of the market, the results show that there is no unstable (chaotic) dynamics in both SMP and PUN return prices, since $\lambda_{max} < 0$, for the entire period of our analysis 2005-2013, so both markets are not chaotic. The result is consistent with the mean reversion findings though **Hurst exponent**. We have also found that a larger negative $\lambda_{max}$ leads to reduced volatility (and vice versa), since the system become more stable so less volatile. This interaction between stability and volatility was tested on both unconditional and conditional volatilities, the later generated, for both markets, from ARIMA/EGARCH models fitted on SMP and PUN data. An interesting result is that in both markets the leverage effect and the asymmetric impact conditions in the conditional volatility are observed. This justifies the selection of EGARCH model.

To further enhance our results, we utilized the Tsallis entropy, a nonlinear "composite" concept that measures the complexity of nonlinear time series , taking into account both stability and volatility. Tsalis entropy has been applied as an alternative approach to rolling Lyapunov algorithm, due to inherent weakness of the later (non-robustness and heavy computational burden). Indeed Tsalis entropy successfully captured the higher moments in the entire spectrum of our analysis and thus could be compared against the unconditional volatility of both time series; the results reveal that Tsalis entropy can be used as a leading indicator in the evolution of the unconditional volatility of both of the above mentioned time series.

The correlation matrix of all these quantities, estimated on a yearly base, have revealed results that are consistent with theory and very significant. To the best of our knowledge, this empirical work is the first that shows a complete 'causal-map' constructed by considering the interdependence of a critical number of market-system related quantities, shedding lights on the way aspects of degree of competition, market stability, price volatility and market's degree of structural complexity are combined or "coupled" together, to generate the dynamic behavior of the wholesale prices of these two different in maturity and other criteria (like energy mixture, number of interconnections, liquidity etc.) market structures, the Greek and Italian electricity markets. The authors consider that this empirical work makes a contribution to the status of the current literature.

**Acknowledgments:** The authors would like to thank IPTO's top management for their support towards implementing this research paper. The opinions expressed in this paper do not necessarily reflect the position of IPTO.

**Author Contributions:** The main focus and structure of this work was proposed by Dr. George P. Papaioannou, to whom coordination is accredited. The remaining authors contributed equally to this work. More specifically Christos Dikaiakos contributed to modeling and thoroughly reviewed the paper. Anargyros Dramountanis contributed to mathematical modeling and simulations. Dionysios Georgiadis contributed also to simulations regarding the best GARCH model. Panagiotis G. Papaioannou also contributed to mathematical consistency, modeling and simulations as well as to the review of this paper.

**Conflicts of Interest:** The authors declare no conflict of interest.

**Appendix A**

**Models**
**SMP**
All SMP models are seasonally integrated.

| Model | Mean model | Variance model | Conditional Volatility equation | Innovation distribution | LogL | AIC | BIC | ARCH test p-value | Serial correlations remain? |
|---|---|---|---|---|---|---|---|---|---|
| 1 | ARIMA(2,0,2) with Seasonal AR(7,14) and MA(7,14) | GARCH(1,1) | Symmetric | Normal | 3.072 | -6.125 | -6.125 | 0 | Yes |
| 2 | ARIMA(2,0,2) with Seasonal AR(7,14) and MA(7,14) | EGARCH(1,1) | Asymmetric | Normal | 3.084 | -6.148 | -6.087 | 0 | Yes |
| 3 | ARIMA(2,0,2) with Seasonal AR(7,14) and MA(7,14) | GJR-GARCH(1,1) | Asymmetric | Normal | 3.078 | -6.136 | -6.075 | 0 | Yes |
| 4 | ARIMA(2,0,1) with Seasonal AR(7,14,21) and MA(7,14,21) | GARCH(1,1) | Symmetric | Normal | 3.094 | -6.166 | -6.099 | 0 | Yes |
| 5 | ARIMA(2,0,1) with Seasonal AR(7,14,21) and MA(7,14,21) | EGARCH(1,1) | Asymmetric | Normal | 3.108 | -6.194 | -6.127 | 0 | Yes |
| 6 | ARIMA(2,0,1) with Seasonal AR(7,14,21) and MA(7,14,21) | GJR-GARCH(1,1) | Asymmetric | Normal | 3.101 | -6.180 | -6.113 | 0 | Yes |
| 7 | ARIMA(1,0,1) with Seasonal AR(7,14,21) and MA(7,14,21) | GARCH(1,1) | Symmetric | Student t | 3.274 | -6.525 | -6.458 | 0 | Yes |
| 8 | ARIMA(1,0,1) with Seasonal AR(7,14,21) and MA(7,14,21) | EGARCH(1,1) | Asymmetric | Student t | 3.287 | -6.552 | -6.485 | 0 | Yes |
| 9 | ARIMA(1,0,1) with Seasonal AR(7,14,21) and MA(7,14,21) | GJR-GARCH(1,1) | Asymmetric | Student t | 3.277 | -6.532 | -6.465 | 0 | Yes |

**PUN**

| Model | Mean model | Variance model | Conditional Volatility equation | Innovation distribution | LogL | AIC | BIC | ARCH test | |
|---|---|---|---|---|---|---|---|---|---|
| | | | | | | | | p-value | Serial correlations remain? |
| 10 | ARIMA(2,1,3) with Seasonal AR(14) and MA(14) | GARCH(1,1) | Symmetric | Normal | 1.814 | -3.606 | -3.539 | 0 | Yes |
| 11 | ARIMA(2,1,3) with Seasonal AR(7,14) and MA(7,14) | EGARCH(1,1) | Asymmetric | Normal | 1.832 | -3.643 | -3.576 | 0 | Yes |
| 12 | ARIMA(2,1,3) with Seasonal AR(7,14) and MA(7,14) | GJR-GARCH(1,1) | Asymmetric | Normal | 1.824 | -3.626 | -3.559 | 0 | Yes |
| 13 | ARIMA(2,1,3) with Seasonal AR(7,14,21) and MA(7,14,21) | GARCH(1,1) | Symmetric | Student t | 1.940 | -3.858 | -3.791 | 0 | Yes |
| **14** | **ARIMA(2,1,3) with Seasonal AR(7,14,21) and MA(7,14,21)** | **EGARCH(1,1)** | **Asymmetric** | **Student t** | **1.963** | **-3.904** | **-3.837** | **0** | **Yes** |
| 15 | ARIMA(1,1,2) with Seasonal AR(7,14,21) and MA(7,14,21) | GJR-GARCH(1,1) | Asymmetric | Student t | 1.950 | -3.877 | -3.810 | 0 | Yes |
| 16 | ARIMA(1,1,1) with Seasonal AR(7,14,21) and MA(7,14,21) | GARCH(1,1) | Symmetric | Student t | 1.928 | -3.834 | -3.767 | 0 | Yes |
| 17 | ARIMA(1,1,1) with Seasonal AR(7,14,21) and MA(7,14,21) | EGARCH(1,1) | Asymmetric | Student t | 1.949 | -3.876 | -3.809 | 0 | Yes |
| 18 | ARIMA(1,1,1) with Seasonal AR(7,14,21) and MA(7,14,21) | GJR-GARCH(1,1) | Asymmetric | Student t | 1.937 | -3.852 | -3.785 | 0 | Yes |

## 2005

| SMP | AIC | BIC | pValue | |
|---|---|---|---|---|
| Spec 1 | -613.89 | -574.90 | 0.0021 | |
| 2 | -606.31 | -567.30 | 0.0018 | |
| 3 | -614.20 | -575.23 | 0.0022 | |
| 4 | -625.29 | -582.40 | 0.0720 | |
| 5 | -752.00 | -709.18 | 0.5920 | |
| 6 | -629.66 | -590.69 | 0.0819 | |
| 7 | -818.02 | -775.00 | 0.0000 | |
| **8** | **-828.06** | **-785.19** | **0.0000** | Best model |
| 9 | -822.68 | -779.81 | 0.0000 | |

| PUN | AIC | BIC | pValue | |
|---|---|---|---|---|
| Spec 1 | -299.00 | -260.00 | 0.1050 | |
| 2 | -324.80 | -281.90 | 0.0377 | |
| 3 | -301.50 | -262.50 | 0.1080 | |
| 4 | -343.94 | -301.00 | 0.1010 | |
| 5 | -344.80 | -298.06 | 0.0990 | |
| **6** | **-353.50** | **-310.70** | **0.0492** | Best model |
| 7 | -326.56 | -283.60 | 0.0872 | |
| 8 | -353.25 | -306.50 | 0.1103 | |
| 9 | -328.35 | -285.40 | 0.0917 | |

## 2006

| SMP | AIC | BIC | pValue | |
|---|---|---|---|---|
| Spec 1 | -756.60 | -721.60 | 0.0000 | |
| 2 | -810.50 | -771.50 | 0.0000 | |
| 3 | -758.90 | -723.80 | 0.0000 | |
| 4 | -773.50 | -734.50 | 0.0000 | |
| 5 | -789.10 | -746.20 | 0.0000 | |
| **6** | **-780.20** | **-741.20** | **0.0000** | Best model |
| 7 | -959.00 | -920.00 | 0.0000 | |
| 8 | -946.50 | -921.60 | 0.0000 | |
| 9 | -959.50 | -920.50 | 0.0000 | |

| PUN | AIC | BIC | pValue | |
|---|---|---|---|---|
| Spec 1 | -667.90 | -625.00 | 0.0155 | |
| 2 | -680.80 | -637.90 | 0.0905 | |
| 3 | -672.12 | -633.15 | 0.0683 | |
| 4 | -841.57 | -794.80 | 0.0051 | |
| 5 | -837.10 | -790.30 | 0.0125 | |
| 6 | -806.00 | -759.20 | 0.0586 | |
| 7 | -837.90 | -791.10 | 0.0033 | |
| **8** | **-845.73** | **-798.90** | **0.0047** | Best model |
| 9 | -821.70 | -778.80 | 0.0067 | |

## 2007

| SMP | AIC | BIC | pValue | |
|---|---|---|---|---|
| Spec 1 | -627.30 | -588.30 | 0.0000 | |
| 2 | -633.00 | -594.11 | 0.0000 | |
| 3 | -631.54 | -596.40 | 0.0000 | |
| 4 | -647.78 | -608.40 | 0.0000 | |
| 5 | -649.20 | -606.30 | 0.0000 | |
| 6 | -647.50 | -608.50 | 0.0000 | |
| 7 | -675.10 | -632.20 | 0.0000 | |
| 8 | -679.30 | -636.50 | 0.0000 | |
| **9** | **-679.40** | **-636.50** | **0.0000** | Best model |

| PUN | AIC | BIC | pValue | |
|---|---|---|---|---|
| Spec 1 | -804.50 | -761.70 | 0.0000 | |
| 2 | -823.70 | -780.80 | 0.0000 | |
| 3 | -826.70 | -783.80 | 0.0013 | |
| 4 | -839.90 | -793.20 | 0.0011 | |
| 5 | -851.00 | -804.30 | 0.0014 | |
| **6** | **-852.40** | **-809.40** | **0.0011** | Best model |
| 7 | -833.70 | -786.90 | 0.0012 | |
| 8 | -849.40 | -802.70 | 0.0019 | |
| 9 | -850.10 | -807.30 | 0.0012 | |

## 2008

## 2009

## 2010

| SMP | AIC | BIC | pValue | | SMP | AIC | BIC | pValue | | SMP | AIC | BIC | pValue | |
|---|---|---|---|---|---|---|---|---|---|---|---|---|---|---|
| Spec 1 | -666.50 | -627.50 | 0.0000 | | Spec 1 | -416.50 | -377.60 | 0.0000 | | Spec 1 | -598.70 | -559.80 | 0.0000 | |
| 2 | -665.20 | -626.20 | 0.0000 | | 2 | -416.40 | -377.40 | 0.0000 | | 2 | -607.30 | -568.30 | 0.0000 | |
| 3 | -666.60 | -627.70 | 0.0000 | | 3 | -420.30 | -381.30 | 0.0000 | | 3 | -603.00 | -564.10 | 0.0000 | |
| 4 | -688.60 | -645.70 | 0.0000 | | 4 | -424.20 | -381.40 | 0.0000 | | **4** | **-619.60** | **-576.70** | **0.0000** | |
| 5 | -689.10 | -646.20 | 0.0000 | | 5 | -423.60 | -380.80 | 0.0000 | | 5 | -623.40 | -580.50 | 0.1850 | |
| **6** | **-690.20** | **-647.30** | **0.0000** | Best model | 6 | -426.30 | -383.40 | 0.0000 | | 6 | -620.70 | -577.80 | 0.1390 | |
| 7 | -682.10 | -639.00 | 0.0000 | | 7 | -441.70 | -398.80 | 0.0000 | | **7** | **-636.20** | **-597.20** | **0.0880** | |
| 8 | -682.20 | -639.30 | 0.0000 | | 8 | -440.00 | -397.20 | 0.0000 | | 8 | +2.00X10$^{20}$ | +2.00X10$^{20}$ | 0.1780 | |
| 9 | -682.90 | -640.00 | 0.0000 | | **9** | **-442.30** | **-399.40** | **0.0000** | Best model | 9 | -635.20 | -592.40 | 0.0417 | Best model 5% |
| PUN | AIC | BIC | pValue | | PUN | AIC | BIC | pValue | | PUN | AIC | BIC | pValue | |
| Spec 1 | -720.20 | -677.30 | 0.0000 | | Spec 1 | -492.9 | -450.10 | 0.0025 | | Spec 1 | -333.50 | -290.60 | 0.0558 | |
| 2 | -751.60 | -708.60 | 0.0000 | | 2 | -493.10 | -450.30 | 0.0190 | | 2 | -335.90 | -293.00 | 0.1432 | |
| 3 | -706.80 | -667.80 | 0.0061 | | 3 | -492.00 | -449.20 | 0.0200 | | 3 | -339.10 | -300.10 | 0.0831 | |
| 4 | -741.90 | -695.10 | 0.0000 | | 4 | -491.70 | -445.00 | 0.0050 | | **4** | **-331.60** | **-284.80** | **0.0559** | Best model ≈5% |
| 5 | -748.90 | -702.10 | 0.0000 | | 5 | -491.90 | -445.20 | 0.0170 | | 5 | -329.70 | -282.90 | 0.0591 | |
| 6 | -750.00 | -708.00 | 0.0000 | | 6 | -491.90 | -445.21 | 0.0062 | | 6 | -336.98 | -294.10 | 0.0823 | |
| 7 | -738.50 | -691.70 | 0.0000 | | **7** | **-508.00** | **-461.20** | **0.0036** | Best model | 7 | -344.80 | -298.00 | 0.2592 | |
| **8** | **-762.60** | **-715.80** | **0.0117** | Best model | 8 | -507.20 | -460.40 | 0.0037 | | **8** | **-330.50** | **-283.70** | **0.0750** | |
| 9 | -748.70 | -705.80 | 0.0000 | | 9 | -508.00 | -461.20 | 0.0036 | | 9 | -351.70 | -308.80 | 0.2240 | |

**2011**                       **2012**                       **2013**

| SMP | AIC | BIC | pValue | | SMP | AIC | BIC | pValue | | SMP | AIC | BIC | pValue | |
|---|---|---|---|---|---|---|---|---|---|---|---|---|---|---|
| Spec 1 | -921.60 | -882.60 | 0.0022 | | Spec 1 | -731.40 | -692.40 | 0.0000 | | Spec 1 | -690.00 | -651.00 | 0.0000 | |
| 2 | -919.10 | -880.20 | 0.0093 | | 2 | -731.80 | -692.80 | 0.0000 | | 2 | -697.90 | -659.00 | 0.0000 | |
| 3 | -919.60 | -880.60 | 0.0063 | | 3 | -732.00 | -693.00 | 0.0000 | | 3 | -691.15 | -656.00 | 0.0000 | |
| 4 | -952.00 | -909.10 | 0.0000 | | 4 | -742.80 | -699.90 | 0.0000 | | 4 | -695.80 | -652.90 | 0.0000 | |
| 5 | -954.30 | -911.50 | 0.0000 | | **5** | **-748.30** | **-705.40** | **0.0000** | **Best model** | 5 | -695.60 | -652.70 | 0.0000 | |
| 6 | -952.00 | -909.20 | 0.0000 | | 6 | -743.40 | -700.50 | 0.0000 | | 6 | -689.00 | -646.10 | 0.0000 | |
| 7 | -959.70 | -916.90 | 0.0000 | | 7 | -743.00 | -700.10 | 0.0000 | | 7 | -695.60 | -652.60 | 0.0000 | |
| **8** | **-964.90** | **-922.10** | **0.0000** | **Best model** | 8 | -748.80 | -705.90 | 0.0000 | | 8 | -702.50 | -659.60 | 0.0000 | |
| 9 | -964.40 | -921.50 | 0.0000 | | 9 | -743.70 | -700.80 | 0.0000 | | **9** | **-706.30** | **-666.30** | **0.0000** | **Best model** |
| PUN | AIC | BIC | pValue | | PUN | AIC | BIC | pValue | | PUN | AIC | BIC | pValue | |
| Spec 1 | -425.20 | -382.30 | 0.0261 | | Spec 1 | -93.50 | -50.60 | 0.0057 | | Spec 1 | -176.20 | -133.40 | 0.0000 | |
| 2 | -441.10 | -398.30 | 0.0654 | | 2 | -95.90 | -53.00 | 0.0057 | | 2 | -184.50 | -141.60 | 0.0000 | |
| 3 | -401.70 | -358.80 | 0.0285 | | 3 | -95.00 | -52.10 | 0.0068 | | 3 | -179.20 | -129.40 | 0.0000 | |
| 4 | -440.14 | -393.30 | 0.0212 | | 4 | -93.90 | -47.10 | 0.0028 | | 4 | -192.79 | -146.00 | 0.0000 | |
| **5** | **-440.29** | **-393.52** | **0.0366** | **Best model** | 5 | -93.00 | -46.30 | 0.0202 | | 5 | -192.60 | -145.80 | 0.0000 | |
| 6 | -438.90 | -392.20 | 0.0240 | | 6 | -96.70 | -49.90 | 0.0041 | | 6 | -191.39 | -144.60 | 0.0000 | |
| 7 | -413.00 | -366.00 | 0.0140 | | 7 | -92.20 | -45.40 | 0.0000 | | 7 | -196.10 | -149.30 | 0.0000 | |
| 8 | -450.00 | -403.00 | 0.0785 | | **8** | **-100.20** | **-53.40** | **0.0018** | **Best model** | **8** | **-210.50** | **-163.50** | **0.0000** | **Best model** |
| 9 | -415.40 | -368.60 | 0.0197 | | 9 | -102.10 | -59.20 | 0.0040 | | 9 | -202.80 | -156.00 | 0.0000 | |

**SMA 2005-2013**

| SMP | AIC | BIC | pValue |
|---|---|---|---|
| Spec 1 | $-6.000 \times 10^3$ | $-6.000 \times 10^3$ | 0.0000 |

| PUN | AIC | BIC | pValue | |
|---|---|---|---|---|
| 2 | -6.090X10$^3$ | -6.020X10$^3$ | 0.0000 | |
| 3 | -6.080X10$^3$ | -6.020X10$^3$ | 0.0000 | |
| 4 | -6.110X10$^3$ | -6.050X10$^3$ | 0.0000 | |
| 5 | -6.140X10$^3$ | -6.070X10$^3$ | 0.0000 | |
| 6 | -6.134X10$^3$ | -6.060X10$^3$ | 0.0000 | |
| 7 | -6.490X10$^3$ | -6.420X10$^3$ | 0.0000 | |
| **8** | **-6.520X10$^3$** | **-6.450X10$^3$** | **0.0000** | **Best model** |
| 9 | -6.500X10$^3$ | -6.430X10$^3$ | 0.0000 | |
| PUN | AIC | BIC | pValue | |
| Spec 1 | -3.600X10$^3$ | -3.550X10$^3$ | 0.0000 | |
| 2 | -3.660X10$^3$ | -3.500X10$^3$ | 0.0000 | |
| 3 | -3.630X10$^3$ | -3.560X10$^3$ | 0.0000 | |
| 4 | -3.850X10$^3$ | -3.780X10$^3$ | 0.0000 | |
| **5** | **-3.900X10$^3$** | **-3.820X10$^3$** | **0.0000** | **Best model** |
| 6 | -3.850X10$^3$ | -3.780X10$^3$ | 0.0000 | |
| 7 | -3.830X10$^3$ | -3.760X10$^3$ | 0.0000 | |
| 8 | -3.870X10$^3$ | -3.800X10$^3$ | 0.0000 | |
| 9 | -3.850X10$^3$ | -3.780X10$^3$ | 0.0000 | |

# References

1. Karakatsani, N. V., & Bunn, D. W.(2008). Forecasting electricity prices: the impact of fundamentals and time-varying coefficients. International Journal of Forecasting, 24(4), 764–785.
2. Amjady, N., Keynia F. (2009). Day-ahead price forecasting of electricity markets by mutual information technique and cascaded neuro evolutionary algorithm. IEE Trans. Power Syst. 2009,24,306-318.
3. Becker, R., Hurn, A.D. and Paulov, V. (2007). Modelling spikes in electricity prices. Economic Record, 83, 371-382.
4. Christensen, I., Hurn, S., Lindsay, K. (2012). Forecasting spikes in electricity prices. Internationa Journal of Forecasting, 24, 400-411.
5. Lei, M., &Feng, Z. (2012). A proposed grey model for short- term electricity price forecasting in competitive power markets. International Journal of Electrical Power and Energy Systems, 43(1), 531-538.
6. Mandal, P., Haque, A.U., Meng, J., Martinez, R., and Srivastava, A.K. (2012). A hybrid intelligent algorithm for short-term energy price forecasting in the Ontario market. Proceedings of IEEE PES 2012, art. No. 6345461.
7. Simonsen, I., Weron, R., Mo, B. (2004). Structure and stylized facts of a deregulated power market. Munich Personal RePEc Article (MPRA), paper 1443, pp. 1-29.
8. Casdagli M. and Eubank, S., eds. (1992). Nonlinear Modeling and Forecasting. Santa Fe Institute Studies in the Science of Complexity, Proc. Vol. XII, Addison-Wesley, Reading, MA\
9. Weigend A. S. and Gershenfeld, N. A., eds. (1993). Time Series Prediction: Forecasting the future and understanding the past. Santa Fe Institute Studies in the Science of Complexity, Proc. Vol. XV, Addison-Wesley, Reading, MA
10. Ott, E. (1993).Chaos in Dynamical Systems. Cambridge University Press, Cambridge.
11. Abarbanel, H. D. I. (1996) Analysis of Observed Chaotic Data. Springer, New York.
12. Kantz, H. and Schreiber, T. (1997). Nonlinear Time Series Analysis. Cambridge University Press, Cambridge.
13. Schuster, H.-G. (1988). Deterministic Chaos: An introduction. Physik Verlag.
14. Strogatz, S. (2000). Nonlinear Dynamics and chaos. Westerview, Cambridge.
15. Bask, M., Widerberg, A. (2009). Market structure and the stability and volatility of electricity prices. Energy Economics, 31, 278-288.
16. Bask, M., Lundgren, J., Rudholm, N. (2011). Market Power in the expanding Nordic power market. Applied Economics, 43, issue 9
17. Weron R., 2006. Modeling and forecasting Electricity loads and prices: a statistical approach. Chichester: Wiley
18. Serletis A. (2007). Quantitative and Empirical Analysis of Energy Markets, World Scientific
19. Serletis A., A.A. Rosenberg, (2007). The Hurst exponent in energy futures prices. Physica A 380 (2007) 325-332
20. Serletis A., A.A. Rosenberg, (2009). Mean reversion in the US stock market, Chaos, Solutions and Fractals 40 (2009) 2007-2015
21. Papaioannou G. and A. Karytinos, (1995). Nonlinear Time Series Analysis of the Stock Exchange: The Case of an Emerging Market, International Journal of Bifurcation and Chaos 5 (1995), 1557-1584
22. Papaioannou, G., Dikaiakos, C., Dramountanis A., and Papaioannou, P. (2016). Analysis and Modeling for short-to medium-term load forecasting using a hybrid Manifold learning Principal Component model and comparison with classical statistical models (SARIMAX, Exponential Smoothing) and Artificial Intelligence Models (ANN, SVM): the case of Greek Electricity Market. Energies, 9,635.
23. Kantz, H., Scheiber, T. (2004). Nonlinear time series analysis. Cambridge University Press.
24. Bask, M. (2010). Measuring potential market risk. Journal of Financial Stability 6 (2010) 180-186.
25. Choi, J-G., Park J-K., Kim, K-H, Kim, J-C. (1996). A daily peak load forecasting system using a chaotic time series. In: proceedings of the international conference on intelligent systems applications to power systems ISAP, p.283-7.
26. Drezga, I., Rahman, S. (1999). Phase-space based short-term load forecasting for deregulated electric power industry Neural Networks. Int Joint Conf IJCNN, 5:3405-9.
27. Hyndman, R., Koehler, A. B., Ord, J. K., & Snyder, R. D.(2008). Forecasting with exponential smoothing: the state space approach. Springer.

28. Kristoufek, L. (2009). R/S analysis and DFA: finite sample properties and confidence intervals. MPRA Paper No. 16446, online at http://mpra.ub.uni-muenchen.de/16446/
29. Liao, G-C. (2006). Hybrid chaos search genetic algorithm and meta heuristics method for short-term load forecasting. Electr Eng, 88:1-6.
30. Wu, W., Zhou, J-Z., Yang, J-J. (2004). Prediction of spot market prices of electricity using chaotic time series. In: Proceeding of the 3rd international conference on machine learning and cybemetics. Shangai:26-29 August 2004, p.1-6.
31. Wolak, F. (1997). Market Design and Price Behavior in Restructured Electricity Market: an International Comparison, working paper, available at www.leland.stanford.edu/~wolak
32. Weron, R., & Misiorek, A. (2008). Forecasting spot electricity prices: a comparison of parametric and semiparametric time series models. International Journal of Forecasting, 24, 744-763.
33. Brockwell, P. J., & Davis, R. A. (1996). Introduction to time series and forecasting (2nd ed.). New York: Springer-Verlag
34. Ljung, L.(1999). System identification — theory for the user (2nd ed.). Prentice Hall: Upper Saddle River.
35. Shumway, R. H., & Stoffer, D. S.(2006). Time series analysis and its applications (2nd ed.).  Springer.
36. Makridakis; Wheelwright; Hyndman; (1998). Forecasting: Methods and Applications 3rd ed.; WILEY: USA.
37. Hyndman, R., & Athanasopoulos, G. (2013). Forecasting: principles and practice. Online athttp://otexts.org/fpp/.
38. Weron, R., 2014. Electricity price forecasting: a review of the state-of-the-art with a look into the future. Int J Forecast, 30(4)
39. Bose R. and Hamacher K. (2012). Alternate entropy measure for assessing volatility in financial markets. PHYSICAL REVIEW E 86, 056112 (2012).
40. Pincus S. and Kalman R.E. (July 19, 2004). Irregularity, volatility, risk and financial market time series. PNAS, September 21, 2004, vol.101, no.36, 13709-13714.
41. Zhou R., Cai R., Tong G. (2013). Applications of Entropy in Finance: A Review. Entropy 2013, 15, 4909-4931; doi:10.3390/e15114909
42. Zhou R., Zhan Y., Cai R., Tong G. (2015). A Mean-Variance Hybrid-Entropy Model for Portfolio Selection with Fuzzy returns. Entropy 2015,17, 3319-3331; doi:10.3390/e1705319
43. Ormos, M., Zibriczky, D. (2014). Entropy-based Financial Asset pricing. PLoS ONE 9 (12):e115742 doi:101371/journal.pone.0115742
44. Gradojevic N., Goncay R. (2011). Financial applications of nonextensive entropy, IEEE Signal prosecing magazine 116, September 2011.
45. Tsallis C. (1989), J.Stat.Phys.52, 479
46. Tsallis C.(2009) Introduction to Nonextensive Statistical Mechanics: Approaching a Complex World. New York: Springer-Verlag (2009)
47. Queiros S. and Tsallis C. (2005). Bridging a paradigmatic financial model and nonextensive entropy. Europhys. Lett., vol.69, no.6, pp893-899
48. Perello J., Monter M., Palatella L., Simonsen I., Masoliver J. (2006). Entropy of the Nordic electricity market: Anomalous scaling, spikes and mean reversion. J.Stat. Mech.Theor.Exp. 2006,11,P11011
49. Amjady, N., & Hemmati, M.(2009). Day-ahead price forecasting of electricity markets by a hybrid intelligent system. European Transactions on Electrical Power, 19(1), 89–102.
50. Ruiz Maria del Carmen, Guillamon A., Gabaldon A. (2012). A New Approach to measure Volatility in Energy Markets. Entropy 2012,14,74-91
51. Papaioannou, G., Dikaiakos, C., Evangelidis, G., Papaioannou, P., and Georgiadis, D. (2015). Co-movement Analysis of Italian and Greek Electricity Market Wholesale Prices by Using a Wavelet Approach. Energies, 8,11770-11799.
52. Regulatory Authority for Energy (RAE). National Report to the European Commission, Athens, Greece, November 2010. Available online: http://www.rae.gr (accessed on 13 October 2015).
53. Di Cosmo, V. Modelling the Italian electricity price. In Proceedings of the Economic and Social Research Institute Trinity College, IEB Symposium, Dublin, Ireland, 20–23 August 2015.
54. Corrado, C.; Assanelli, M. An overview of Italy's Energy Mix. Gouvern. Eur. Géopol. L′Énergie June 2012. Available online: http://www.ifri.org (accessed on 13 October 2015).
55. Cataldi, A.; Stefano, C.; Zoppoli, P. The Merit-Order effect in the Italian power market: The Impact of solar and wind generation on national wholesale electricity prices. Energy Policy 2015, 77, 79–88.

56. Regulation (EC) No 714/2009. In Proceedings of the European Parliament and the Council on the conditions for access to the network for cross-border exchanges in electricity and repealing Regulation (EC), Regulation of European Parliament and of the Council, Brussels, Belgium, 13 July 2009.
57. Simonsen, I.(2005). Volatility of Power markets. Physica A. 2005:365(1);10-20.
58. Brock W.A., Dechert, W.D., Scheinkman, J., and LeBaron, B. (1996). A test for independence based on the correlation dimension. Econometric reviews, 15(1996), 197-235.
59. Opong, K.K., Mulholland, G., Fox, A.F., and Farahmand, K. (1999). The behavior of some UK equity indices: an application of Hurst and BDS tests. Journal of Empirical Finance, 3,267-282.
60. Yousefpoor, P., Esfahani, M.S., and Nojumi, H. (2008). Looking for systematic approach to select chaos tests. Applied Mathematics and Computation, 198(1), 73-91.
61. Barnett, W., and Serletis, A. (2000). Martingales, nonlinearity and chaos. Journal of Economic Dynamics and Control **24**(5), 703-724.
62. Barnett, W., A., Galiant, A. R., Hinich, M. J., Jungeilges, J., Kaplan, D., and Jensen, M. J. (1997). A single-blind controlled competition among tests for nonlinearity and chaos. Journal of Econometrics **82**(1), 157-192.
63. Barnett, W., A., and Hinich, M. J. (1992). Empirical chaotic dynamics in economics. Annals of Operations Research **37**(1), 1-15.
64. Brock W.A. (1986). Distinguishing random and deterministic systems. Journal of Economic theory, 40(1), 168-195.
65. Brock W.A., Hsieh, D.A., and LeBaron, B. (1993) Nonlinear Dynamics, chaos and instability: Statistical Theory and Economic evidence. MIT Press, Cambridge, MA.
66. Adrangi, B., Chatrath, A., Dhanda, K.K., and Raffiee, K. (2001). Chaos in oil prices? Evidence from future markets. Energy Economics, 23(4), 405-425.
67. Hsieh, D.A. (1991). Chaos and nonlinear dynamics: application to financial markets. Journal of Finance, 46(5), 1839-1877.
68. Dakhlaoui, I., Aloui, C. (2013). The US Oil spot market: a deterministic chaotic process or a stochastic process? Journal of Energy Markets, Vol.6, No.1, Spring 2013, 51-93.
69. Kugiumtzis, D., Lillekjendlie, B., Christophersen, N. (1994). Chaotic time series I, Modeling, Identification and Control 15, 205.
70. Theiler, J. (1990). Statistical precision in dimension estimation. Phys, Rev. A, 41, 3038.
71. Takens, F. (1981). Detecting Strange Attractors in Tarbulence. Lecture Notes in Math., Springer, New York, Vol. 898.
72. Mane, R. (1981). Ergodic Theory and Differentiable Dynamics. Springer – Valag Berlin Heidelberg.
73. Sauer, T., Yorke, J. and Casdagli, M. (1991). Embedology, J. Stat. Phys. 65, 579.
74. Broomhead D.S. and King G.P. (1986). Extracting qualitative dynamics from experimental data. Physica D 20, 217
75. Albano, A.M., Muench, J., Schwartz, C., Mees, A.I. & Rapp, P.E. (1998).Singular value decomposition and the Grassberger-Procaccia algorithm. Phys. Rev. A, 38, 3017.
76. Casdagli, M. (1991). Chaos and deterministic versus stochastic nonlinear modeling, J. Roy. Stat. Soc. 54, 303
77. Kugiumtzis, D. (1996). State space reconstruction parameters in the analysis of chaotic time series – the role of the time window length. Physica D, 95,13.
78. Fraser A. M. and Swinney, H. L. (1986). Independent coordinates for strange attractors from mutual information, Phys. Rev. A 33, 1134.
79. Kennel, M.B., Brown, R. & Abarbanel, H.D.I. (1992). Determining embedding dimension for phase-space reconstruction using geometrical construction. Phys. Rev. A, 45, 3403.
80. Fraser A. M. and Swinney, H. L. (1986). Independent coordinates for strange attractors from mutual information, Phys. Rev. A 33, 1134.
81. Oum, Y., Oren, S., and Deng, S. (2006). Hedging quantity risks with standard power options in a competitive wholesale electricity market. Special Issue on Applications of Financial Engineering in Operations, Productions, Services, Logistics and Management Naval Research Logistics, 53:697-712.
82. Vehvilainen I., Applying mathematical finance tools to the competitive Nordic electricity market. Helsinki University of Technology, Institute of Mathematics, Research Reports. Teknillisen korkeakoulun laitoksen tutkimusraporttisarja, Espoo 2004.
83. Karatzas, I., Shreve, S. (1998). Methods of Mathematical Finance. Springer.
84. Hurst, H. E. (1965). Long-Term Storage: An Experimental Study. Constable, London.

85. Weron, R. and Przybylowicz, B. (2000). Hurst analysis of electricity price dynamics. Physica A, 283, 462-468.
86. E. E. Peters, Fractal Market Analysis: Applying Chaos Theory to Investment and Economics, Wiley, New York, 1994.
87. Di Mateo, T. (2007). Multi-scaling in finance. Quantitative Finance, 7:21-36.
88. Barabasi, A.L., and Vicsek, T. (1991). Multifractality of self-affine fractals. Physical Review A, 44:2730-2733.
89. Di Mateo, T., Aste, T., Dacorogna, M.M. (2005). Long-term memories of developed and emerging markets: Using the scaling analysis to characterize their stage of development. Journal of Banking & Finance, 29:827-851.
90. Pozzi, F., Aste, F., and Di Mateo, T. (2012). Exponential smoothing weighted correlations. The European Physical Journal, June 2012, 85:175
91. Peng, C-K, Buldyrev, S.V., Havlin, S. et al., (1994). Mosaic organization of DNA nucleotides. Phys. Rev. E49:1685-1689
92. Stanley, H.E., Amaral, L.A.N., Carning, D, Gopokrishnan, P., Lee, Y., Liu, Y. (1999). Econophysics: Can physicists contribute to the science of economics? Physica A, 269 (1999), 156-169.
93. Weron, R. (2002b) Measuring long-range dependence in electricity prices. In: Empirical Science of Financial Fluctuations, H. Takayasu (ed.).Springer Tokyo, pp. 110-119.
94. Geweke, J. and Porter-Hudak, S. (1983). The estimation and application of long memory time series models, Journal of Time Series Analysis 4, 221-238.
95. Weron, R. (2002a) Estimating long-range dependence: Finite samples properties and confidence intervals, Physica A 312, 285-299.
96. Anis, A. A. and Lloyd, E.H.(1976) The expected value of the adjusted rescaled Hurst range of independent normal summands, Biometrica 63, 283-298.
97. Shreve, S. (2004). Stochastic Calculus for Finance II. Continuous-Time Models. Springer Finance.
98. Eydenland, A., Wolyniec, K. (2003). Energy and Power Risk Management. John Wiley & Sons, Inc., Hoboken, New Jersey.
99. Benth, F. E., Benth, J. S., & Koekebakker, S.(2008). Stochastic modeling of electricity and related markets. Singapore: World Scientific.
100. Black, F., Scholes, M. (1973). J. Pol. Econ., 3,637.
101. Merton, R. (1973). Bell. J. Econ. Man. Sci., 4,141.
102. Eckmann J.-P. and Ruelle, D. (1985). Ergodic theory of chaos and strange attractors, Rev. Mod. Phys. 57, 617
103. Stoop R. and Parisi, J. (1991). Calculation of Lyapunov exponents avoiding spurious elements, Physica D 50, 89
104. Kantz, H. (1994). A robust method to estimate the maximal Lyapunov exponent of a time series, Phys. Lett. A 185, 77
105. Rosestein, M. T., Collins, J.J. & De Luca, C.J. (1993). A practical method for calculating largest Lyapunov exponents from small data sets. Physica D, **65,** 117. (1994). Reconstruction expansion as a geometry-based framework for choosing proper delay times. Physica D, **73**, 82.
106. Banks, J., Dragan, V., and Jones, A. (2003). Chaos: A Mathematical Introduction. Cambridge University Press.
107. Iseri, M., Calgar, H., and Calgar, N. (2008). A model proposal for the chaotic structure of Instabul stock exchange. Chaos, Solitons and Fractals **36**(5), 1392-1398.
108. Kyrtsou C, and Terraza M., (2002). Stochastic chaos of ARCH effects in stock series? International Review of Financial Analysis 11(4), 407-431.
109. Orzesko, W. (2008). The new method of measuring the effects of noise reduction in chaotic data. Chaos, Solitons and Fractals 38, 1355-1368.
110. Wolf, A., Swift, J. B., Swinney, H. L., and Vastano, J.A.(1985). Determining Lyaunov exponents from a time series. Physica D 16, 285-317.
111. Rosenstein, M.T., Collins, J. J., and De Luca, C. J. (1993). A practical method for calculating largest Lyapunov exponents from small data sets. Physica D **65**(1), 117-34.
112. Nychka, D., Ellner, S., Gallant, R., and McCaffrey, D. (1992). Finding chaos in noisy systems. Journal of the Royal Statitstical Society B **54**, 399-426.
113. McCaffrey, D., Ellner, S., Gallant, R., and Nychka, D. (1992). Estimating the Lyapunov exponent of a chaotic system with nonparametric regression. Journal of the American Statistical Association **87**(419), 682-695.

114. Gencay, R., and Dechert, W. D. (1992). An algorithm for the n-Lyapunov exponents of an n-dimensional unknown dynamical system. Physica D **59**(1), 142-157.
115. Brock, W. A., and Sayer, C. L.(1988). Is the business cycle characterized by deterministic chaos? Journal of Monetary Economics **22**(1), 71-90.
116. Nychka, D., Elmer, S., Bailey, B. (1997). Chaos with confidence: asymptotics and application of local Lyapunov exponents. American Mathematical Society, 115-133.
117. Bailie, R.T., Bollerslev, T., Mikkelsen, H.O., (1996). Fractionally integrated generalized autoregressive conditional heteroskedasticity. J. Econ. 74, 3-30.
118. BenSaïda Almed (2015). A practical test for noisy chaotic dynamics, SoftwareX 3-4 (2015) 1-5
119. BenSaida A., Litimi H. (2013). High level chaos in the exchange and index markets. Chaos solitons Fractals , 2013; 54:90-95.
120. Petrella, A., Sapio, A. (2012). Assessing the impact of forward trading, retail liberalization, and white certificates on the Italian wholesale electricity prices. Energy Policy 40 (2012) 307-317.
121. C. Tsalis, Journal of Statistical Physics 52, 479 (1988).
122. Balasis, G., Daglis, I.A., Papadimitriou, C., Kalimeri, M., Anastasiadis, A. and Eftaxias C. (July 17 2008). Dynamical complexity in Dst time series using non-extensive Tsallis entropy. Geophysical Research Letters, Vol. 35, L14102, doi:10.1029/2008GL034743, 2008
123. Gradojevic,N., Garic M. (2015). Predicting Systemic Risk with Entroping Indicators.
124. Sheraz, M., Dedu, S., Preda V. (2014). Entropy Measures for Assessing Volatile Markets. 2nd International Conference 'Economic Scientific Research – Theoretical, Empirical and Practical Appoaches', ESPERA 2014, 13-14 November 2014, Bucharest, Romania.
125. Gencay, R., Gradojevic N. (September 2006). Crash of 87. Was it Expected? Aggregate Market Fears and Long Range Depedence.
126. Uritskaya, O., Serletis, A. (2008). Quantifying multiscale inefficiency in electricity markets. Energy Economics, 30 (2008) 3109-3117.
127. Weron, R., Przybylowicz, B. (2000). Hurst analysis of electricity price dynamics. Physica A, 283, 462-468
128. Di Mateo, T., Aste, T., Dacorogna, M.M. (2003). Scaling behaviors in differently developed markets. Physica A: Statistical Mechanics and its Applications, 324:183-188